\documentclass{osa-article}
\usepackage{amsmath,bm,color,graphicx,braket,enumitem}
\journal{osajournal}

\articletype{Research Article}

\newcommand{\Tr}{\mathop{\mathrm{Tr}} \nolimits}
\newcommand{\openone}{\leavevmode\hbox{\small1\normalsize\kern-.33em1}}

\newcommand{\matriz}[1]{\mathsf{#1}}
\newcommand{\var}[1]{\mathrm{Var} ( #1 )}

\journal{osajournal}

\articletype{Review Article}

\begin{document}

\title{Quantum concepts in optical polarization}

\author{Aaron~Z.~Goldberg,\authormark{1,2},  
Pablo de la Hoz,\authormark{3,4} 
Gunnar~Bj\"{o}rk,\authormark{5}
Andrei~B.~Klimov,\authormark{6}
 Markus~Grassl,\authormark{7,2}, 
Gerd~Leuchs\authormark{2,8}, and 
Luis~L.~S\'{a}nchez-Soto\authormark{2,4,*}}

\address{
\authormark{1} Department of Physics, University of Toronto, Toronto ON, M5S 1A7, Canada\\
\authormark{2} Max-Planck-Institut f\"ur die Physik des Lichts, 91058 Erlangen, Germany \\
\authormark{3} School of Physics and Astronomy, University of St Andrews, St Andrews KY16 9SS, UK \\
\authormark{4}Departamento de \'Optica, Facultad de F\'{\i}sica,
Universidad Complutense, 28040~Madrid, Spain \\
\authormark{5} Department of Applied Physics,  Royal Institute of Technology (KTH), SE106~91  Stockholm, Sweden\\
\authormark{6} Departamento de F\'{\i}sica,  Universidad de  Guadalajara, 44420~Guadalajara, Jalisco, Mexico\\
\authormark{7} International Centre for Theory of Quantum Technologies, University of Gda\'{n}sk, 80~308 Gda\'{n}sk, Poland \\
\authormark{8} Institute of Applied Physics, Russian Academy of Sciences, 603950 Nizhny Novgorod, Russia}

\email{\authormark{*}lsanchez@fis.ucm.es} 


\begin{abstract}
We comprehensively review the quantum theory of the polarization properties of light. In classical optics, these traits are characterized by the Stokes parameters, which can be geometrically interpreted using the Poincar\'e sphere. Remarkably, these Stokes parameters can also be applied to the quantum world, but then important differences emerge: now, because fluctuations in the number of photons are unavoidable, one is forced to work in the three-dimensional Poincar\'e space that can be regarded as a set of nested spheres. Additionally, higher-order moments of the Stokes variables might play a substantial role for quantum states, which is not the case for most classical Gaussian states. This brings about important differences between these two worlds that we review in  detail. In particular, the classical degree of polarization produces unsatisfactory results in the quantum domain. We compare alternative quantum degrees and put forth that they order various states differently. Finally, intrinsically nonclassical states are explored and their potential applications in quantum technologies are discussed.
\end{abstract}

\pagestyle{plain}

\section{Introduction}

Polarization, the vectorial aspect of light, is of paramount importance for a proper understanding of the physical world and continues to be the subject of much fundamental research today. Manipulating polarization is also crucial for  applications: in many instances, it is a key measurement variable; whereas, in other cases, it  is a source of noise whose control is imperative. The subject is so relevant that several monographs~\cite{Shurcliff:1962aa,Clarke:1971aa,Kliger:1990aa,Collett:1993aa,Azzam:1996aa,Huard:1997aa,Brosseau:1998aa,Pye:2001aa,Damask:2004aa,Collett:2005aa,Goldstein:2011aa,Ghatak:2012aa,Gil:2016aa} and review papers~\cite{McMaster:1961aa,Brosseau:2006aa,Gil:2007aa,Brosseau:2010aa,Brown:2011aa} are entirely devoted to it; the interested reader can find therein extensive information, including historical surveys of our understanding of polarized light.

Far from its source, any freely-propagating monochromatic electromagnetic field can be considered to a good approximation as a plane wave, with its electric field lying in a plane perpendicular to the direction of propagation. This simple observation is at the root of the notion of polarization: the endpoint of the electric field of such a wave traces in time a well-defined curve that is, in general, an ellipse.

The polarization ellipse is a simple amplitude description of polarized light, but it cannot be directly measured. In 1852, Stokes~\cite{Stokes:1852rt} pointed out that polarization can be specified by four intensity parameters~\cite{Wolf:1954aa,McMaster:1954ve,Walker:1954aa,Barakat:1987aa} that  can be easily measured~\cite{Hecht:1970aa,Berry:1977aa,Boyer:1979aa,Schaefer:2007ly,Azzam:2016aa}. In addition, they lead in a natural way to the Poincar\'e sphere~\cite{Poincare:1889aa}, in which the polarization state is characterized by two angles  directly related to the parameters of the polarization ellipse. This provides us with an elegant geometrical picture in which to analyze the effect of polarization transformations.

These arguments apply only to ideal plane waves. In practice, however, the fields with which one deals in optics exhibit some randomness. The chaotic nature of the light emission process requires then a statistical description. Actually, the rapid time fluctuations of the field cannot be discerned by any detector and one should consider instead the correlations of the field at different space-time points. From this viewpoint, polarization is closely related to coherence theory~\cite{Wolf:2007aa,Refregier:2007aa,Martinez-Herrero:2009aa}. In a naive picture, while a definite ellipse represents complete polarization, partial polarization arises by the rapid and random succession of different   ellipses. In more quantitative terms, the Stokes parameters become random variables and one must deal with a probability distribution on the Poincar\'e sphere.

On the experimental side, polarization of light is a robust characteristic that can be efficiently manipulated using modest equipment without introducing more than marginal losses. It is thus not surprising that this is often the preferred degree of freedom  for encoding information, as one can convince oneself by looking at some recent cutting-edge experiments, including quantum key distribution~\cite{Muller:1993aa}, quantum dense coding~\cite{Mattle:1996wd}, quantum teleportation~\cite{Bouwmeester:1997nx}, rotationally invariant states~\cite{Radmark:2009oq}, phase superresolution~\cite{Resch:2007kl}, and weak measurements~\cite{Dixon:2009hc}. This seems to call for a full theory of polarization in quantum optics.

The fact that the Stokes parameters can be immediately translated into the quantum realm was noticed in the seminal work of Fano~\cite{Fano:1949aa}, and discussions on the resulting Stokes operators can be found in old textbooks (see, e.g.,\cite{Jauch:1955aa,Akhiezer:1965aa}), including their connection with the spin of the photon~\cite{Falkoff:1951aa}.  At this quantum level, no field state can have definite values of the three Stokes operators, for they do not commute and any sharp simultaneous measurement of these quantities is thus precluded.  In physical terms, this means that there is no state with a well-defined polarization ellipse, much in the same way as one cannot assign a definite trajectory to a particle.  The unavoidable fluctuations imply that the points on the Poincar\'e sphere lose their meaning. This establishes a first major difference with the classical description and is at the origin of many nonclassical features, the most tantalizing of which is perhaps polarization squeezing~\cite{Chirkin:1993dz,Korolkova:2002fu,Luis:2006ye,Mahler:2010fk,Chirkin:2015aa}.

On the other hand, classical polarization is often restricted to the mean values of the Stokes parameters. This is justified since most classical light has Gaussian statistics.  However, non-Gaussian states are of utmost relevance in quantum optics, so that higher-order moments of the Stokes operators come into play. This opens the quantum world to polarization properties that have not been addressed in the classical domain. The classic degree of polarization cannot include these new phenomena, so it must be generalized to account for these higher-order polarization effects.

A number of results are dispersed in the literature (see~\cite{Luis:2016aa} for a recent review), but we think that a comprehensive account of polarization in quantum optics is missing. This is precisely the goal of this paper. To this end, and to be as self-contained as possible, we begin in Section~\ref{sec:classpol} with a short overview of the basic concepts of the classical theory. In Section~\ref{sec:quanpol} we extend those concepts into the quantum domain, introducing the basic tools of the SU(2) symmetry and underlining the differences with the classical case. Section~\ref{sec:phsp} exploits this symmetry to present the quantum formulation in phase space, which is nothing but the Poincar\'e sphere. This formulation is statistical in nature and offers logical connections between the quantum and classical descriptions, thus enabling a natural comparison between the two.

The advantages of encoding quantum information via polarization ultimately relies on the ability to create, manipulate, and measure polarization states. All of these tasks require a step-by-step verification in the experimental procedures; this is essentially the scope of polarization tomography, which is the subject of Section~\ref{sec:tomo}.

In Section~\ref{sec:desiderata}, we put forward criteria and desiderata for any  measure of polarization.  We examine several proposals and discuss their benefits and shortcomings, showing how they may be modified to avoid potential shortcomings. In particular, we apply the results obtained to various nonclassical states, whose description lies outside any classical framework.  

Section~\ref{sec:complementarity} discusses the connection between quantum complementarity and the phenomenon of partial polarization. Actually, a proper quantum understanding of interference leads us to a new way of looking at optical polarization.

In Section~\ref{sec:unpol} we revisit the notion of unpolarized states.  In classical optics, the field components of unpolarized light are modeled by zero-mean, uncorrelated, stationary Gaussian random processes~\cite{Barakat:1989fp}, which in geometrical terms means that they reduce to the origin of the Poincar\'e sphere.  This is an incomplete characterization, for it overlooks higher-order moments~\cite{Singh:2013ab}.  At the quantum level, the SU(2) invariance fixes once and for all the structure of the density matrix~\cite{Agarwal:1971zr,Prakash:1971fr,Karasev:1993aa,Soderholm:2001ay} and, as a result, all the moments of the Stokes variables. However, one can broaden the idea of unpolarized states up to a given order: a state that lacks polarization information up to that order will be called $M$th-order unpolarized. We explore these states and show how they motivate different levels of what is called hidden polarization~\cite{Klyshko:1992wd,Klyshko:1997yq,Klyshko:1998wd}.  We also exhibit states with extremal higher-order fluctuations and their potential metrological applications. Finally, our conclusions are summarized in Section~\ref{sec:conclusions}.

\section{Polarized light in classical optics}
\label{sec:classpol}

\subsection{Polarization ellipse}

To facilitate comparison with the quantum version, we first briefly survey the basic aspects of polarized light in classical optics. The topic is treated in any textbook~\cite{Born:1999yq} and in the more specific monographs already quoted~\cite{Shurcliff:1962aa,Clarke:1971aa,Kliger:1990aa,Collett:1993aa,Azzam:1996aa,Huard:1997aa,Brosseau:1998aa,Pye:2001aa,Damask:2004aa,Collett:2005aa,Goldstein:2011aa,Ghatak:2012aa,Gil:2016aa}.

Given a point in space, the state of polarization of a light beam that propagates in a fixed direction, say $z$, is given by the temporal evolution of the electric field of the wave, which lies in a plane perpendicular to the propagation direction. We shall be concerned with monochromatic plane waves of frequency $\omega$ and wave vector $\mathbf{k}$.  Let $\mathbf{E}( z, t)$ be the electric field at a point $z$, at time $t$, of  the wave; the components of the electric field are
\begin{equation}
  \label{eq:Efield} 
  E_{H} (z,t) = {E}_{0H} 
  \exp [ - i  ( \omega t - kz + \delta_{H} )  ] \, ,
  \qquad \qquad
  E_{V} (z,t) ={E}_{0V}  
  \exp [ - i  ( \omega t - kz + \delta_{V} ) ] \, .
\end{equation}
The subscripts $H$ (horizontal) and $V$ (vertical) refer to two Cartesian components transverse to $z$ and the coefficients $ {E}_{0H}$ and ${E}_{0V}$ denote the real amplitudes of the corresponding components with phases $\delta_{H}$ and $\delta_{V}$, respectively. Note that the measurable fields are given by the real parts of the complex expressions. It should also be remarked that the monochromatic plane wave used in the following discussions cannot be strictly realized in the experiment; the formalism, however, holds also for quasimonochromatic fields in the paraxial approximation, replacing $k$ and $\omega$ by their respective mean values $\bar{k}$ and $\bar{\omega}$. The theory cannot deal, though, with the polarization of multimode fields~\cite{Karassiov:2007gb}.

To obtain the curve that the tip of the electric field vector describes in time, we eliminate the time variable in Eq.~(\ref{eq:Efield}). After some direct calculations we get
\begin{equation}
  \label{eq:polelip} 
  \left ( \frac{E_{H}}{E_{0H}} \right )^{2} +
  \left ( \frac{E_{V}}{E_{0V}} \right )^{2} 
  -  2 \left ( \frac{E_{H}}{E_{0H}} \right ) 
  \left (  \frac{E_{V}}{E_{0V}} \right )  
  \cos \delta =  \sin^{2} \delta \, ,
\end{equation}
where $\delta = \delta_{V} - \delta_{H}$ is the relative phase between both oscillations and the real representation of the field is used. This is the equation of an ellipse, which degenerates into a straight line or a circle for some particular values of $\delta$. Although we have eliminated the temporal variable, the fields' components $E_{H}$ and $E_{V}$ continue to be time-space dependent, but for monochromatic radiation the amplitudes and phases are constant for all time. This means the polarization ellipse remains fixed as the polarized beam propagates in a linear medium.

\begin{figure}
  \begin{small}
\centerline{\includegraphics[height=7.5cm]{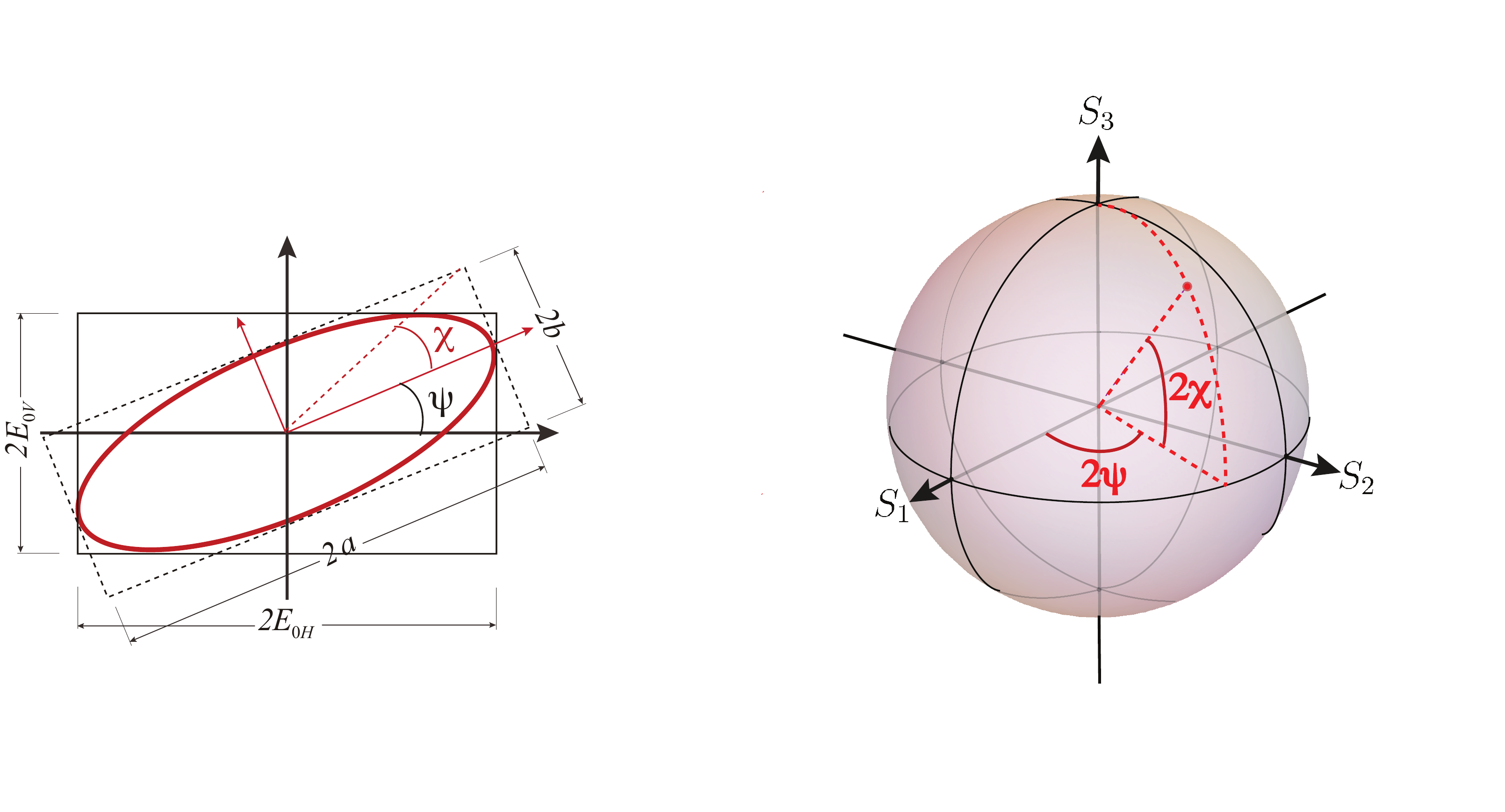}}
  \end{small}
  \caption{ (Left) Polarization ellipse showing the orientation angle $\psi$ and the ellipticity $\chi$, which are functions of the semi-major and semi-minor axes, $a$ and $b$. (Right) Poincar\'e sphere and the parametrization of Stokes parameters  for polarized light in the form of spherical coordinates.}
  \label{fig:ellipse}
\end{figure}

In general, \eqref{eq:polelip} describes a rotated ellipse, with the semi-axes $a$ and $b$ ($a \ge b$). The angle of the semi-major axis, measured counter-clockwise from the positive horizontal axis, is the orientation angle $\psi$ ($0 \leq \psi \leq \pi$), as sketched in Fig.~\ref{fig:ellipse}.  The degree to which the ellipse is oval is described by a shape parameter called ellipticity $\chi$ ($- \pi/4< \chi \le \pi/4$), defined as $\tan \chi = \mp b/a$, the sign distinguishing the two senses in which the ellipse may be described. These angles depend on the amplitude and relative phase:
\begin{equation}
  \tan (2 \psi) = \frac{2E_{0H}E_{0V}}{E_{0H}^2-E_{0V}^2} \, \cos \delta \, , 
  \qquad \qquad
  \sin (2 \chi ) =  \frac{2E_{0H}E_{0V}}{E_{0H}^2+E_{0V}^2} \, \sin \delta \, .
  \label{eq:nosval}
\end{equation}
The orientation angle $\psi$ is zero when  $\delta$ is  $\pi$ or $3\pi/2$: in these situations, \eqref{eq:polelip} describes an ellipse in its standard form. In terms of the amplitudes, the orientation is also zero if $E_{0H}$ ($E_{0V}$) is zero, and  we have vertical (horizontal) linearly polarized light.  For the extreme cases in which $b=0$ we have $\chi=0$ and the light is linearly polarized. In contrast, when $b=a$ we have $\chi=\pm \pi/4$ and the wave is circularly polarized.

\subsection{The polarization matrix and the Stokes parameters}

For our purposes in what follows, it will prove convenient to recast \eqref{eq:Efield} in the form 
\begin{equation}
  \label{eq:Efield2} 
  E_{H} (z,t) = \mathcal{E}_{0}  a_{H} 
    \exp [ - i  ( \omega t - kz )  ]  \, ,
  \qquad \qquad
  E_{V} (z,t) = \mathcal{E}_{0} \, a_{V}  
  \exp [ - i  ( \omega t - kz ) ] \, , 
\end{equation}
where we have absorbed the phases $\delta_{H}$ and $\delta_{V}$ in the dimensionless complex amplitudes $a_{H}$ and $a_{V}$, and we have used a common field amplitude $\mathcal{E}_{0}$ that can be identified as the \textit{electric  field per photon} in the terminology of quantum optics~\cite{Scully:2012aa}. 

The complete polarization information at any plane $z$ is thus conveyed by the  amplitude
\begin{equation}
  \label{eq:Jones}
 \mathbf{A}_{HV} = 
  \left(
    \begin{array}{c}
      a_{H} \\
      a_{V} 
    \end{array}
  \right )
\end{equation}
which is usually called the Jones vector~\cite{Jones:1941aa}.   The subscript $HV$ stresses the basis used to decompose the field amplitudes. Obviously, this vector can be expressed in any other polarization basis, which is obtained from the linear one $\{H, V\}$ by a unitary transformation. In particular, a convenient choice is the circularly polarized basis $\{ + , - \}$ such that
\begin{equation}
  {\mathbf{A}}_{\pm}  = \frac{1}{\sqrt{2}} 
\left ( 
\begin{array}{cc}
1 & i \\
1 & -i
\end{array}
\right ) 
 {\mathbf{A}}_{HV} \, .
\end{equation}

Since only bilinear quantities in the field amplitudes can be measured, it is advantageous to consider the so-called polarization (or coherence) matrix\cite{Fano:1954aa,Wolf:1959aa,Parrent:1960aa,Barakat:1963aa,Barakat:1985aa}
\begin{equation}
  \matriz{J} _{\pm} =  \mathbf{A}_{\pm} \otimes 
\mathbf{A}_{\pm}^{\dagger} 
  =
  \left(
    \begin{array}{cc}
     a_{+}^{\ast} a_{+} &   a_{+} a_{-}^{\ast} \\
     a_{+}^{\ast}  a_{-} &   a_{-}^{\ast} a_{-}
    \end{array}
  \right ) \, ,
\end{equation}
where $\dagger$ stands for the Hermitian conjugate and $\otimes$ represents the Kronecker product.  The elements of the main diagonal of $\matriz{J}_{\pm}$ are real and nonnegative, for they are the intensities of the $+$ and $-$ polarization components (in units of $\mathcal{E}_{0}^{2}$). In consequence, its trace is equal to the average intensity of the wave.  The nondiagonal elements are complex conjugate to each other and, thus, $\matriz{J}_{\pm}$ is a Hermitian matrix. By a direct application of the Cauchy-Schwarz inequality we can show that the determinant of $\matriz{J}_{\pm}$ is nonnegative. Actually, the nonnegativity of its eigenvalues constitutes a complete set of necessary and sufficient conditions for a Hermitian matrix $\matriz{J}_{\pm}$ to be a polarization matrix; i.e., to represent the state of polarization of a light field.

The polarization matrix can be expanded in a basis of the vector space of $2 \times 2$ complex matrices. A natural basis is the one constituted by the identity ($\sigma_{0}= \openone$) plus the three Pauli matrices $\sigma_{i}$ (henceforth, the Latin indices run from 1 to 3, and Greek indices from 0 to 3). The Pauli matrices are Hermitian ${\sigma}_{i} = {\sigma}_i^{\dagger}$, trace orthogonal $\Tr ( \sigma_{i} \, \sigma_{j} )= 2 \delta_{ij}$ and satisfy  $\sigma_i^2= \openone$.  In addition, they are unitary and traceless. The corresponding expansion gives real coefficients~\cite{Fano:1954aa}
\begin{equation}
  S_{\mu} = \frac{1}{2} \Tr ( \matriz{J}_{\pm} \, {\sigma}_{\mu} ) \, ,
\end{equation}
which are known as the Stokes parameters. This can be also compactly expressed as
 \begin{equation}
  \label{eq:stokesalt}
    S_{\mu} = \frac{1}{2}  \mathbf{A}_{\pm}^{\dagger} \, 
\sigma_{\mu} \, \mathbf{A}_{\pm} \, 
\end{equation}
This relation can be inverted, so we can write the Stokes parameters in terms of the elements of the polarization matrix as
\begin{eqnarray}
  \label{eq:stokes}
  & S_{0} = \frac{1}{2} 
    ( a_{+}^{\ast} a_{+} + a_{-}^{\ast} a_{-} ) \, , &     \nonumber\\
  &  & \\
  & S_{1} =\frac{1}{2} ( a_{+}^{\ast} a_{-} + a_{+} a_{-}^{\ast}) \, , 
    \qquad 
    S_{2} = \frac{i}{2}  (a_{+} a_{-}^{\ast} -  a_{+}^{\ast} a_{-} ) \, ,
    \qquad
    S_{3} = \frac{1}{2}  (a_{+}^{\ast} a_{+} - a_{-}^{\ast} a_{-} ) \,  .   &
              \nonumber
\end{eqnarray}

A word of caution is in order here. Our definition (\ref{eq:stokes}) differs in two ways from the standard one in classical optics~\cite{Born:1999yq}. It contains an extra factor 1/2 and the parameters $S_{1}$ and $S_{3}$ are interchanged.  Both modifications are unessential, as the overall structure remains invariant, but they smooth the way for a proper quantum definition.  

The positivity of $\matriz{J}_{\pm}$ immediately implies that
\begin{equation}
  S_{0}^{2} \ge S_{1}^{2} + S_{2}^{2} + S_{3}^{2} \, ,
  \label{eq:strel}
\end{equation}
and the equality holds when $\det \, \matriz{J}_{\pm} = 0$, which is exactly the case for monochromatic light. This suggests the introduction of the polarization state as a point on a spherical surface, called the Poincar\'e sphere~\cite{Poincare:1889aa}, with the coordinates $(S_{1}, S_{2}, S_{3})$. The position of the point on this sphere is characterized by the orientation and ellipticity angles $\psi$ ($0 \leq \psi \leq \pi$) and $\chi$ ($- \pi/4 \leq \chi \leq \pi/4$) such that (see Fig.~\ref{fig:ellipse}) 
\begin{equation}
  \label{S1} 
  S_{1} = S_{0} \cos(2 \chi) \sin(2\psi),  \qquad
  S_{2} =  S_{0} \cos(2 \chi) \cos (2\psi)  \, , \qquad 
  S_{3} =  S_{0}  \sin(2 \chi)  \, .
\end{equation}
The vector $ \mathbf{S} = (S_{1}, S_{2}, S_{3})^{\top}$ ($\top$ being the transpose) is known as the Stokes vector.  With our choice of the circular basis, all linear polarization states lie on the equator of the Poincar\'e sphere, while the circular polarization states are in the north and south pole (circular polarized to the right and the left, respectively). This agrees with the standard use~\cite{Jones:2016aa}. Elliptically polarized states are represented everywhere else on the surface of the sphere.

In practice, for quasimonochromatic light or fields whose components may fluctuate in time in a complicated manner, the amplitudes $a_{\pm}$ depend on time and one should treat them as random variables. One must then take time  averages of the matrix elements of $\matriz{J}_{\pm}$, which turns out to be a $2 \times 2$ covariance matrix whose elements are the second-order moments of the field amplitudes. Under the assumption that they are stationary and ergodic, the time average can also be understood as an ensemble average over different realizations. In that case, one can also study the equivalent version of $\matriz{J}_{\pm}$ in the frequency domain, which is called the cross-spectral density matrix~\cite{Mandel:1995qy}.

To conclude, we mention that a generalization of the Stokes parameters of a random electromagnetic beam has been introduced by Ellis and Dogariu~\cite{Ellis:2004aa} in the space-time domain and has also been studied by Korotkova and Wolf~\cite{Korotkova:2005aa} in the space-frequency domain. Whereas the usual Stokes parameters depend on one spatial variable, one can naturally  introduce two-point Stokes parameters~\cite{Tervo:2009aa}, which are defined in the space-frequency domain by the cross-spectral density matrix that characterizes the correlations at two points. The two-point Stokes parameters depend on two spatial variables and contain additional information about the coherence properties. Another relevant approach is the model of spatial-angular Stokes parameters~\cite{Luis:2005ab,Luis:2005aa},  which are introduced for generalized rays including spatial and angular dependence and allow their evolution during propagation to be considered.

\subsection{Polarization transformations}
\label{subsec:poltrans}

As stressed before, the polarization of the field is determined by its Jones vector $\mathbf{A}$ (henceforth the circular basis is always assumed and, accordingly, we drop the corresponding subscript). Its Euclidean norm
\begin{equation}
  \mathfrak{I} = \Vert \mathbf{A} \Vert^{2} = 
  \mathbf{A^\dagger} \mathbf{A} 
  = \lvert a_{+} \rvert^{2} + \lvert a_{-} \rvert^{2}  = 2 S_{0} \, .
  \label{eq:Nclas}
\end{equation}
is the intensity, measured in units of $\mathcal{E}_{0}^{2}$.

If we rewrite the Stokes parameters in the equivalent form
\begin{eqnarray}
  \label{eq:stokesalt2}
  & S_{0} = \frac{1}{2} 
    ( \lvert a_{+} \rvert^{2} + \lvert a_{-} \rvert^{2} ) \, , &     
    \nonumber\\
  &  & \\
  & S_{1} = \frac{1}{2} ( \lvert a_{45^{\circ}} \rvert^{2} 
   -  \lvert a_{135^{\circ}} \rvert^{2} ) \, , 
    \qquad 
    S_{2} = \frac{1}{2}  ( \lvert a_{H} \rvert^{2} - \lvert a_{V} \rvert^{2} ) \, ,
    \qquad
    S_{3} = \frac{1}{2}  ( \lvert a_{+} \rvert^{2} - \lvert a_{-} \rvert^{2} )  \,  ,  &
      \nonumber
\end{eqnarray}
we can attach a clear operational meaning to them.  Apart from the factor $1/2$, the parameter $S_{0}$ is proportional to the total intensity, while $S_{3}$ represents the excess in intensity between the right- and left-handed circularly polarized components.  The parameter $S_{1}$ has a similar interpretation with respect to linearly polarized components at $45^{\circ}$ and $135^{\circ}$. Finally, the parameter $S_{2}$ is equal to the intensity excess of horizontal over the vertical polarized components.

\begin{figure}
  \centerline{\includegraphics[width=.80\columnwidth]{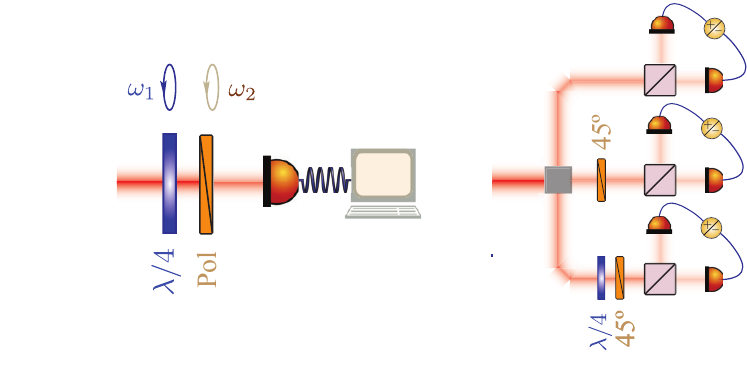}}
  \caption{Typical Stokes polarimetric setups. (Left) Rotating wave plate polarimeter consisting of a rotating quarter wave plate, a linear polarizer and a photodector. (Right) Division-of-amplitude polarimeter, which splits the beam into three separate beams. Each of these beams is analyzed to provide a Stokes parameter. The brown boxes are polarizing beam splitters.}
  \label{fig:polar}
\end{figure}

The assessment of the Stokes parameters can be performed with a variety of methods~\cite{Chen:2020aa} that can be roughly divided into two groups. One relies on a measurement of the complex amplitudes $a_{+}$ and $a_{-}$ (or, equivalently, $a_{H}$ and $a_{V}$) with a coherent dual polarization receiver. The other relies on the measurement of  intensities, which can be performed with different configurations, two of the most popular being sketched in Fig.~\ref{fig:polar}. The first intensity scheme is based on a rotating wave plate and a polarizer, while the second requires splitting the incoming field into three beams, so that each beam is analyzed, obtaining in this way intensities in the bases $+/-$, $H/V$, and $45^{\circ}/135^{\circ}$.

Let us now consider linear transformations of the field, which are represented by complex $2 \times2$ matrices $\matriz{T}$, such that 
\begin{equation}
\label{eq:transT}
    \mathbf{A}^\prime = \matriz{T} \, \mathbf{A} \, .
\end{equation} 
For energy conserving, lossless transformations, the intensity is preserved; i.e., $\mathbf{A}^{\prime \dagger} \mathbf{A}^{\prime} = \mathfrak{I}$. This describes phase plates, which are ubiquitous in polarimetry. Accordingly, the matrix $\matriz{T}$ is unitary and will then be denoted  as $\matriz{U}$.  In fact, we impose that $\matriz{U} \in$ SU(2); i.e., the group of $2 \times 2$ complex unitary matrices with unit determinant~\cite{Cornwell:1984aa}. We can write any $\matriz{U} \in $ SU(2) in terms of a unit vector $\pmb{\mathfrak{n}} $, in the direction of the rotation axis [specified by the spherical angles $(\Theta, \Phi)$], and a rotation angle  $\vartheta$,
\begin{eqnarray} 
\matriz{U} (\pmb{\mathfrak{n}}, \vartheta ) & = & 
\exp(-i \vartheta \;  \pmb{\mathfrak{n}} \cdot \pmb{\sigma}/2) \nonumber \\
& = & \left(
\begin{array}{cc}
\cos (\vartheta /2) - i \sin (\vartheta /2) \cos \Theta & 
- i \sin (\vartheta /2) \sin  \Theta \;  e^{-i \Phi}  \\
 - i \sin (\vartheta /2) \sin  \Theta  \; e^{i \Phi} & 
 \cos (\vartheta /2) + i \sin (\vartheta/2) \cos \Theta 
\end{array}
 \right)  \, .
\label{eq:Ut}
\end{eqnarray}

When the complex amplitudes undergo the unitary transformation \eqref{eq:transT} the Stokes vector $\mathbf{S} \in \mathbb{R}^{3} $ also transforms linearly; that is,
\begin{equation}
    \label{eq:transR}
    \mathbf{S}^\prime = \matriz{R} (\matriz{U})  \, \mathbf{S} \, ,
\end{equation}
where the $3\times 3$ matrix $\matriz{R}(\matriz{U}) \in$ SO(3) is a rotation. Actually, one can check that~\cite{Cornwell:1984aa}
\begin{equation}
  \matriz{U} \, (\mathbf{S} \cdot \pmb{\sigma} ) \, \matriz{U}^\dagger = 
  \left ( \matriz{R} (\matriz{U}) \; \mathbf{S} \right )  \cdot \pmb{\sigma} \, ,
  \label{eq:doucov}
\end{equation}
where $\pmb{\sigma} = (\sigma_{1}, \sigma_{2}, \sigma_{3})^{\top}$. More explicitly, $\matriz{R} (\matriz{U})$ can be chosen as
\begin{equation}
  R(\matriz{U})_{jk} = \frac{1}{2} \, \Tr ( \sigma_{j} \, \matriz{U} \, \sigma_{k} \,
  \matriz{U}^{-1} ) \, .
  \label{eq:cor23}
\end{equation}
The correspondence $\matriz{U} \mapsto \matriz{R}(\matriz{U})$ constitutes a representation of SU(2), and every rotation is an image of some element $\matriz{U}$. Note that both the matrices $\matriz{U}$ and $-\matriz{U}$ lead to the same rotation: $\matriz{R}(\matriz{U}) = \matriz{R}(- \matriz{U})$, so the above correspondence produces a double covering of SO(3) by SU(2).

An arbitrary SU(2) transformation can be generated by any pair of Pauli matrices; for example,  $\sigma_{3}$ and $\sigma_{2}$. Using (\ref{eq:cor23}), one can immediately show that
\begin{equation}
\begin{array}{lll}
\matriz{U}(\mathbf{e}_{3}, \phi) = e^{- i \phi \sigma_{3}/2} = 
\left ( 
\begin{array}{cc}
e^{-i \phi /2} & 0 \\
0 & e^{i \phi /2} 
\end{array}
\right )  & 
\mapsto & 
\matriz{R} (\mathbf{e}_{3}, \phi) = \left ( 
\begin{array}{ccc}
\cos \phi & \sin \phi & 0 \\
- \sin \phi & \cos \phi & 0 \\
0 & 0 & 1  
\end{array}
\right ) \, , \\
& & \\
\matriz{U}(\mathbf{e}_{2}, \theta) = e^{- i \theta \sigma_{2}/2} = 
\left ( 
\begin{array}{cc}
\cos (\theta/2)  & \sin (\theta/2)  \\
- \sin (\theta/2) & \cos (\theta/2)
\end{array}
\right )  & 
\mapsto & 
\matriz{R} (\mathbf{e}_{2}, \theta) = \left ( 
\begin{array}{ccc}
\cos \theta & 0 & \sin \theta \\
0 & 1 & 0 \\
- \sin \theta & 0 & \cos \theta 
\end{array}
\right ) \, .\\
\end{array}
\end{equation}
Therefore, $\sigma_{3}$ generates differential phase shifts between the amplitudes, whereas $\sigma_{2}$ generates rotations around the direction of propagation. It then follows that any energy-conserving polarization transformation can be realized with linear optics: phase plates and rotators.

In many setups one also has to consider energy-nonconserving transformations. They are associated with dichroic devices, which attenuate the field components at different rates. This can be represented by the matrix 
\begin{eqnarray} 
 \matriz{H} (\pmb{\mathfrak{n}}, \eta ) & = & 
e^{-\rho}  \exp( \eta \;  \pmb{\mathfrak{n}} \cdot \pmb{\sigma}/2) \nonumber \\
& = & e^{-\rho} \left(
\begin{array}{cc}
 \cosh (\eta /2) + \sinh (\eta /2) \cos \Theta & 
 \sinh (\eta /2) \sin  \Theta \;  e^{-i \Phi}  \\
  \sinh (\eta /2) \sin  \Theta  \; e^{i \Phi} & 
 \cosh (\eta /2) - \sinh (\eta/2) \cos \Theta 
\end{array}
 \right)  \, ,
 \end{eqnarray}
where $e^{-\rho} = e^{-( \eta_1 + \eta_2 )/2}$ and $e^{\eta} = e^{(\eta_2 - \eta_1)}$   are the isotropic and the relative amplitude transmittances of the dichroic device, with $e^{-\eta_1}$ and $e^{-\eta_2}$ its principal transmittances, major and minor, respectively, and $\pmb{\mathfrak{n}}$ is the axis of diattenuation~\cite{Kim:2000aa}.  Since the isotropic transmittance $e^{-\rho}$ reduces both components at the same rate and does not affect the polarization, it can be neglected as far as polarization is concerned. Therefore, $\matriz{H} (\pmb{\mathfrak{n}},\eta)$ is a Hermitian matrix of unit determinant; that is, it belongs to the group SL($2, \mathbb{C})$  of $2 \times 2$ complex matrices with unit determinant. In fact, it can be denoted by
\begin{equation}
    \matriz{H} ( \pmb{\mathfrak{n}},\eta )=\matriz{U} (\pmb{\mathfrak{n}},0 ) e^{\eta \sigma_3/2} \matriz{U} (\pmb{\mathfrak{n}},0 )^\dagger ,
\end{equation} 
so any polarization transformation, whether or not it conserves energy, can thus be realized with phase plates, rotators, and a single diattenuator.

The matrix $\matriz{H}$ induces a linear transformation on the Stokes parameters~\cite{Han:1997aa,Moreva:2006fv,Savenkov:2009aa,Simon:2010aa}, which transform as a four-vector; namely, 
\begin{equation}
\label{eq:Mueclas}
S_\mu^\prime = \matriz{M}_{\mu \nu} (\matriz{H}) \, S_\nu \,,
 \end{equation}
 where the $4 \times 4$ matrix is known as the Mueller matrix~\cite{Soleillet:1929aa,Mueller:1948aa} and summation over repeated indices is understood henceforth. Following a similar procedure as before, it can be expressed as~\cite{Barut:1977}
 \begin{equation}
  \matriz{M}(\matriz{H})_{\mu\nu} = \frac{1}{2} \, \Tr ( \sigma_{\mu} \, \matriz{H} \, \sigma_{\nu} \,
  \matriz{H}^\dagger ) \, ,
  \label{eq:Mueller from Jones}
\end{equation} 
and it turns out that it is a boost along the axis $\pmb{\mathfrak{n}}$~\cite{Franssens:2015aa,Tudor:2016aa}.  For example, $\matriz{H}_3\equiv \exp ( \eta\sigma_3/2 )$, which describes the attenuation of the $a_+$ and $a_-$ field amplitudes by $e^{- \rho} \,  e^{\eta/2}$ and $e^{- \rho} \, e^{-\eta/2}$, respectively, corresponds to a boost along the $S_3$ axis with rapidity $\eta$.  The most general polarization  transformation can be expressed by the product 
\begin{equation}
\matriz{T} = \matriz{U} \matriz{H} \, ,
\label{eq:SL2C polar decomp}
\end{equation}
which is called the polar decomposition~\cite{Halmos:1982aa}.  Up to the scaling factor $t=\det(\matriz{T})$, general polarization transformations belong to SL($2, \mathbb{C}$) and describe the attenuation of a Stokes vector in addition to its rotations.

It was first noted by Barakat~\cite{Barakat:1963} that transformation matrices $\matriz{T}$ preserve the quadratic form 
\begin{equation}
   M^{2} = S_0^2-S_1^2-S_2^2-S_3^2 ,
\end{equation} 
and are homomorphic to the proper ortochronous Lorentz group, of which SL($2, \mathbb{C}$) is the universal covering~\cite{Barakat:1981}. The invariant $M^{2}$ is similar to the mass of particle, while the Stokes four-vector $S_{\mu}$ is the corresponding four-momentum. The correspondence $\matriz{T} \mapsto \matriz{M}$ constitutes a representation of SL(2,$\mathbb{C}$), and every Lorentz transformation is an image of some element $\matriz{T}$, up to a scaling factor.  

A final note on nondeterministic polarization transformations is warranted. Following such transformations, the coherency matrix may not be rank-one, and thus it may contain more information than the Jones vector $\mathbf{A}$. Fortunately, nondeterministic polarization transformations can be represented by convex combinations of deterministic ones, through \cite{Kim:1987}
\begin{equation}
    \matriz{J}^\prime = 
   \matriz{T}_i \, \matriz{J} \, \matriz{T}_i^\dagger \, .
\end{equation} 
This allows for a description of polarization transformations using only the linear optical elements mentioned above.

\subsection{Degree of polarization}

Before discussing the classical quantification of  polarization, some clarification may be in order. There exist many definitions of the degree of polarization. They differ not only in their mathematical definitions, but also in their basic assumptions. One may treat light as a beam (possibly multimode)~\cite{Qasimi:2007aa}, and sometimes with additional assumptions about its form, such as the Gaussian-Schell model~\cite{Yao:2008aa}; as a scalar field \cite{Vahimaa:2004aa} or as a vectorial field, either two-dimensional (that is appropriate for the far field of a source), or three-dimensional (appropriate for the near field) \cite{Gil:2015aa}. One can define degrees of polarization for all these cases, but quite naturally, one cannot expect them to agree, or even quantify the same physical property. In what follows, we shall only discuss a two-mode model, describing far-field light that is indistinguishable in every respect except for its transverse-field degree of freedom. Thus, our two-mode description is equivalent to a two-dimensional, vectorial description of light and it is only such degrees of polarization we will discuss in the following. 

If the relation between the field components $E_{+}$ and $E_{-}$ (or, equivalently, between the components $E_{H}$ and $E_{V}$) is completely deterministic, the field is fully polarized. For such a pure state (borrowing the terminology from quantum optics), the polarization matrix  satisfies
\begin{equation}
  \matriz{J}_{\mathrm{pol}}^{2}  = \matriz{J}_{\mathrm{pol}} \, . 
\end{equation}
On the other hand, if the components of the field are fully uncorrelated, the off-diagonal elements are zero. If, in addition, the energy is distributed evenly between the horizontal and vertical components, the coherence matrix is proportional to the unit matrix: 
\begin{equation}
  \matriz{J}_{\mathrm{unpol}} = \frac{1}{2} \mathfrak{I}  \, \openone \, ,
\end{equation}
and we say that this field is unpolarized.

This leads to the important decomposition of an arbitrary $\matriz{J}$ into fully polarized and unpolarized parts~\cite{Wolf:1959aa,Parrent:1960aa}
\begin{equation}
  \label{eq:decpu}
  \matriz{J} =\mathfrak{I}  \; 
  [ (1 - \mathbb{P} ) \, \matriz{J}_{\mathrm{unpol}} + 
  \mathbb{P}  \,  \matriz{J}_{\mathrm{pol}} ] \, ,
\end{equation}
where $\mathbb{P}$, called the degree of polarization, can be physically interpreted as the fraction of the total energy contained in the fully polarized part.

Alternatively, $\mathbb{P}$ can be written in a slightly different yet equivalent way~\cite{Born:1999yq}
\begin{equation}
  \label{eq:defPur}
  \mathbb{P}  = \sqrt{\frac{2 \Tr ( \matriz{J}^{2} )}{\Tr^2 (\matriz{J} )} - 1} =   \sqrt{1 - \frac{4 \det (\matriz{J})}{\Tr^2 (\matriz{J})}}    \, ,
\end{equation}
as can be checked by a direct calculation. In the first form, the degree of polarization seems to be intimately linked to $\Tr (\matriz{J}^{2} )$, which, following again a quantum notation, is called the purity. In the second form, it can be immediately related to the eigenvalues of $\matriz{J}$: if we denote them by $\lambda_{+}$ and $\lambda_{-}$, ($\lambda_{+} > \lambda_{-}$), then $  \Tr^{2}( \matriz{J}) = (\lambda_{+} + \lambda_{-})^{2} = \mathfrak{I}^2$ and $\det (\matriz{J}) = \lambda_{+} \lambda_{-}$, so that
\begin{equation}
\label{eq:scram}
  \mathbb{P}= \frac{\lambda_{+} - \lambda_{-}}
  {\lambda_{+} +  \lambda_{-}} \, .
\end{equation}
The action of any unitary transformation on $\matriz{J}$ does not affect its trace. We can thus regard the intensity of a partially polarized field as the sum of two uncorrelated fields components with intensities $\lambda_+$ and $\lambda_-$. If the light is thermal, the lack of correlation implies the statistical independence of both polarization components and the corresponding intensities.

Another equivalent definition of $\mathbb{P}$ is
\begin{equation}
  \label{eq:defPcl}
  \mathbb{P}= \frac{\sqrt{S_{1}^{2} + S_{2}^{2} + S_{3}^{2}}}{S_{0}} \, ,
\end{equation}
representing the intensity-normalized length of the Stokes vector. Stokes vectors on the surface of the Poincar\'e sphere represent totally polarized states (pure states) and Stokes vectors inside the sphere represent partially polarized states (mixed states). The maximally mixed state is at the origin and represents classical unpolarized light.

In the relativistic picture presented in Section~\ref{subsec:poltrans}, pure states correspond to $M=0$. As $M$ increases, the system becomes more and more mixed, until being completely random for $M=S_{0}$. 

To conclude, it is opportune to recall that the concept of the von Neumann entropy $\mathcal{S}$ can be transferred in a direct way to electromagnetic waves by~\cite{Fano:1957aa,Barakat:1983aa,Gase:1994aa,Barakat:1996aa} 
\begin{equation}
  \mathcal{S} = - \Tr (\matriz{J} \, \ln \matriz{J} ) \, .
\end{equation}
This quantity is a measure of the difference in the amount of information between a pure state and a mixed state (both with the same intensity). Using the eigenvalues of $\matriz{J}$, this entropy can be written as~\cite{Barakat:1983aa}
\begin{equation}
\mathcal{S} = - \sum_{i= \pm} \lambda_{i} \, \ln \lambda_{i} \, ,
\end{equation} 
or, equivalently, as
\begin{equation}
 \mathcal{S} = -  \left \{ \tfrac{1}{2} ( 1 + \mathbb{P} ) 
    \ln \left [ \tfrac{1}{2}   ( 1 + \mathbb{P} ) \right ] 
    + \tfrac{1}{2} ( 1 - \mathbb{P} ) 
    \ln \left [ \tfrac{1}{2}   ( 1 - \mathbb{P} ) \right ] \right \} \, .
\end{equation}
Therefore, $\mathcal{S}$ is unequivocally  characterized by $\mathbb{P}$ and
decreases monotonically with $\mathbb{P}$. The maximum $\mathcal{S} = \ln 2$
corresponds to $ \mathbb{P} = 0$; whereas, the minimum $\mathcal{S} = 0$ is
reached for $ \mathbb{P} = 1$.

\section{Polarized light in quantum optics}
\label{sec:quanpol}

\subsection{Stokes operators and the polarization sector}
\label{sec:polsec}

As heralded in the Introduction, the Stokes parameters are fitting analytical tools for treating polarization in the quantum domain because they can easily be translated into truly quantum observables.  We again begin with  a monochromatic plane wave, propagating in the $z$ direction. The quantum field is now characterized by two complex amplitude operators, denoted by $\hat{a}_{+}$ and $\hat{a}_{-}$ (we shall use carets to denote operators). They are the quantum equivalent of the amplitudes $a_{+}$ and $a_{-}$ in (\ref{eq:Efield2}) and obey the bosonic commutation rules (with $\hbar =1$ throughout)
\begin{equation}
  [ \hat{a}_{\mathfrak{s}}, \hat{a}_{\mathfrak{s}^{\prime}}^{\dagger} ] = 
  \delta_{\mathfrak{s} \mathfrak{s}^{\prime}} \openone \, ,  
  \qquad \qquad
  \mathfrak{s}, \mathfrak{s}^{\prime} \in \{+, -\} \, ,
  \label{eq:bosonic commutation relations} 
\end{equation}
which bring about the existence of unavoidable quantum noise precluding their sharp simultaneous measurement. The Stokes operators are then a direct extension of their classical counterparts~\cite{Collett:1970ys,Karassiov:1993lq,Shumovsky:1998aa,Luis:2000ys,Agarwal:2013aa}
\begin{eqnarray}
  \label{eq:Stokop}
  & \hat{S}_{0} = \tfrac{1}{2}  
    ( \hat{a}^{\dagger}_{+} \hat{a}_{+} + 
    \hat{a}^{\dagger}_{-} \hat{a}_{-} ) \, , & \nonumber \\
  & & \\
  &  \hat{S}_{1} = \tfrac{1}{2} 
    ( \hat{a}^{\dagger}_{+}  \hat{a}_{-} + 
    \hat{a}^{\dagger}_{+} \hat{a}_{-} ) \, ,  
    \qquad
    \hat{S}_{2} =  \tfrac{i}{2} 
    ( \hat{a}_{+} \hat{a}^{\dagger}_{-} - 
    \hat{a}^{\dagger}_{+} \hat{a}_{-} ) \, , 
    \qquad
    \hat{S}_{3}  = \tfrac{1}{2}  
    ( \hat{a}^{\dagger}_{+} \hat{a}_{+} - 
    \hat{a}^{\dagger}_{-} \hat{a}_{-} ) \, , & \nonumber
\end{eqnarray}
so that the components of the Stokes vector $\hat{\mathbf{S}} = (\hat{S}_{1}, \hat{S}_{2}, \hat{S}_{3})^{\top}$  satisfy the commutation relations of angular momentum 
\begin{equation} 
[\hat{S}_k,\hat{S}_\ell]= i \epsilon_{k\ell m}  \hat{S}_m \, , 
\qquad \qquad 
[\hat{S}_{0}, \hat{\mathbf{S}} ] = \pmb{0} \,  ,
  \label{eq:CR}
\end{equation}
where $\epsilon_{k\ell m}$ is the Levi-Civita fully antisymmetric tensor; i.e., $\epsilon_{k\ell m}$ is $1$ if $(k, \ell, m)$ is an even permutation of $(1, 2, 3)$, $- 1$ if it is an odd permutation, and $0$ in any index is repeated. Note that $\hat{S}_{0} = \hat{N}/2$, with $\hat{N} = \hat{N}_{+} + \hat{N}_{-}$being the operator for the total number of excitations. Mathematically,  the operators (\ref{eq:Stokop}) are the Jordan-Schwinger representation~\cite{Jordan:1935aa,Schwinger:1965kx} of  SU(2) in terms of bosonic amplitudes. This construction is by no means restricted to polarization, but encompasses many different instances of two-mode problems, such as, e.g., strongly correlated systems, Bose-Einstein condensates, and Gaussian-Schell beams, where the modes can even be spatially separated~\cite{Chaturvedi:2006vn}.

In classical optics, the total intensity is a nonfluctuating quantity, so the Poincar\'e sphere appears as a smooth surface with radius equal to the intensity. In contradistinction, in quantum optics we have that
\begin{equation}
  \hat{\mathbf{S}}^{2} = 
  \hat{S}_{1}^{2} + \hat{S}_{2}^{2} + \hat{S}_{3}^{2} =
  S (S+1) \hat{\openone} ,
\end{equation}
with  $S = S_{0} = N/2$. As fluctuations in the number of photons are unavoidable (leaving aside photon-number states), we are forced to work in a three-dimensional Poincar\'e space that can be regarded as a set of nested spheres with radii proportional to the different photon numbers that contribute to the state: they have been aptly termed as  \emph{Fock layers}~\cite{Donati:2014aa}. One can also introduce \emph{normalized} Stokes operators~\cite{Zukowski:2017aa} for which the effects of intensity fluctuations are removed, making them more sensitive when detecting entanglement. 

The second equation in~(\ref{eq:CR}) expresses in the quantum language that polarization and intensity are separate concepts: the form of the ellipse (polarization) does not depend on its size (intensity).  This fact brings about remarkable simplifications. First, it means that we must handle each subspace with a fixed number of photons $N$ separately. In other words, in the previous onionlike picture of Fock layers, each shell has to be addressed independently.  This can be highlighted if instead of the Fock states $\{ |n_{+}, n_{-} \rangle \}$, which are an orthonormal basis of the Hilbert space of these two-mode fields, we employ the relabeling
\begin{equation}
  | S, m \rangle \equiv |n_{+} = S+m,n_{-} =S-m \rangle \, .
  \label{eq:Spinequiv}
\end{equation}
The relabeled states form the angular momentum basis of common eigenstates of $\{\hat{\mathbf{S}}^2,\hat{S}_3\}$. They span a $(2S + 1)$-dimensional subspace, $\mathcal{H}_{S}$, wherein they act in the standard way
\begin{equation}
  \hat{S}_{z} \, | S, m \rangle  =    
  m | S, m \rangle \, , 
  \qquad \qquad 
   \hat{S}_{\pm} \, | S, m \rangle  =  
  \sqrt{ S (S+1 ) -  m (m \pm 1) } \, 
  |S, m \pm   1 \rangle \, ,
\end{equation}
with $\hat{S}_{\pm} = \hat{S}_{1} \pm i \hat{S}_{2}$ being the raising and lowering operators. The label $S$ always indicates the use of this basis.

Second, for any arbitrary function of the Stokes operators $f (\hat{\mathbf{S}} )$, we have $[ f (\hat{\mathbf{S}} ), \hat{N}] = 0$, so the matrix elements of the density matrix $\hat{\varrho}$ connecting subspaces with different photon numbers do not contribute to $\left\langle f (\hat{\mathbf{S}} )\right\rangle$. Then, it is clear that the moments of any energy-preserving observable (such as $\hat{S}_\mathbf{n}$) do not depend on the coherences between different subspaces.  The only accessible information from any state described by the density matrix $\hat{\varrho}$ is thus its block-diagonal form
\begin{equation}
  \hat{\varrho}_{\mathrm{pol}} =  \bigoplus_S w_{S} \, \hat{\varrho}^{(S)} =
  \sum_{S=0}^{\infty} \sum_{m,m'=-S}^S w_{S} \, \hat{\varrho}^{(S)}_{m m'} \;
  | S,m \rangle \langle S,m^{\prime } | \, ;
  \label{eq:PolSec}
\end{equation}
the blocks off of the diagonal are zero matrices, and the diagonal block matrices are given by $\hat{\varrho}^{(S)}$, which are the density matrices in the $S$th subspaces ($2S$ runs over all the possible photon numbers, i.e. $S = \tfrac{1}{2}, 1, \tfrac{3}{2}, 2, \dots$). We have included the factor $w_{S}$, which is the photon-number distribution, so that all the density matrices $\hat{\varrho}^{(S)}$ are normalized to unit trace. The form $\hat{\varrho}_{\mathrm{pol}}$ is called the polarization sector~\cite{Raymer:2000zt,Marquardt:2007bh,Muller:2012ys} and also the polarization density matrix~\cite{Karassiov:2006hq,Karassiov:2007aa}. Since any $\hat{\varrho}$ and its associated block diagonal form $\hat{\varrho}_{\mathrm{pol}}$ cannot be distinguished in polarization measurements,  we henceforth drop the subscript pol. 

Finally, note that the SU(2) transformations are represented in the subspace   $\mathcal{H}_{S}$ by the operator $\hat{U} (\pmb{\mathfrak{n}},\vartheta) = \exp (- i \vartheta \hat{\mathbf{S}} \cdot \pmb{\mathfrak{n}})$. As we have seen before, the action of this unitary operator induces a rotation, as indicated in Eq.~(\ref{eq:doucov}).

\subsection{Uncertainty relations and polarization squeezing}
\label{sec:uncrel}

The Stokes operators satisfy the standard uncertainty relations of the $\mathfrak{su}(2)$ algebra; viz.,
\begin{equation}
  \label{eq:polsquez1}
  \var{\hat{S}_{k}} \,  \var{\hat{S}_{\ell}}  \ge 
  \lvert \epsilon_{k \ell m} \; \langle \hat{S}_{m} \rangle  \rvert^{2} \, ,
\end{equation}
where $\var{\hat{X}} = \langle \hat{X}^{2} \rangle - \langle \hat{X} \rangle^{2}$ stands for the variance. The noncommutability of these operators precludes the simultaneous sharp measurement of the physical quantities they represent.  Note that the lower bound in Eq.~(\ref{eq:polsquez1}) is state dependent, and, in particular, some of the uncertainty relations  may become trivial; all three variance bounds vanish simultaneously when $\langle \hat{\mathbf{S}} \rangle = 0$. To bypass this problem it is often convenient to use uncertainty relations in terms of sum of variances~\cite{Maccone:2014aa,Zheng:2020aa}, which in our case reads
\begin{equation}
  \var{\hat{\mathbf{S}}}  =
  \var{\hat{S}_{1}}   + \var{\hat{S}_{2}} + \var{\hat{S}_{3}}  
  \geq 2  \langle \hat{S}_0 \rangle \, .
\end{equation}
The states satisfying the equality  $\var{\hat{\mathbf{S}}} = 2 \langle \hat{S}_0 \rangle$ are precisely the SU(2) coherent states~\cite{Arecchi:1972zr,Perelomov:1986ly,Gazeau:2009aa} (see Appendix~\ref{sec:cs} for a brief account of their properties), so they can be rightly considered as the most classical states allowed by the quantum theory.  They live in the subspace $\mathcal{H}_{S}$ and are defined in the standard angular momentum basis by
\begin{equation}
  |S, \mathbf{n} \rangle = \sum_{m=-S}^{S} 
  c_{m} (\mathbf{n}) \, |S,m \rangle \, ,
\end{equation}
where the coefficients $c_{m} (\mathbf{n} )$ follow a binomial distribution peaked around the direction given by the unit vector $\mathbf{n}$ of spherical angles $(\theta, \phi)$: 
\begin{equation}
  c_{m} ( \mathbf{n} )=
  \binom{2S}{S+m}^{1/2} 
  [ \sin ( \theta / 2 )  ]^{S-m} \, 
  [ \cos ( \theta/2)]^{S+m} \, \exp [ - i  ( S+m ) \phi] \, .
  \label{eq:coeff}
\end{equation}

Another issue with the relations (\ref{eq:polsquez1}) is that they are not explicitly SU(2) invariant, which can lead to confusing conclusions.  A way of attaining the desirable SU(2) invariance is by using specific components of the Stokes operators. To this end, we first define the mean-polarization direction by (assuming $\langle \hat{\mathbf{S}} \rangle \neq 0$) 
\begin{equation}
\mathbf{n}_{\parallel} =  \frac{\langle \hat{\mathbf{S}} \rangle}
  {\lvert \langle \hat{\mathbf{S}}\rangle \rvert} \, ,
\end{equation}
and two other orthogonal vectors $\{ \mathbf{n}_{\perp 1}, \mathbf{n}_{\perp 2} \}$ that, together with $ \mathbf{n}_{\parallel}$, define an orthonormal reference frame. If we denote by $ \hat{S}_{\mathbf{n}} =  \hat{\mathbf{S}} \cdot \mathbf{n}$ the projection of the Stokes vector onto the direction $\mathbf{n}$, the commutation relations (\ref{eq:CR}) then read $ [ \hat{S}_{\perp 1}, \hat{S}_{\perp 2} ] = i  \hat{S}_{\parallel}$, which gives only one nontrivial uncertainty relation, namely 
\begin{equation}
  \label{eq:URpp}
  \var{\hat{S}_{\perp 1}} \; \var{\hat{S}_{\perp 2}} \ge 
  \lvert \langle \hat{S}_{\parallel} \rangle \rvert \, ,
\end{equation}
and two trivial ones $\var{\hat{S}_{\perp 1}} \; \var{\hat{S}_{\parallel}} \ge 0$ and $\var{\hat{S}_{\perp 2}} \; \var{\hat{S}_{\parallel}} \ge 0$. The equality in Eq.~(\ref{eq:URpp}) is reached by the eigenvectors of $\hat{S}_{\perp 1} + i \kappa \hat{S}_{\perp 2}$, for any real $\kappa$; these are the so-called intelligent states~\cite{Aragone:1974aa, Aragone:1976aa}.  The SU(2) coherent states are the only states satisfying the three equalities simultaneously, as $\var{\hat{S}_{\parallel}} = 0$ for them.

Another way of ensuring SU(2) invariance is to use the real symmetric $3 \times 3$ covariance matrix for the Stokes variables~\cite{Barakat:1989fp,Rivas:2008ys,Bjork:2012zr}, defined as
\begin{equation}
 \Lambda_{k \ell}  = 
  \tfrac{1}{2} \langle \{ \hat{S}_{k}, \hat{S}_{\ell} \} \rangle - 
  \langle \hat{S}_{k} \rangle \langle \hat{S}_{\ell} \rangle \, ,
\label{eq:gamma}
\end{equation}
where $\{ \cdot , \cdot \}$ is the anticommutator. Note that while the Stokes operators are all Hermitian, the noncommutability makes mixed, nonsymmetric products (such as $\hat{S}_{k} \hat{S}_{\ell}$) non-Hermitian,  also precluding their direct measurement. The symmetrization included in the definition (\ref{eq:gamma}) prevents this problem. In terms of the matrix $\Lambda$, we have 
\begin{equation}
  \var{\hat{S}_{\mathbf{n}}} = 
  \mathbf{n}^{\top} \, \Lambda \, \mathbf{n} \, .
\end{equation}
By construction, $\Lambda_{k \ell} = \Lambda_{\ell k}$, so  $\var{\hat{S}_{\mathbf{n}}}$ is a symmetric quadratic form in $\mathbf{n}$. In consequence, the minimum of $ \var{\hat{S}_{\mathbf{n}}}$ with respect to the direction $\mathbf{n}$ exists and is unique. If we incorporate the constraint $\mathbf{n}^{\top} \cdot \mathbf{n} =1$ as a Lagrange multiplier $\lambda$, this minimum is given by
\begin{equation}
\Lambda \mathbf{n} = \lambda \, \mathbf{n} \, .
\end{equation} 
The admissible values of $\lambda$ are thus the eigenvalues of $\Lambda$ (which are real and non-negative) and the directions minimizing $ \var{\hat{S}_{\mathbf{n}}}$ are the corresponding eigenvectors.

As for any second-rank tensor, we can readily define three invariants: the determinant, the sum of the principle minors, and the trace. In terms of the   eigenvalues $\lambda_{i}$ $i \in \{1, 2, 3 \}$, we can form state-dependent uncertainty relations; viz.,
\begin{align}
0 & \leq \lambda_{1} \lambda_{2} \lambda_{3} 
\leq \tfrac{1}{27} \langle  \hat{S}_{0}^{3} (\hat{S}_{0} + 2)^{3} \rangle \, , \nonumber \\
\hat{S}_{0}^{2} & \leq \lambda_{1} \lambda_{2} + \lambda_{2}\lambda_{3} + 
\lambda_{3}\lambda_{1}  \leq \tfrac{1}{3} \langle 
\hat{S}_{0}^{3} (\hat{S}_{0} + 2)^{3} \rangle \, ,\\
2 \langle \hat{S}_{0} \rangle & \leq \lambda_{1} + \lambda_{2} + \lambda_{3}   \leq \langle  \hat{S}_{0} (\hat{S}_{0} + 2) \rangle \, . \nonumber
\end{align}
Reference~\cite{Shabbir:2016aa} discusses in detail the structure of these relations and their possible saturation.

There is an alternative approach focusing on the sum of the variances of two Stokes operators. The resulting relations 
\begin{equation}
  \label{eq:ursu}
  \var{\hat{S}_{k}} +  \var{\hat{S}_{\ell}} \geq C 
\end{equation}
are referred to as planar uncertainty relations~\cite{He:2011aa,Puentes:2013aa}. The lower bound $C$ is state dependent, but it can be explicitly calculated for $S= 1/2$, $1$, and $3/2$: the results are $1/4$, $7/16$, and $0.600933$, respectively. For large photon numbers, numerical calculations suggest that $C \simeq \tfrac{3}{2} \langle \hat{S}_{0} \rangle^{2/3}$~\cite{Dammeier:2015aa}.

The concept of squeezing is closely linked to the uncertainty relations above.  Squeezing occurs whenever the fluctuations of one of the Stokes operators is below the shot-noise level, which is fixed by SU(2) coherent states. But, unlike in the bosonic case, in which the coherent state variances are equal in any direction, in the case of an SU(2) coherent state the variances of the Stokes operators depend on the direction $\mathbf{n}$. Actually, the parallel component satisfies $\var{\hat{S}_{\parallel}} =0$, so squeezing is primarily determined by the fluctuations of the orthogonal components and alternative squeezing criteria depend on the particular functions used.

For an SU(2) coherent state $\var{\hat{S}_{\bf{n_\bot}}} = S/2$. It is thus sensible to establish that squeezing takes place when the variance of $\hat{S}_{\bf{n_\bot}}$ is less than $S/2$, and the associated squeezing parameter is~\cite{Kitagawa:1993aa}
\begin{equation}
  \xi_{S}^2= \frac{2}{S} \inf_{\bf{n_\bot}} \;
  \var{\hat{S}_{\bf{n_\bot}}}  \, . 
  \label{eq:spsqKU}
\end{equation}
Obviously, $\xi_{S}^2=1$ for the SU(2) coherent states, whereas we may have $\xi_{S}^2<1$; that is, the fluctuation in one direction may be reduced.

In the context of interferometry, a suitable degree of squeezing is the ratio of the phase sensitivity of a general state to that of the SU(2) coherent states~\cite{Wineland:1992aa,Wineland:1994aa,Hillery:1993aa,Agarwal:1994aa,Brif:1996aa}.  For an SU(2) coherent  state, the phase sensitivity is $1/N$  and a direct calculation gives 
\begin{equation}
  \xi_{R}^2= \frac{S} {2 \lvert \langle \bf{S}_{\parallel}\rangle \rvert^2}
\inf_{\bf{n_\bot}}   \var{S_{\bf{n_\bot}}} \, . 
  \label{eq:spsqW}
\end{equation}
There are a few other squeezing parameters that are discussed in great detail in a recent comprehensive review~\cite{Ma:2011aa}.

The idea of polarization squeezing  can be extended to the simultaneous fluctuations of two Stokes components, say $k$ and $\ell$, as
\begin{equation}
  \var{\hat{S}_{k}} + \var{\hat{S}_{\ell}} < 
  \sqrt{\langle \hat{S}_{k} \rangle^{2} + \langle \hat{S}_{\ell}
    \rangle^{2}} \, ,
\end{equation}
which is referred to as planar squeezing~\cite{He:2011aa,Puentes:2013aa}.   Alternative results concerning the simultaneous squeezing of two or three~\cite{Prakash:2011aa} Stokes operators have been obtained.

Finally, we mention that uncertainty relations can be assessed using measures of uncertainty other than variance; the most popular alternatives are entropic measures~\cite{Wehner:2010aa}. This  leads to the idea of entropic spin squeezing, which has been considered by several authors~\cite{Abdel-Aty:2002aa,Civitarese:2013aa}. 

When a state spans several photon numbers, we are forced to scrutinize multiple Fock layers. When this happens, we bring to bear an averaged Stokes vector 
\begin{equation}
\langle\mathbf{\hat{S}}\rangle = \sum_{S=0}^\infty w_{S} \;
\Tr [ \hat{\varrho}^{(S)}\mathbf{\hat{S}} ] \, . 
\end{equation}
As a result of this parsing, the squeezing of the state can be much smaller than the corresponding squeezing in the individual Fock layers~\cite{Muller:2016aa}.

Polarization squeezing has been observed in numerous experiments~\cite{Bowen:2002kx,Heersink:2003aa,Glockl:2003aa,Heersink:2005ul,Dong:2007fu,Corney:2008uq,Shalm:2009mi,Iskhakov:2009pi,Andersen:2016aa}. The Kerr effect in fibers is probably one of the most efficient and will be discussed with more detail below in Sec.~\ref{sec:Kerr}. The squeezing can be achieved by a single pass of optical pump pulses on the two polarization axes of a polarization maintaining optical fiber. When compensating for the birefringence inside the fiber the two orthogonal polarized squeezed beams interfere at the output of the fiber. The resulting polarization squeezing can then be determined by a Stokes measurement, as described in Sec.~\ref{sec:tomo}. The simplicity of the setup and very good spatial and spectral overlap of the two interfering beams led to a measured squeezing of around 7 dB.

\subsection{The dark plane}

It is always possible to establish a basis in which only one of the Stokes operators (\ref{eq:Stokop}) has a nonzero expectation value, say \mbox{$\langle \hat{S}_{k} \rangle = \langle \hat{S}_{\ell}\rangle=0$} and $\langle \hat{S}_{m} \rangle \neq 0$.  The only uncertainty inequality thus reads $\var{\hat{S}_{k}} \, \var{\hat{S}_{\ell}} \geq \lvert \langle\hat{S}_{m} \rangle \rvert^{2}$. Polarization squeezing can then be defined as~\cite{Korolkova:2002fu,Bowen:2002kx,Schnabel:2003vn}
\begin{equation}
  \var{\hat{S}_{k}} < \lvert \langle\hat{S}_{m} \rangle \rvert 
  < \var{\hat{S}_{\ell}} \, .
  \label{eq:polsq1}
\end{equation}
The choice of the conjugate operators $\{\hat{S}_{k},\hat{S}_{\ell} \}$ is by no means unique; there exists an infinite set of operators $\{ \hat{S}_\perp(\theta), \hat{S}_\perp(\theta+\pi/2) \}$ that are perpendicular to the state's  classical excitation direction $\langle \hat{S}_{m} \rangle$, for which $\langle \hat{S}_\perp(\theta) \rangle = 0$ for all $\theta$. All these pairs exist in the $S_{k}$-$S_{\ell}$ plane, which is called the \emph{dark plane} because it is the plane of zero mean intensity. We can express a generic $\hat{S}_{\perp}(\theta)$ as $\hat{S}_{\perp}(\theta) = \hat{S}_{k} \, \cos \theta + \hat{S}_{\ell}
\, \sin \theta$, $\theta$ being an angle defined relative to
$\hat{S}_{k}$.  Condition (\ref{eq:polsq1}) is then equivalent to
\begin{equation}
  \var{\hat{S}_\perp (\theta_{\mathrm{sq}})} < 
\lvert \langle \hat{S}_{0} \rangle \rvert  < 
  \var{\hat{S}_\perp ( \theta_{\mathrm{sq}} + \pi/2 )} \, ,
  \label{eq:polsq2}
\end{equation}
where $\hat{S}_\perp(\theta_{\mathrm{sq}} )$ is the squeezed parameter and $\hat{S}_\perp( \theta_{\mathrm{sq}} + \pi/2 )$ the antisqueezed parameter.

Many experiments use circularly polarized light, which fulfills $\langle \hat{S}_{1} \rangle = \langle \hat{S}_{2} \rangle = 0$, $\langle \hat{S}_{3} \rangle = \alpha^{2}$. In this case the dark plane is exactly the $S_1$-$S_2$ plane and $\langle \hat{a}_{+} \rangle = \alpha$ and $\langle \hat{a}_{-} \rangle = 0$.  Expressing the fluctuations of $\hat{\mathbf{S}}$ in terms of the noise of the circularly polarized modes $\delta \hat{a}_{\pm}$ and assuming $ \lvert \langle \delta \hat{a}_{\pm} \rangle \rvert \ll \alpha$, we find~\cite{Corney:2008uq}
\begin{equation}
  \delta \hat{S}_\perp (\theta) =  \alpha \, \delta\hat{X}_{-} (\theta) = 
  \alpha [ \delta \hat{X}_{H} (\theta ) +   
  \delta \hat{X}_{V} (\theta +  \pi/2) ] \, , 
\label{eq_polsq_darkmode}
\end{equation}
where $\hat{X}_{H} = (\hat{a}_{H} e^{-i \theta} + \hat{a}_{H}^{\dagger} e^{{i \theta}} )/\sqrt{2}$ is the rotated quadrature for the $H$ mode, and an analogous expression for the $V$ mode. On the other hand, since $\delta \hat{N} = \alpha ( \delta\hat{a}_{+} + \delta \hat{a}^\dagger_{+} ) = \alpha \, \delta\hat{X}_{+}$, the intensity exhibits no dependence on the dark mode.  In consequence, the condition (\ref{eq:polsq2}) can be recast as
\begin{equation}
  \var{\hat{X}_{-} (\theta )}  < 1 \, ;
  \label{eq_sq_equivalence}
\end{equation}
that is, polarization squeezing is equivalent to vacuum quadrature squeezing in
the orthogonal polarization mode. This is also seen by considering that the sphere can locally be replaced by its tangent plane since $S \simeq \alpha^{2}$; i.e., for bright states, the Poincar\'e sphere has a large enough radius such that the curvature is locally negligible and the projection in the $S_1$-$S_2$ dark plane is equivalent to a rescaled canonical $x$-$p$ quadrature
phase space. 

\subsection{Higher-order fluctuations}

Squeezing refers  to the behavior of the second-order moments of the Stokes operators. As indicated before, higher-order fluctuations play a crucial role in the quantum domain. To deal with them, it is convenient to use the so called irreducible tensorial sets~\cite{Fano:1959ly,Silver:1976aa,Blum:1981ya,Varshalovich:1988ct,Manakov:2002aa}, a basic concept in the quantum theory of angular momentum. For a fixed spin $S$, these operators (also called polarization operators) are defined as
\begin{equation}
  \label{Tensor}
  \hat{T}_{Kq}^{(S)} = \sqrt{\frac{2 K +1}{2 S +1}}
  \sum_{m,  m^{\prime}= -S}^{S} C_{Sm, Kq}^{Sm^{\prime}} \,
  |  S , m^\prime \rangle \langle S, m | \, ,
\end{equation}
with $C_{S_{1}m_{1}, S_{2} m_{2}}^{Sm}$ denoting the Clebsch-Gordan coefficients~\cite{Varshalovich:1988ct} that couple a spin $S_{1}$ and a spin $S_{2}$ to a total spin $S$ and vanish unless the usual angular momentum coupling rules are satisfied: $m_1+m_2=m$, $ 0 \leq K \leq 2S$, and $ -K\leq q \leq K$. 

According to the properties of the Clebsch-Gordan coefficients, $K$ takes the values $0, 1,2,\dots, 2S$, giving rise to $(2S+1)^2$ polarization operators that constitute a basis for the space of linear operators acting on $\mathcal{H}_{S}$. This is guaranteed by the property 
\begin{equation}
  \Tr \, [ \hat{T}_{Kq}^{(S)}\hat{T}_{K'q'}^{(S)\dagger} ] = 
  \delta_{S S'}\delta_{K K'}\delta_{q q^{\prime}} \, .
  \label{eq:Tortho}
\end{equation}
Polarization operators are, in general, non-Hermitian.  But, due to
symmetry properties, for every fixed $S$ they satisfy the relation
\begin{equation}
  \hat{T}_{Kq}^{(S)\dagger} = (-1)^q \; \hat{T}_{K-q}^{(S)} \, .
  \label{eq:Therm}
\end{equation} 
Most importantly, they have the correct transformation properties under SU(2) transformations~\cite{Varshalovich:1988ct}.

Although the definition of $\hat{T}_{Kq}^{(S)}$ might look a bit unfriendly, the essential observation for what follows is that the operators $\hat{T}_{Kq}^{(S)}$ are proportional to the $K$th powers of the generators of SU(2), so they are intimately linked to the moments of the Stokes variables.  Actually, one can recast Eq.~(\ref{Tensor}) as
\begin{equation}
  \label{eq:multi}
  \begin{array}{lll}
    \hat{T}_{00}^{( S )}   =  \displaystyle
    \frac{1}{\sqrt{2 S + 1}} \hat{\openone} \, , & & \\
     & & \\
    \hat{T}_{10}^{( S )}   =   \displaystyle
    \frac{\sqrt{3}}{\sqrt{( 2 S + 1 ) (S+1) S}} \,    \hat{S}_{z} \, ,
    \qquad  \qquad &
    \hat{T}_{1\mp 1}^{( S )}  =  \displaystyle 
    \frac{\sqrt{3}}{\sqrt{( 2 S + 1 ) (S+1) S}}  \, 
    \hat{S}_{\pm} \, , \\
     & & \\
    \hat{T}_{20}^{{(S)}}  =  \textstyle{\sqrt{\frac{C}{6}}}  
   (3\hat{S}_{z}^{2}- \hat{S}^{2}) \, , &
    \hat{T}_{2\mp 1}^{( S )}  =   \textstyle{\sqrt{\frac{C}{2}}} \, 
    \{ \hat{S}_{z},  \hat{S}_{\pm} \}  \, ,  &
    \hat{T}_{2 \mp 2}^{( S )}  =   \textstyle{\sqrt{\frac{C}{2}}}   \, 
   \hat{S}_{\pm}^{2} \, , 
  \end{array}
\end{equation}
with $C=30/[(2S + 3)(2 S + 1) (2S-1) (S+1)]$, and so on.

The expansion of the density matrix $\varrho^{(S)}$ in polarization operators reads
\begin{equation}
  \label{rho1}
  \hat{\varrho}^{(S)} =  \sum_{K= 0}^{2S} \sum_{q=-K}^{K} 
  \varrho_{Kq}^{(S)} \,   \hat{T}_{Kq}^{(S)} \, ,
\end{equation}
where the corresponding expansion coefficients
\begin{equation}
  \varrho_{Kq}^{(S)}=\Tr [
    \hat{\varrho}^{(S)}\hat{T}_{Kq}^{(S)\dagger} ]
  \label{eq:Multipoles}
\end{equation}
are known as state multipoles and contain all the information about the state, but arranged in a manifestly SU(2)-invariant form. Apart from their theoretical relevance, we will show in Sect.~\ref{sec:dv} that they can be experimentally determined using simple measurements. 

In order to represent a physical state, the density operator must have unit trace, be Hermitian, and be positive definite. These conditions impose some restrictions on the expansion coefficients.  The unit trace fixes the value of the monopole, the only spherical tensor that is not traceless
\begin{equation}
  \varrho_{00}^{(S)} = \frac{1}{\sqrt{2S+1}} \, ,
  \label{eq:monop}
\end{equation}
so, in a way, the monopole is trivial.  Hermiticity imposes the symmetry $\varrho_{K-q}^{(S)} = ( -1)^{q} \, \varrho_{Kq}^{(S) \ast}$. The  positive semidefiniteness of $\hat{\varrho}^{(S)}$ forces constraints on the multipoles, which can be expressed as
\begin{equation}
  \sum_{q=- K}^{K}  \lvert \varrho_{Kq}^{(S)} \rvert^{2} \leq  C_{K}^{(S)} \, .
  \label{eq:cons}
\end{equation}
For the simplest case of $S=1/2$, we have $C_1^{(1/2)}=1/2$, and for $S=1$, we have $C_{1}^{(1)}=1$  and   $C_{2}^{(1)}=2/3$. The general structure of the allowed ranges of the multipoles is quite complicated and can be seen in Refs.\cite{Band:1971aa} and ~\cite{Kryszewski:2006aa}.  

The dipole $\varrho^{(S)}_{1q}$ is the first-order moment of $\hat{\mathbf{S}}$ and thus corresponds to the classical picture of polarization, in which the state is represented by the average value of $\hat{\mathbf{S}}$.  A complete characterization of the state demands the knowledge of all the multipoles to all orders. This implies measuring the probability distribution of $\hat{\mathbf{S}}$ in all directions, and then performing an integral inversion, which turns out to be a hard task.  However, in most realistic cases, only a finite number of multipoles are needed and then the reconstruction of the $K$th multipole entails measuring along only $(2K + 1) \ll (2S+1)^2$ independent directions, as we shall see in Section~\ref{sec:tomo}.

\subsection{Quantum polarization transformations}

The Stokes operators obey the same transformation rules as their classical counterparts in \eqref{eq:Mueclas}
\begin{equation}
    \hat{S}_\mu^\prime=  \matriz{M}_{\mu\nu} \, \hat{S}_\nu \, ,
\end{equation}
where we have assumed again summation over repeated indices.  SU(2) transformations leave $\hat{S}_0$ unchanged while rotating the vector of  Stokes operators $\hat{\mathbf{S}}$, and are immediately seen to describe Mueller matrices corresponding to phase shifters and wave plates. To describe other optical elements, such as polarizers, transformations beyond SU(2) are required ~\cite{Goldberg:2019}.

Deterministic polarization transformations that, unlike SU(2) transformations, do not conserve energy must include loss.  Consider the Lorentz boost $\matriz{H}_3$ describing diattenuation of the two classical field amplitudes $a_\pm$. A quantum description of this process $\hat{a}_{\pm} \mapsto e^{-\rho\pm\eta/2}\hat{a}_{\pm}$ does not preserve the commutation relations listed in \eqref{eq:bosonic commutation relations}, and so does not represent a valid trace-preserving quantum transformation. Rather, one must introduce loss channels $\hat{b}_\pm$ into which some of the light from modes $\hat{a}_\pm$ can be coupled, per
\begin{equation}
    \hat{a}_{\pm} \mapsto e^{-\rho\pm\eta/2}\hat{a}_{\pm}+\sqrt{1-e^{-2\rho\pm\eta}}\hat{b}_\pm.
\end{equation}
Combining this transformation with the act of ignoring the initially-unpopulated loss modes $\hat{b}_\pm$ leads to the desired transformation of the polarization modes $\hat{a}_\pm$. A left-circular polarizer, for example, corresponds to the boost $\eta \mapsto\infty$ that maintains $\rho=\eta/2$. The entirety of the $-$ polarization component is transmitted into an inaccessible mode beyond the polarization Hilbert space.

This analysis is readily extended to a general SL($2, \mathbb{C}$) transformation
\begin{equation}
   \left(
    \begin{array}{c}
      \hat{a}_{+} \\
      \hat{a}_{-} 
    \end{array}
  \right ) \mapsto \matriz{T} 
  \left(
    \begin{array}{c}
      \hat{a}_{+} \\
      \hat{a}_{-} 
    \end{array}
  \right ),
  \label{eq:SL2C on operators}
\end{equation} 
where $\matriz{T}$ is given by \eqref{eq:SL2C polar decomp}, since all SL($2,\mathbb{C}$) polarization transformations can be realized using rotation operations supplemented by a single diattenuation operation. More insight is gained by considering these general transformations in an enlarged Hilbert space. Again considering a vacuum mode $\hat{b}$, the unitary transformation [$\matriz{U}\in$SU(3)]
\begin{equation}
   \left(
    \begin{array}{c}
      \hat{a}_{+} \\
      \hat{a}_{-} \\
      \hat{b}
    \end{array}
  \right ) \mapsto \matriz{U}\left(
    \begin{array}{c}
      \hat{a}_{+} \\
      \hat{a}_{-} \\
      \hat{b}
    \end{array}
  \right ) \, ,
\end{equation} 
achieves \eqref{eq:SL2C on operators} while maintaining global conservation of energy. The $2 \times 2$ matrices representing SL($2, \mathbb{C}$) correspond to a projection of a $3\times 3$ matrix representation of SU(3), and the study of such projections from the perspective of random matrices~\cite{Zyczkowski:2000} may provide insight into the statistics of quantum polarization transformations. All deterministic polarization transformations can thus be realized as photon-number-conserving operations $\matriz{U}$ on an enlarged Hilbert space.

Convex combinations of deterministic quantum polarization transformations suffice to represent all classical polarization transformations. Depolarization, for example, which corresponds to a loss in the degree of polarization while maintaining $\langle \hat{S}_{0} \rangle$, can be characterized by a weighted sum of SU(2) operations acting on $\hat{\varrho}$~\cite{Rivas:2013}. It can be cast into an SU(2)-invariant master equation~\cite{Klimov:2008depol}, and leads to the decay of the higher-order multipoles at a rate that increases quadratically with multipole rank $K$~\cite{Rivas:2013}. All other quantum channels lead to polarization transformations that are more sophisticated than their classical counterparts.

\section{Phase-space representation of polarization states}

\label{sec:phsp}

\subsection{The Husimi $Q$-function}

In the conventional formulation of quantum optics, a system is described in the language of Hilbert space. However, for many purposes it proves advantageous to use a phase-space formulation, which is  surveyed in a number of books~\cite{Schleich:2001aa,QMPS:2005aa,Schroek:1996aa} and review papers~\cite{Hillery:1984aa,Hansen:1984aa,Lee:1995aa,Ozorio:1998aa,Polkovnikov:2010aa,Weinbub:2018aa}. The idea is to exploit the Weyl correspondence between ordinary $c$-number functions in phase space and quantum operators in Hilbert space.  The SU(2) symmetry inherent to the polarization structure greatly simplifies the task of finding this correspondence. Actually,  Stratonovich~\cite{Stratonovich:1956qc} and Berezin~\cite{Berezin:1975mw}  worked out quasiprobability distributions on the sphere satisfying all the pertinent requirements; this  construction  was later generalized by others~\cite{Agarwal:1981bh,Brif:1998if,Heiss:2000kc,  Klimov:2000zv,Klimov:2008yb} and has demonstrated to be extremely useful in visualizing the properties of spinlike systems~\cite{Dowling:1994sw,  Atakishiyev:1998pr,Chumakov:1999sj,Chumakov:2000le,Klimov:2002cr}.

We do not need this complete machinery (a brief account can be found in  Appendix~\ref{app:phsp}); for our goals it is enough if we concentrate on the Husimi $Q$-function~\cite{Husimi:1940aa},  defined in complete analogy to its counterpart for continuous variables, namely~\cite{Agarwal:1981bh}
\begin{equation}
  \label{eq:defQS}
  Q ( S, \mathbf{n} ) =  \langle S, \mathbf{n} | \hat{\varrho}^{(S)} |
  S, \mathbf{n} \rangle \, ,
\end{equation}
where $\hat{\varrho}^{(S)}$ is the density matrix in the $S$th subspace of the polarization sector (\ref{eq:PolSec}). This $Q$-function is only defined in a subspace with fixed spin $S$. Since the SU(2) coherent states $|S, \mathbf{n} \rangle$ are the only states saturating the uncertainty relation (\ref{eq:polsquez1}), the definition of $Q ( S, \mathbf{n} )$ is quite appealing, for it comprises the projection onto the states that have the most definite polarization allowed by the quantum theory.  The function $Q(S, \mathbf{n} )$ is everywhere nonnegative and properly normalized
\begin{equation}
\frac{2S+1}{4 \pi} \int d\mathbf{n} \; Q( S, \mathbf{n}) = 1 \, ,
\end{equation}
with $d \mathbf{n} = \sin \theta d\theta d\phi$ being the invariant differential element of solid angle. In consequence, it can be interpreted as a genuine probability distribution over the $S$th Fock layer. 

Most states require the full polarization sector as in \eqref{eq:PolSec}. For the total polarization matrix $\hat{\varrho}$, the  $Q$-function can be obtained by summing over all the Fock layers (with the proper normalization)~\cite{Sanchez-Soto:2013cr}
\begin{equation}
  \label{eq:Q12}
  Q ( \mathbf{n})=
  \sum_{S=0}^{\infty} (2S + 1) \;
  Q (S, \mathbf{n} ) \, ,
\end{equation}
which is normalized according to 
\begin{equation}
  \frac{1}{4 \pi}\int_{\mathcal{S}_{2}} d\mathbf{n} \, 
  Q ( \mathbf{n} ) = 1 \, ,
\end{equation}
the integral now being extended to the unit sphere $\mathcal{S}_{2}$. 

A point to be stressed is that (\ref{eq:Q12}) involves only diagonal elements between states with the same number of excitations. Because of the lack of off-diagonal contributions of the form $\langle S, \mathbf{n} | \hat{\varrho} | S^\prime, \mathbf{n} \rangle$ with $S \neq S^\prime$, the total $Q$-function is an average of the $Q$-functions over the Fock layers. The role of the sum over $S$ is to remove the total intensity from the description of the state~\cite{Klimov:2006aa}.

\begin{figure}
  \centerline{\includegraphics[height=5cm]{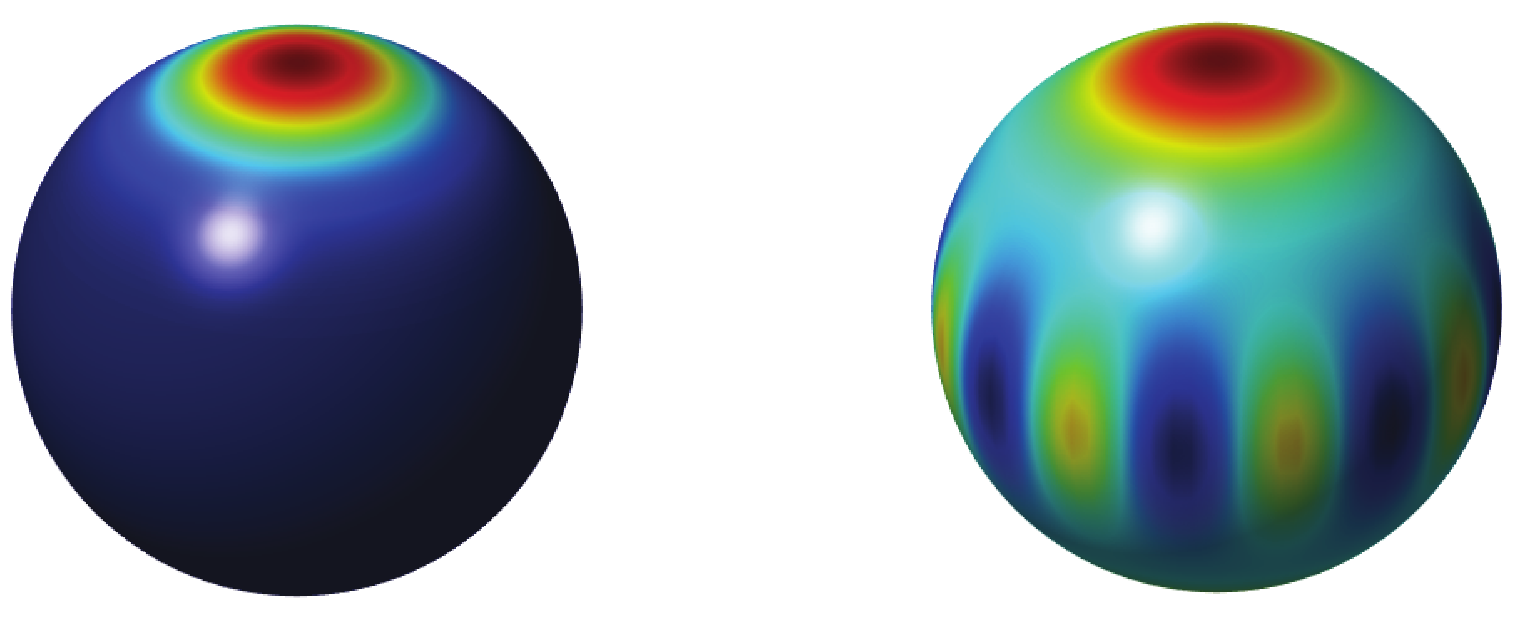}}
  \caption{Density plots of the Husimi $Q$-function for: (left) an SU(2)
    coherent state centered in the north pole, (right) a NOON
    state. In both cases $S=4$. In all the density plots in the paper we use the same colormap that ranges from dark blue (corresponding to the numerical value 0) to bright red (corresponding to the numerical value 1).}
  \label{fig:cohNoon}
\end{figure}

For the distinguished case of pure states, which can be expanded in the
angular momentum basis with coefficients $\Psi_{Sm}$, (i.e., $  | \Psi \rangle = \sum_{S=0}^\infty \sum_{m=-S}^S  \Psi_{S m}\,  | S ,  m \rangle$), the Husimi  $Q$-function takes the form 
\begin{equation}
  Q_{\Psi} ( \mathbf{n} ) =  \sum_{S=0}^\infty (2S+1) \;
  \lvert \langle S, \mathbf{n} | \Psi \rangle \rvert^{2} =
  \sum_{S=0}^\infty (2S+1)  \; \left | 
  \sum_{m=-S}^S c_m( \mathbf{n} )  \, \Psi_{Sm} \right |^2 \, ,
  \label{eq:Qpsgen}
\end{equation}
where the coefficients $c_{m} ( \mathbf{n}) $ are given in Eq.~(\ref{eq:coeff}). 

Let us examine a few illustrative examples. We first consider states with fixed $S$.  For an SU(2) coherent state $|S, \mathbf{n}_{0} \rangle$ one immediately finds
\begin{equation}
  Q_{\mathbf{n}_{0}} ( \mathbf{n} ) =  
  \left [ 
    \tfrac{1}{2} (1 + \mathbf{n} \cdot \mathbf{n}_{0} ) 
  \right ]^{2S} \, ,
  \label{eq:QCS}
\end{equation}
which, as expected, is a distribution strongly peaked around the direction $\mathbf{n}_{0}$.  

The second example is a NOON state~\cite{Lee:2002aa,Dowling:2008aa} 
\begin{equation}
  |\mathrm{NOON} \rangle = \frac{1}{\sqrt{2}}
  (|N,0\rangle - |0,N\rangle) =  \frac{1}{\sqrt{2}}
  (|S,S\rangle - |S,-S\rangle) \, ,
  \label{Eq: NOON}
\end{equation}
expressed first in the Fock and then in the angular momentum basis. We have now 
\begin{equation}
  Q_{\mathrm{NOON}} ( \mathbf{n} ) =  
\tfrac{1}{2} \left [ \sin^{4S} (\theta/2) +  \cos^{4S} (\theta/2) -
 2 \sin^{2S} (\theta/2) \cos^{2S} (\theta/2) \cos(S \phi) \right ] \, .
\end{equation}
This exhibits $2S$ minima equidistantly placed around the equator of the  Poincar\'{e} sphere, as we can see in Fig.~\ref{fig:cohNoon}.

Let us consider two further examples of states spanning the whole polarization sector. The first one is a quadrature coherent state in both modes 
$ |\alpha_{+}, \alpha_{-} \rangle$. Without loss of generality we take the state to be $|\alpha_{+}, 0_{-} \rangle$, for any other state of this family can be generated from this one via an SU(2) transformation.  The decomposition into invariant subspaces $\mathcal{H}_{S}$ reads
\begin{equation}
\label{eq:2mcsdec}
  |\alpha_{+}, 0_{-} \rangle = e^{- \lvert \alpha \rvert^{2}/2}
  \sum_{S=0}^{\infty} \frac{\lvert \alpha \rvert^{2S}}
  {\sqrt{( S+m)! (S-m)!}} e^{i (S+m) \delta} \; |S, m \rangle \,  .
\end{equation}
The final result is
\begin{equation}
  Q_{ \alpha_{+}, 0_{-}} ( \mathbf{n} ) =  
  [ 1 + \lvert \alpha \rvert^2 \cos^2 (\theta/2) ] \;  
  \exp [ - \lvert \alpha^2 \rvert \sin^2 (\theta/2)  ]\, .
  \label{eq:twomodecs}
\end{equation}

The last example is a two-mode squeezed vacuum state~\cite{jmo:1987aa,josab:1987aa}
\begin{align}
\label{eq:TMSVdec}
  |\mathrm{TMSV} \rangle  & =  \hat{S} ( r )  \; |0_{+}, 0_{-} \rangle
  \nonumber \\ 
  & = \frac{1}{\cosh r} \sum_{N} (-1)^{N} (\tanh r)^{N} |N, N \rangle = 
   \frac{1}{\cosh r} \sum_{S} (-1)^{S} (\tanh r)^{S} |S, 0 \rangle   \, , 
\end{align}
again expressed in both Fock and angular momentum bases. The two-mode squeezing operator is $ \hat{S} (r) = \exp [ r ( \hat{a}_{+} \hat{a}_{-} - \hat{a}_{+}^\dagger \hat{a}_{-}^\dagger )/2 ]$ and the squeezing parameter $r$ has been chosen to be real without loss of generality. The resulting Husimi $Q$- function is
\begin{equation}
  \label{Wtwom}
  Q_{ \mathrm{TMSV}}  ( \mathbf{n} ) = \frac{1}{\cosh^{2} r}
  \frac{1 - \tanh^{2} r}{[1 + \tanh^{2} r +   2 \tanh r \, 
    \cos (2 \theta)  ] ^{3/2}} \, .
\end{equation}

\begin{figure}
  \centerline{\includegraphics[height=5cm]{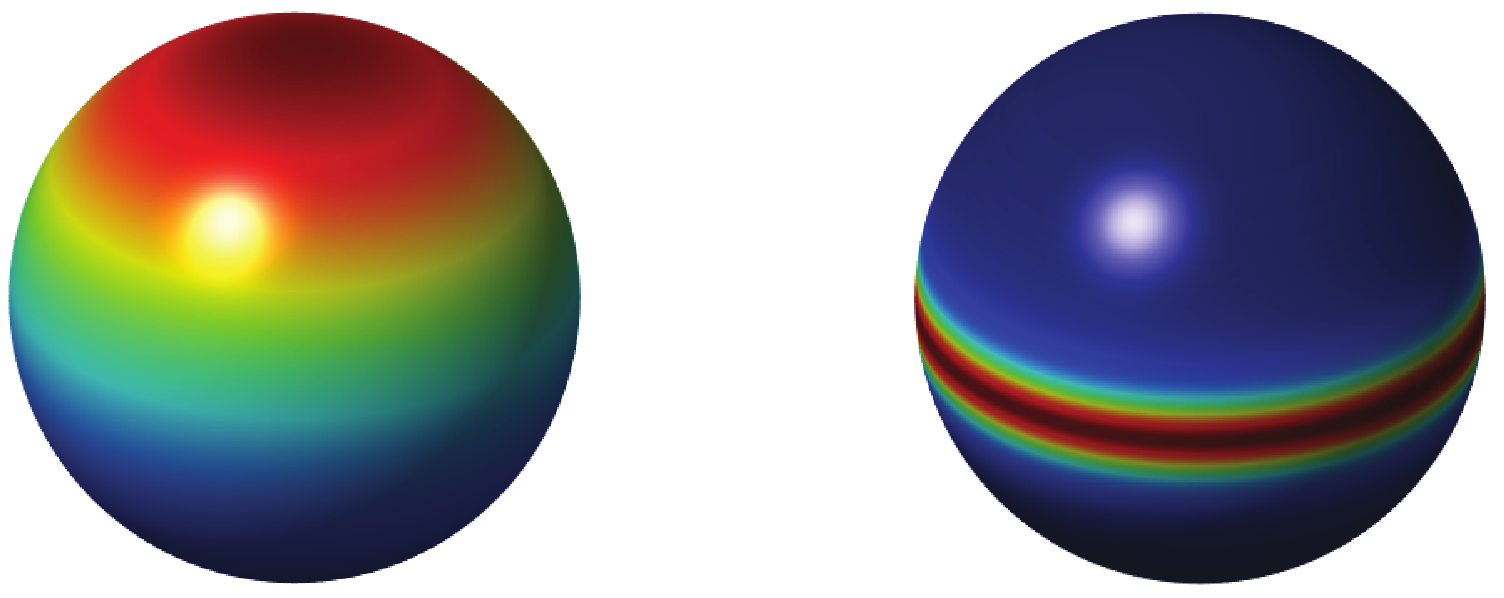}}
  \caption{Density plots of the Husimi $Q$-function for: (left) a two-mode
    coherent state  in which only the $+$ mode is excited, (right) a two-mode squeezed vacuum state with squeezing parameter $r = 0.9$.}
  \label{fig:Wcohsqu}
\end{figure}

In Fig.~\ref{fig:Wcohsqu} we have plotted the $Q$-functions for these last two
states. For the state $|\alpha_{+}, 0_{-} \rangle$, the $Q$-function has a caplike structure highly concentrated around the north pole, as the classical intuition suggests.  For the two-mode squeezed vacuum (we take $r = 0.9$), we see the presence of a Gaussian peak centered at the equator ($\theta = \pi/2$) and independent of $\phi$. This means that we can see this state as arising from a superposition of states with equal amplitudes in each polarization mode, but all possible relative phases $\phi$. These contributing states are concentrated around the equator, and therefore we see a napkin ring there.

To conclude, we recall that polarization can also be understood as arising from the superposition of two perpendicular harmonic modes of the same frequency.  To translate this picture into the quantum realm, we assume that a state can be characterized by the standard two-mode Husimi $Q$-function
\begin{equation}
  \label{prodw1}
  Q ( \alpha_{+}, \alpha_{+} )  = \langle \alpha_{+}, \alpha_{-} | \:
  \hat{\varrho} \:| \alpha_{+}, \alpha_{-} \rangle \, .
\end{equation}
These two pictures can easily be related, because the polarization ellipse needs only three independent quantities to be fully characterized: the amplitudes of each mode and the relative phase between them.  We therefore introduce the parametrization
\begin{equation}
\label{eq:Hopffib}
  \alpha_{+} = r e^{i \zeta} \; e^{i \phi/2}  \cos (\theta/2) \, ,
  \qquad \qquad
  \alpha_{-} = r e^{i \zeta} \; e^{- i \phi/2} \sin (\theta /2)  ,
\end{equation}
where $\zeta$ is a global (irrelevant) phase, the radial variable 
$r^2 = \lvert \alpha_{+} \rvert^2 +  \lvert \alpha_{-} \rvert^2$ represents the total intensity (here considered as a continuous variable), and the
parameters $\theta$ and $\phi$ can be interpreted as the polar and
azimuthal angles, respectively, on the Poincar\'e sphere: $\theta$ describes the relative amount of intensity carried by each mode and $\phi$ is the relative phase between them. In mathematical terms, (\ref{eq:Hopffib}) is an example of a Hopf fibration~\cite{Hopf:1931aa}, a remarkable nontrivial principal fiber bundle that occurs in different situations in theoretical physics in various guises~\cite{Urbantke:2003aa}. 

Both Husimi functions $Q (\alpha_{+}, \alpha_{-} )$ and $Q(\mathbf{n})$ should be closely related. In fact, the latter can be understood as a marginal of the former, as has been worked out~\cite{Klimov:2006aa,Luis:2005ab}. To this end, it is essential to realize that the two-mode quadrature coherent states are expressed in terms of the SU(2) coherent states by~\cite{Atkins:1971aa}
\begin{equation}
  |\alpha_{+}, \alpha_{-} \rangle = e^{- r^{2}/2} 
  \sum_{S=0}^{\infty} \frac{r^{2S} e^{2 i S \zeta }}{\sqrt{(2S)!}} 
  |S, \mathbf{n} \rangle \,. 
  \label{eq:quadang}
\end{equation}
By integrating $Q(\alpha_+,\alpha_-)$ over the intensity variable $r$ we  get the same result as in Eq.~(\ref{eq:Q12}). We will later discuss experimental methods for measuring the Husimi $Q$-function.

As is clear from Eqs.~\eqref{eq:2mcsdec} and \eqref{eq:TMSVdec}, the two-mode coherent states $| \alpha_{+}, \alpha_{-} \rangle $ are separable in the basis $|S,m \rangle$; whereas, the two-mode squeezed vacuum $| \mathrm{TMSV} \rangle $ is nonseparable (or entangled). Despite their classical flavor, SU(2) coherent states are typically entangled in our chosen modal decomposition. Naturally, entanglement properties are closely related to the Hilbert space structure and the nature of observables chosen~\cite{Zanardi:2004aa,Harshman:2007aa,Sanchidrian:2018aa}. It is worth recalling that similar entanglement structures exist in polarization optics, also at the classical level~\cite{Simon:2010ab,Qian:2011kx,Kagalwala:2013aa,Toppel:2014aa,Korolkova:2019aa}.

\subsection{Husimi $Q$-function and higher-order fluctuations}

The Husimi $Q$-function completely encompasses the information that can be obtained from a quantum state; knowledge thereof is tantamount to tabulating the values of all the multipoles $\varrho_{Kq}^{(S)}$. This can be stressed if we rewrite $Q(S, \mathbf{n})$ as~\cite{Klimov:2002cr} (see also Appendix~\ref{app:phsp})
\begin{equation}
  \label{eq:QSU2rj}
  Q (S,  \mathbf{n} ) =  \sqrt{\frac{4 \pi}{2S+1}} 
  \sum_{K=0}^{2S} \sum_{q=-K}^{K}  C_{SS,K0}^{SS} \, \varrho_{Kq}^{(S)} \,
  Y_{Kq}^{\ast} ( \mathbf{n} ) \, ,
\end{equation}
where $ Y_{Kq}^{\ast} ( \mathbf{n})$ are the spherical harmonics. The Clebsch-Gordan coefficient $C_{SS,K0}^{SS}$ has a simple analytical form~\cite{Varshalovich:1988ct}
\begin{equation}
  \label{eq:Cesp}
  C_{SS,K0}^{SS} = \frac{\sqrt{2S+1} \; (2S)!}
  {\sqrt{(2S-K)! \, (2S+1 + K)!}} \, .
\end{equation}

When a state lives in a complete polarization sector, then, by substituting Eq.~(\ref{eq:QSU2rj}) into the general definition Eq.~(\ref{eq:Q12}), the total -function appears as a sum 
\begin{equation}
  \label{eq:QsumK}
  Q ( \mathbf{n} ) =  \sum_{K=0}^{\infty} 
  Q_{K} ( \mathbf{n} ) \, ,
\end{equation}
where each partial multipole component is
\begin{equation}
  \label{eq:QSU2K}
  Q_{K} ( \mathbf{n} ) =
  \sum_{S=\lfloor K/2 \rfloor}^{\infty}  \sqrt{4 \pi (2S+1)} 
  \sum_{q=-K}^{K} C_{SS,K0}^{SS} \, \varrho_{Kq}^{(S)} \; 
  Y_{Kq}^{\ast} ( \mathbf{n} )  \, .
\end{equation}
Here, the floor function $\lfloor x \rfloor$ is the largest integer less than or equal to $x$.  The partial components $Q_K$ inherit the properties of $Q$, but they exclusively contain information about the $K$th moments of the Stokes variables. In this way, Eq.~(\ref{eq:QsumK}) is the appropriate tool for arranging the successive moments.

\begin{figure}
  \centerline{\includegraphics[height=5cm]{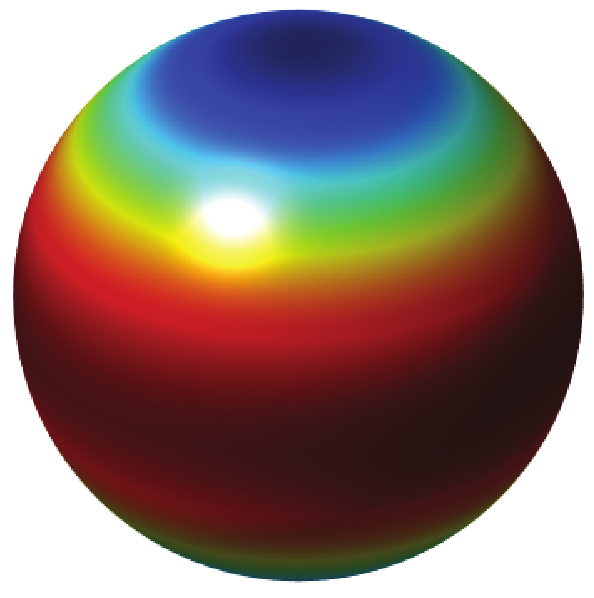}}
  \caption{Density plot of Husimi $Q$-function for the state $|1, 0 \rangle$. Here, the dipolar component ($K=1$ vanishes), so this state lacks first-order information. The features of the Husimi function are then due to the quadrupolar ($K=2$) components of the state.}
  \label{Fig:Qpartial}
\end{figure}

We illustrate this viewpoint with the simple example of the state $| 1_{\mathrm{+}}, 1_{\mathrm{-}} \rangle$. This represents the photon pairs generated in type-II optical parametric down-conversion~\cite{Rubin:1994aa} and is generally viewed as a highly nonclassical state. In the angular momentum basis, the state is $|1, 0 \rangle$  and its $Q$-function can immediately be calculated:
\begin{equation}
  \label{eq:Qt11}
  Q ( \mathbf{n} ) = \sin^{2} \theta  \, .
\end{equation}
The $Q$-function does not depend on $\phi$ and its shape has an equatorial bulge, revealing that the state is highly delocalized on the Poincar\'e
sphere, in agreement with its nonclassical character. The partial components, according to (\ref{eq:QSU2K}), are
\begin{equation}
  \label{eq:Qp11}
  Q_{0} ( \mathbf{n} ) = \frac{1}{3} \, ,
  \qquad \qquad
  Q_{1} ( \mathbf{n} ) = 0 \, ,
  \qquad \qquad
  Q_{2} ( \mathbf{n} ) = \left ( \frac{2}{3} - \cos^{2} \theta \right )  \, .
\end{equation}
The sum of these three terms gives, of course, the result Eq.~(\ref{eq:Qt11}), but there is more information encoded in Eq.~(\ref{eq:Qp11}): the dipolar contribution is absent, which means that this state conveys no first-order information. This is the reason why this was the first state in which hidden polarization was detected~\cite{Usachev:2001ve,Bushev:2001xq}. Figure~\ref{Fig:Qpartial} shows the  $Q$-function for this state.

The expansion \eqref{eq:QSU2rj} can be inverted using the orthonormality of the spherical harmonics. In this way, the multipoles appear as~\cite{Agarwal:1998aa,Bouchard:2017aa}
\begin{equation}
  \varrho_{Kq}^{(S)} = \sqrt{\frac{4\pi}{2S+1}}\frac{1}{C_{SS,K0}^{SS}}
   \int_{\mathcal{S}_2}  d\mathbf{n} \;  Q( \mathbf{n}) \; 
   {Y}_{Kq} (\mathbf{n}) \, .
  \end{equation}
When expressed in the Cartesian basis these multipoles appear in a very transparent way. For example, the three dipole $(\varrho_{1q})$ and the five quadrupole $(\varrho_{2q})$ terms are given, respectively, by
\begin{equation}
  \wp_{q} = \langle n_{q} \rangle \, , \qquad  \qquad 
\mathcal{Q}_{qq^{\prime}}=\langle 3 n_{q} n_{q^{\prime}} - 
\delta_{q q^{\prime}} \rangle \, ,
  \end{equation} 
where the expectation values of a function $f (\mathbf{n})$ are calculated with respect the $Q$-function
\begin{equation}
 \langle f ( \mathbf{n} )\rangle= 
\frac{\int_{\mathcal{S}_2}\; d\mathbf{n} \;
f(\mathbf{n}) \; Q( S, \mathbf{n})}
{\int_{\mathcal{S}_2} d\mathbf{n} \; Q(S, \mathbf{n})} \, .
\end{equation} 
Therefore, the state multipoles appear as the standard ones in electrostatics, but with charge density replaced by $Q(S, \mathbf{n})$  and distances by directions~\cite{Jackson:1999aa}. They are the $K$th directional moments of the state and, therefore, the multipoles resolve progressively finer angular features. The extension to the complete polarization sector is direct.

\subsection{Propagation in a Kerr medium}
\label{sec:Kerr}

The optical Kerr effect refers to the intensity-dependent phase shift that light experiences during its propagation through a third-order nonlinear medium. This leads to a remarkable non-Gaussian operation that has attracted considerable interest due to potential applications in a variety of areas, such as quantum nondemolition measurements~\cite{Braginskii:1968aa,Unruh:1979aa,Milburn:1983aa,Imoto:1985aa,Grangier:1998aa,Sanders:1989aa,Xiao:2008aa}, generation of quantum superpositions~\cite{Milburn:1986aa,Yurke:1986aa,Tombesi:1987aa,Gantsog:1991aa,Wilson-Gordon:1991aa,Tara:1993aa,Luis:1995aa,Chumakov:1999aa,Korolkova:2001aa}, and quantum logic~\cite{Turchette:1995aa,Semiao:2005aa,Matsuda:2007aa,You:2012aa}.  

Special mention must be made of the role that this cubic nonlinearity has played in the generation of squeezed light, which is precisely our interest here. Optical fibers are the paradigm for that purpose~\cite{Levenson:1985aa,Schmitt:1998aa}, although, due to the typically small values of the nonlinearity in silica glass, one needs long propagation distances and high powers to observe nonlinear effects, which brings other unwanted results~\cite{Shelby:1985aa,Elser:2006aa}.

Let us consider the following Hamiltonian
\begin{equation}
\hat{H} = \upchi \hat{a}_{+}^{\dag} \hat{a}_{+} \, 
\hat{a}_{-}^{\dag} \hat{a}_{-} \, ,
\label{Hkerr}
\end{equation}
where $\upchi$ is an effective coupling constant that depends on the third-order nonlinear susceptibility.  This describes the cross-Kerr effect in which a nonlinear phase shift of an optical polarization mode (say, $+$) is induced by the other mode ($-$)~\cite{Agrawal:2001aa}. 

For any state described by the density operator $\hat{\varrho}$, the evolution can be formally written as  $\hat{\varrho} (t) = \exp (-i t  \hat{H} )  \, \hat{\varrho} (0) \, \exp ( i t  \hat{H} )$. By expanding this equation in the two-mode Fock basis, the evolution may, in principle, be tracked. Taking the example of an initially pure, two-mode coherent state $| \Psi (0) \rangle = | \alpha_{+}, \alpha_{-} \rangle$, the resulting time-evolved state is~\cite{Agarwal:1989aa}
\begin{equation}
  \label{eq:evolexaCohst}
 |\Psi (t) \rangle 
  = \exp [ - ( |\alpha_{+}|^2 + |\alpha_{-}|^2)/2 ]   
  \sum_{n_{+}, n_{-} =0}^\infty
  \frac{\alpha_{+}^{n_{+}} \alpha_{-}^{n_{-}} }{\sqrt{n_{+}!\, n_{-}!}} 
  \, \exp (- i \upchi t  n_{+} n_{-} ) |n_{+}, n_{-} \rangle \, .
\end{equation}
The term $\exp (- i \upchi t n_{+} n_{-} )$ arises because of the coupling between the  modes and prevents the state from being factorized into single-mode states; i.e., the state becomes entangled. This exact expression is only of practical use for few-photon states. 

Phase-space methods are especially adapted to this problem. If we employ the two-mode Husimi function $ Q(\alpha_{+},  \alpha_{-})$ and the basic techniques outlined, e.g., in Ref.~\cite{Gardiner:2004aa}, the quantum dynamics can be mapped to the following second-order differential equation,%
\begin{eqnarray}
i \partial_{t} Q &=&\upchi ( |\alpha_{-} |^{2}+1 ) 
\left(\alpha_{+}^{\ast} \frac{\partial Q}{\partial \alpha_{+}^{\ast}} 
- \alpha_{+} \frac{\partial Q}{\partial \alpha_{+}}\right) + 
\upchi ( |\alpha_{+}|^{2}+1 )
\left( \alpha_{-}^{\ast} \frac{\partial Q}{\partial \alpha_{-}^{\ast}} -
\alpha_{-} \frac{\partial Q}{\partial \alpha_{-} }\right)  \nonumber \\ 
&-&  \upchi \left( \alpha_{+}^{\ast} \alpha_{-}^{\ast} 
\frac{\partial }{\partial \alpha_{-}^{\ast}} 
\frac{\partial }{\partial \alpha_{+}^{\ast}} 
- \alpha_{+} \alpha_{-} \frac{\partial }{\partial \alpha_{-}}
\frac{\partial }{\partial \alpha_{+} }\right) Q \, . 
 \label{Q2E} 
\end{eqnarray}
\begin{figure}
  \centering
  \includegraphics[height=5.5cm]{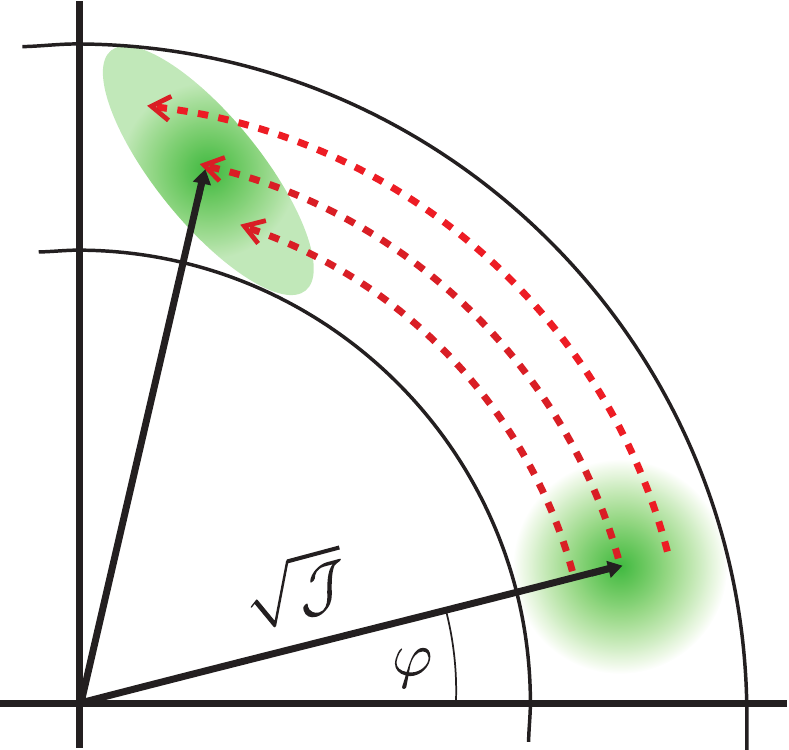}
  \caption{\label{fig:Squeezing} Schematic representation of the effect of a Kerr medium in the phase
    space of a single mode, assuming
    an initially-coherent state.}
\end{figure}

When both modes are initially in strongly-excited coherent states, which defines the semiclassical limit, the $Q$-function dynamics is approximately described by the first two terms of in Eq.~(\ref{Q2E}) (at least, for times $\upchi t \ll 1$). If we introduce polar coordinates for each amplitude $\alpha_{\pm} = \sqrt{\mathcal{I}_{\pm}} e^{i \varphi_{\pm}}$, where $\varphi_{\pm}$ is the polar angle in phase space, and $\mathcal{I}_{\pm}$ is the mode intensity, we find  
\begin{equation}
\partial_{t} Q = \upchi ( \mathcal{I}_{-} + 1 ) \frac{\partial Q}{\partial
\varphi_{+}} + \upchi ( \mathcal{I}_{+} +1 ) 
\frac{\partial Q}{\partial \varphi_{-} } \, .
\label{eq:W_p}
\end{equation}
Since the operator $\partial/\partial \varphi_{\pm}$ generates rotations in phase space,~(\ref{eq:W_p}) reflects that the amplitudes in each mode experience different rotations, with angles proportional to the intensity components of the other mode~\cite{Heersink:2005ul,Rigas:2013aa}. The result is schematized in Fig.~\ref{fig:Squeezing}: the shaded area indicates the region in phase space occupied by the state. For an initial coherent state this area is a circle; the top of the circle corresponds to higher intensity and therefore is more phase shifted than the bottom, resulting in an elliptical noise distribution.

Equation~(\ref{eq:W_p}) can be readily solved:
\begin{eqnarray}
  \label{eq:Wigner-T}
  Q (\mathcal{I}_{+},\varphi_{+} ; \mathcal{I}_{-} ,\varphi_{-} |t) =  
  Q (\mathcal{I}_{+}, \varphi_{+} + (\mathcal{I}_{-} + 1) \upchi t;
  \mathcal{I}_{-} , \varphi_{-}  + (\mathcal{I}_{+}+ 1) \upchi t|0) \, .
\end{eqnarray}
The cross-dependence of the phases on the amplitudes of the other field leads to the mode correlation. These intermodal correlations can be assessed, e.g., in terms of the linear entropy~\cite{Rigas:2013aa}. For an initial two-mode coherent state $|\sqrt{\mathcal{I}_{0+}}, \sqrt{\mathcal{I}_{0-}} \rangle$ the $Q$-function acquires the form 
\begin{equation}
Q (\mathcal{I}_{+}, \varphi_{+} ; \mathcal{I}_{-} ,\varphi_{-} |t)=
\exp \left [ - \lvert  \mathcal{I}_{+} e^{i \varphi_{+} + i (\mathcal{I}_{-} +1) \upchi t} - \mathcal{I}_{0+} \rvert^{2}\right ]
\exp \left [ - \lvert  \mathcal{I}_{-} e^{i \varphi_{-} + i (\mathcal{I}_{+} +1) \upchi t} - \mathcal{I}_{0-} \rvert^{2}\right ]  \, . 
\label{Wt_0}
\end{equation}
At $ t  = 0$ the $Q$-function  is made of two independent Gaussians, while as time goes by the induced mode correlations lead to a non-Gaussian state.

The problem can also be treated in terms of the SU(2) $Q$-function, since in a Fock layer with fixed $S$, the Hamiltonian \eqref{Hkerr} reduces to~\cite{Corney:2008uq}
\begin{equation}
H  = \upchi S_{z}^{2} \, ,
\end{equation}
apart from an unessential constant.  The evolution equation for the SU(2) $Q$-function takes the form~\cite{Klimov:2002cr}
\begin{equation}
\partial _{t} Q = - \upchi ( 2 S \cos \theta \, \partial_{\phi} - 
\sin \theta \, \partial_{\phi} \partial_{\theta}) Q \, ,  
\label{Qee}
\end{equation}%
This equation can be exactly solved by expanding in the basis of the harmonic functions $Y_{Kq}(\theta ,\phi )$. Nevertheless, in case of large photon number $S \gg 1$, we can perform the  semiclassical expansion and find an approximate solution of (\ref{Qee}) by simply neglecting the term with the second derivative; i.e., reducing (\ref{Qee}) to the form%
\begin{equation}
\partial_{t} Q \simeq - 2 \upchi S \cos \theta \, \partial_{\phi} Q \, ,
\end{equation}
whose solution is 
\begin{equation}
Q(\theta ,\phi |t) = Q (\theta ,\phi -\upchi Nt\cos \theta ),
\end{equation}
which describes evolution of every point of the initial distribution along the classical trajectory. Note, however, that this SU(2) approach is valid only for initial states with a fixed number of photons in each mode. 

\section{Polarization tomography}
\label{sec:tomo}

Quantum tomography is the attempt to infer an unknown quantum state from the distinct outcomes of a collection of measurements performed on a finite set of identical copies of the system~\cite{lnp:2004uq,Teo:2015aa}.  What makes polarization special is that  the  density operator contains much more than polarization information; reconstruction of the polarization sector suffices for polarization tomography. 

\begin{figure}
  \centering{\includegraphics[height=5.5cm]{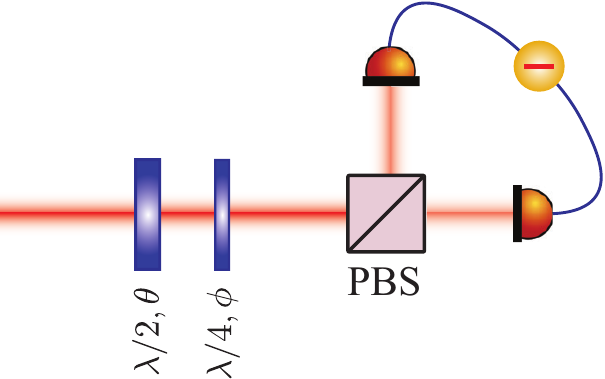}}
  \caption{Typical setup for an efficient Stokes tomography.  The
    modes are  combined on a polarizing beam splitter (PBS)
    and interference between the modes can be adjusted with the
    combination of a half-wave plate ($\lambda/2, \theta$) and 
    a quarter-wave plate ($\lambda/4, \phi$). The polarization 
    states are spatially separated into orthogonal
    components that can be processed independently.}
  \label{fig:Stokessetup}
\end{figure}

A basic setup for any polarization tomography is sketched in Fig.~\ref{fig:Stokessetup}. The state to be characterized is analyzed using a general polarization measurement apparatus consisting of a half-wave plate  $(\lambda /2 ,\theta)$ followed by a quarter-wave plate ($\lambda /4, \phi$) and a polarizing beam splitter (PBS). In physical terms, the wave plates
transform the input polarization, performing the operation
\begin{equation}
\label{eq:dissph}
\hat{D}(\mathbf{n}) =  e^{- i \phi \hat{S}_{3}} e^{- i \theta \hat{S}_{2}}
= \exp \left [ \tfrac{1}{2} \theta (\hat{S}_{+} e^{-i \phi} - 
\hat{S}_{-} e^{i \phi} )\right] \, ,
\end{equation} 
which represents a displacement over the sphere by the unit vector $\mathbf{n}$.  This allows the measurement of different Stokes parameters by projecting onto the basis $| S, m \rangle$.   The two outputs of the photocurrent's sum directly gives the eigenvalue of $\hat{S}_{0}$, while their difference gives the observable $ \hat{S}_{\mathbf{n}}$~\cite{Marquardt:2007bh}. From a practical viewpoint there are two very different tomographic regimes.  

\subsection{Discrete-variable regime}
\label{sec:dv}

In the discrete-variable regime of single, or few, photons one is interested in two-mode states, which for many purposes can be regarded as spin systems. Consequently, the polarization states can be determined from correlation functions of different orders~\cite{White:1999fk,Kwiat:2000rw,James:2001vn,Thew:2002pd,Barbieri:2003ij,Bogdanov:2004bs,Moreva:2006fv,Barbieri:2007uq,Adamson:2010ys,Sansoni:2010zr,Altepeter:2011ly}. Given the small dimensionality of the Hilbert space involved, the state reconstruction can be readily performed.

Altogether, the setup yields the probability distribution for $\hat{S}_{\mathbf{n}}$, from which we can infer the moments
\begin{equation}
\mathfrak{m}_{\ell}^{(S)} (\mathbf{n} ) = \Tr [
\hat{S}_{\mathbf{n}}^{\ell} \, \hat{\varrho}^{(S)} ] \, .
\end{equation}  
For simplicity, we restrict ourselves to a single Fock layer with a fixed number of photons $S$, but everything can be smoothly extended to the whole polarization sector. The moments can be expressed in terms of the multipoles as
\begin{equation}
  \label{eq:lqh}
  \mathfrak{m}_{\ell}^{(S)} (\mathbf{n} ) =   
  \Tr  [ \hat{S}_{3}^{\ell} \,  \hat{D} (\mathbf{n} ) \, 
  \hat{\varrho}^{(S)}  \,  \hat{D}^\dagger  (\mathbf{n})  ]   =  
  \Tr \left [   \hat{S}_{3}^{\ell} \, \sum_{K=0}^{2S} \sum_{q, q^{\prime}=-K}^{K} 
    \varrho_{Kq}^{(S)} \, D_{qq^\prime}^{K} (\mathbf{n}  ) \,
    \hat{T}_{Kq}^{(S)} \right ] \,  ,
\end{equation}
where $D_{q^{\prime} q}^{K} (\mathbf{n} ) $ is a Wigner rotation matrix. This trace can be computed using the machinery of angular momentum, and then the moments connect to the multipoles in quite an elegant way
\begin{equation}
  \label{eq:mrho}
  \mathfrak{m}_{\ell}^{(S)} (\mathbf{n} ) = \sqrt{\frac{4\pi }{2 S + 1}}
  \sum_{K=0}^{\ell} \sum_{q=-K}^{K} \varrho_{Kq}^{(S)}  \,
  f_{K\ell}^{(S)}  \, Y_{Kq} ( \mathbf{n} )  \, ,
\end{equation}
where  $f_{K\ell}^{(S)} = \sum_{m} m^{\ell} C_{Sm,K0}^{Sm}$ ($K \leq \ell$). Given the orthonormality of $Y_{Kq} (\mathbf{n} )$, we can  invert Eq.~(\ref{eq:mrho}) to obtain
\begin{equation}
  \varrho_{Kq}^{(S)} =   \sqrt{\frac{2 S +1}{4\pi}}  
 \frac{1}{ f_{K\ell}^{(S)}}  
\int_{\mathcal{S}^{2}} d\mathbf{n} \;  
\mathfrak{m}_{\ell}^{(S)} (\mathbf{n} )  \; Y_{Kq}^{\ast} (\mathbf{n}) \, .
\end{equation}
The reconstruction of the polarization state thus requires the knowledge of \textit{all} the multipoles; this implies measuring \textit{all} the moments in \textit{all} directions, which proves to be very demanding~\cite{Muller:2012ys}.

\begin{figure}
  \centering{\includegraphics[width=0.65\columnwidth]{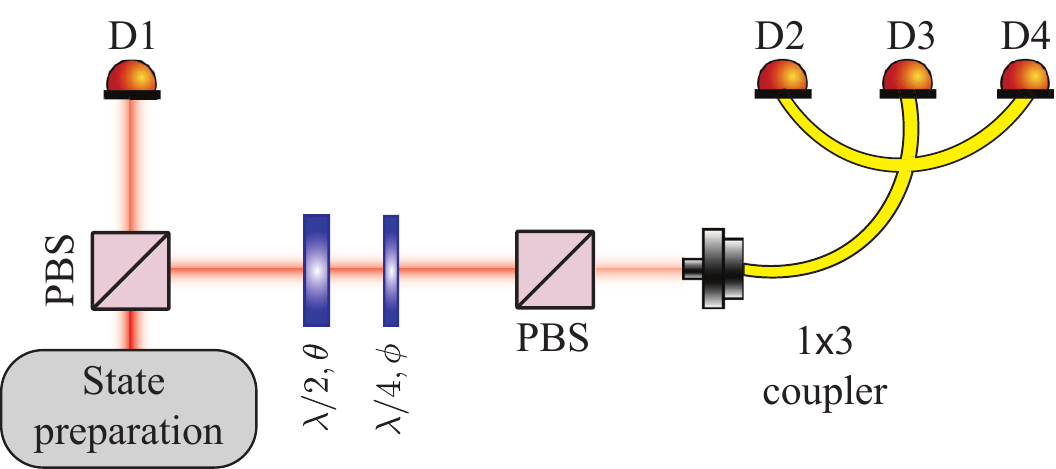}}
  \caption{Experimental scheme for the multipole reconstruction of a three-photon state. The detection of a single photon at detector D1 heralds the state.  Fourfold coincidences with detectors D1, D2, D3, and D4 allow for the required projection measurements. The scheme also works for states with higher photon number.}
  \label{fig:4fold}
\end{figure}

Nonetheless, one can approach the problem in a much more economical way. To determine the $K$th multipole, it is enough to perform a Stokes measurement in $2K+1$ independent directions. The proposal proceeds recursively: first, we measure the first-order moments along the three coordinate axis (or other equivalent ones) and reconstruct $ \varrho_{1q}^{(S)}$  from a linear inversion.  The measurement of the second moments gives us 
\begin{equation}
  \mathfrak{m}_{2} (\mathbf{n} ) = \frac{1}{2 S + 1} f_{02}^{(S)} + 
\sqrt{\frac{4\pi }{2S+1}}    f^{(S)}_{22}  \sum_{q=-2}^{2} 
  \varrho_{2q}^{(S)} \: Y_{2q} (\mathbf{n} ) \, ,
\end{equation}
with $ f_{02}^{(S)} = \tfrac{1}{3}S(S+1)(2S+1)$, $f_{22}^{(S)} = \tfrac{4}{5!} (2S+1) \sqrt{S(2S-1)(S+1)(2S+3)}$, and $f_{12}^{(S)} =0$. We need to fix five optimal directions to invert that system. For example, we can choose the directions as those that maximize the minimum angle between the lines and thus in some sense spread the measurements as much as possible over the Poincar\'{e} sphere~\cite{Conway:1996ys}. The system can then be solved, yielding $ \varrho_{2q}^{(S)}$, and thus all the information needed to characterize the polarization to second order is known.

The process can be continued in this way up to any desired order. Choosing the appropriate directions is, in general, a tricky question if one wants to ensure linear independence, but it has been thoroughly studied~\cite{Hofmann:2004aa,Filippov:2010kx}. In practice, methods such as maximum likelihood estimation are much more efficient for performing that inversion~\cite{lnp:2004uq}. This strategy has been experimentally verified for photon pairs generated in spontaneous parametric down-conversion; i.e., the states $|1_{H},1_{V} \rangle$~\cite{Bjork:2012zr}. A sketch of the experimental setup is shown in Fig.~\ref{fig:4fold}.

A similar scheme has been discussed in Ref.~\cite{Schilling:2010aa}, but instead of  Stokes moments, one measures  $N$th-order intensity moments.  In this way, an optimal measurement of arbitrary-order coherences between the two orthogonally polarized amplitudes can be achieved.

\begin{figure}
  \centering{\includegraphics[width=0.95\columnwidth]{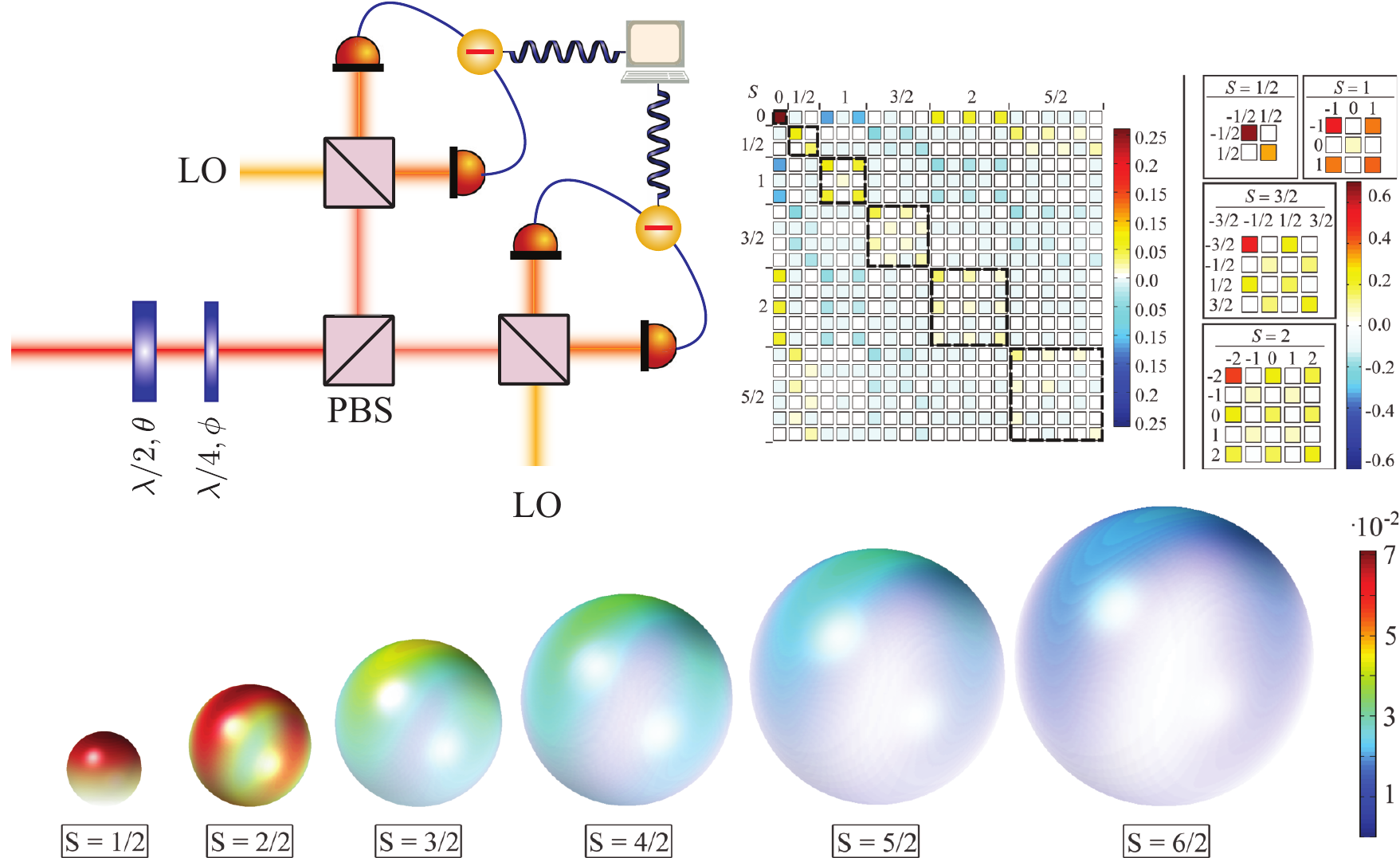}}
  \caption{(Top left) Experimental setup for the characterization of two-mode fields. The polarization states are separated into orthogonal components followed by homodyne tomography. (Top right) Measured polarization sector (black blocks) for the case of a polarization squeezed state with coherent amplitude $\alpha = 1.13$ and squeezing parameter $r= 0.41$. (Bottom) econstructed Husimi functions $Q(S, \mathbf{n})$ of the Fock layers indicated in the insets for the same state.}
  \label{fig:parsing}
\end{figure}

When a state spans a whole polarization sector, one should parse it into Fock layers. The most convenient way is to perform two-mode tomography by characterizing each mode after the polarizing beam splitter with homodyne tomography~\cite{Lvovsky:2009ys}. In this case, it is more convenient to work in the $\{ H, V\}$ basis, so the state to be characterized is of the form $| \Psi \rangle = | \psi_{H}\rangle \otimes | \psi_{V} \rangle$. In Ref.~\cite{Muller:2016aa} this parsing has been done for the case of $| \psi_{H} \rangle = \hat{S} (r ) \hat{D} (\alpha_{H}) | 0_{H}\rangle$ and $| \psi_{V} \rangle = \hat{S} (r ) | 0_{V} \rangle$, with the single-mode squeezing operator $\hat{S}(r)=\exp\left[r\left(\hat{a}^{\dagger2}-\hat{a}^2\right)/2\right]$  and displacement operator $\hat{D}(\alpha)=\exp (\alpha \hat{a}^\dagger-\alpha^{\ast} \hat{a})$. The results, shown in Fig.~\ref{fig:parsing}, confirm in a crystal-clear manner that the Husimi $Q$-function can be experimentally sampled. While these measurements are well known in the context of cold atoms~\cite{Evrard:2019aa}, their use in quantum optics is not widespread.

A similar method was recently proposed in Ref.~\cite{Bayraktar:2016aa}. This  method is based on two  polarization concepts: measurement projections onto the Stokes operator eigenvectors and division of the polarization multipole orders, characterized by $S$. The method assumes photon-number-resolving detectors, and measures not only the Stokes parameters for every $S$, but also the correlations between them. 

\subsection{Continuous-variable regime}

The photodetection in the typical setup described in Fig.~\ref{fig:Stokessetup} can be modeled by the projection operators  $\hat{\Pi}_{m}^{(S)} = |S, m \rangle \langle S, m |$ so that for each direction $\mathbf{n}$ we detect the tomographic probabilities 
\begin{equation}
  \label{cojo}
  w^{(S)}_{m} (\mathbf{n} ) =  
  \Tr [ \hat{\varrho} \,  \hat{\Pi}_m^{(S)} (\mathbf{n}  ) ] =  
  \Tr [ \hat{\varrho} \; \hat{D} (\mathbf{n} ) \,  \hat{\Pi}_m^{(S)}
  \,  \hat{D}^\dagger ( \mathbf{n}   ) ] \, ,
\end{equation}
which correspond to the probabilities of simultaneously detecting $n_H = S+m$ photons in the horizontal mode  and $n_V= S-m$ photons in the vertical one for each direction $\mathbf{n}$.  When the total number of photons is not measured and only the difference $m$ is observed, the available projections reduce to
$ \hat{\Pi}_{m} = \sum_{S = |m|}^\infty |S, m \rangle \langle S, m |$.

The reconstruction in each invariant subspace  $\mathcal{H}_{S}$  can now be carried out exactly since each subspace is essentially equivalent to a spin $S$~\cite{Brif:1999kx,Amiet:1999vn,DAriano:2003ys,Klimov:2002zr,Karassiov:2004xw,Karassiov:2005ss}. One can proceed in a variety of ways, but perhaps the simplest one is to look for an integral representation of the tomograms~\cite{Marquardt:2007bh}
\begin{equation}
  \label{eq:intreptom}
  w_{m}^{(S)} (\mathbf{n} ) = \frac{1}{2 \pi} \int_{0}^{2\pi}
  d\omega \, \Tr [ \hat{\varrho}^{(S)} \, 
  \exp ( i \omega \hat{\mathbf{S}}_{\mathbf{n}} ) ] \, e^{- i m \omega}   \, .
\end{equation}
The probabilities appear as the Fourier transform of the characteristic function for the observable $\hat{\mathbf{S}}_{\mathbf{n}}$. After some manipulations, we find that
\begin{equation}
  \label{uf} 
  \hat{\varrho}^{(S)} =  \frac{1}{4 \pi} \sum_{m=-S}^{S}
  \int_{\mathcal{S}_2} d \mathbf{n}^\prime \  
  w_m^{(S)}  (\mathbf{n}^\prime) \, 
  \mathcal{K} ( \hat{\mathbf{S}}_{\mathbf{n}^\prime} - m ) \, ,
\end{equation}
where the kernel $\mathcal{K} (x) $ is
\begin{equation}
  \label{kersing}
  \mathcal{K}  ( x ) = \frac{ 2 S +1}{4 \pi^2} \int_{0}^{2\pi} d \omega
  \; \sin^2 (\omega / 2 ) \, e^{-i \omega  x} \, .
\end{equation}
Although Eq.~(\ref{uf}) is a formal solution, it is handier to map this density matrix onto the corresponding Husimi $Q$-function. For that purpose, we only need to calculate the matrix elements of the kernel $\mathcal{K} (x)$. The most direct way to proceed is to note that
\begin{equation}
  \label{evalK}  
\langle S, \mathbf{n} | 
 \mathcal{K}   ( \hat{\mathbf{S}}_{\mathbf{n}^\prime} - m ) | S, \mathbf{n}   \rangle  =  \frac{2S+1}{4 \pi^2} \int_{0}^{2\pi} d\omega \;
  \sin^2 ( \omega/2 )  \;  e^{- i m \vartheta} 
 \left [ \cos (\omega /2 ) - i \sin  ( \omega / 2 ) \cos \mathcal{X} \right ]^{2S} \, ,
\end{equation}
where $\cos \mathcal{X} = \mathbf{n} \cdot \mathbf{n}^\prime$.  In the continuous-variable regime we take $S \gg~1$, so  the integral in Eq.~(\ref{evalK}) reduces to $d^2 \delta (x) /dx^2$ evaluated at $x = S \, \mathbf{n} \cdot \mathbf{n}^\prime - m$. Now, $m$ can be
taken as a quasicontinuous variable, and we integrate by parts to obtain
\begin{equation}
  Q (S, \mathbf{n} ) = \frac{2S+1}{4\pi^2} \int_{-\infty }^{\infty} \!
  dm \int_{\mathcal{S}_2} \!  d\mathbf{n}^\prime \, \frac{d^2 w_{m}^{(S)}
    ( \mathbf{n})}{dm^2}  \, \delta ( S \, \mathbf{n} \cdot
  \mathbf{n}^\prime - m ) \, .
\end{equation}
This means that, in the limit of large photon numbers, the inversion reduces to an inverse Radon transform~\cite{Deans:1983aa} of the measured tomograms,
which greatly simplifies the computation of $Q(S,\mathbf{n})$.

\begin{figure}
  \centering
  \includegraphics[width=0.95\columnwidth]{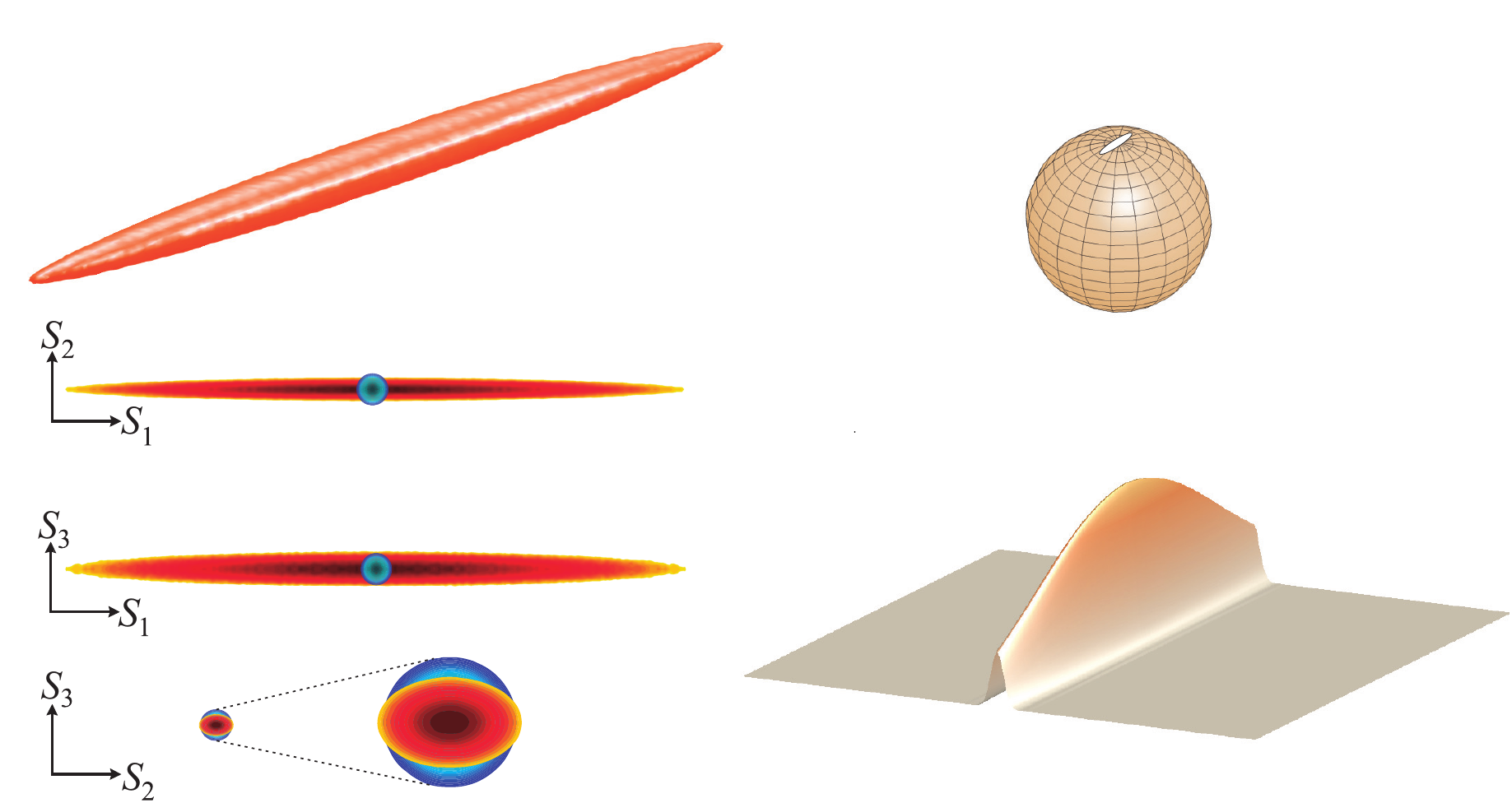}
  \caption{(Left) Isocontour surface (of the level $1/e$ from the
    maximum) of the Husimi function $Q (S, \mathbf{n} )$ for a
    bright squeezed state. We also show cross sections of the $Q$
    function through the three coordinate planes. In blue we show the
    isotropic section for a coherent state, which we scale to unity for
    all the plots. (Right) Reconstructed total Husimi function 
     $Q (\mathbf{n})$ obtained by summing over all the (continuous) Fock layers.}
  \label{fig:pol3D}
\end{figure}

In Fig.~\ref{fig:pol3D} we show an isocontour surface of $Q(S,\mathbf{n} )= \mathrm{constant}$ in the Poincaré space having $S_1$, $S_2$ and $S_3$ as the orthogonal axes for bright squeezed states, as described in Ref.~\cite{Marquardt:2007bh}. The ellipsoidal shape of the state is clearly visible. The antisqueezed direction of the ellipsoid is dominated by excess noise. We also sketch density plots of the projections on the coordinate planes of the previous Husimi $Q$-function.   The projections on the planes $S_1$-$S_2$ and $S_2$-$S_3$ show an additional spreading of the squeezed state in the $S_3$ direction caused by the imperfect polarization contrast in the measurement setup that mixes some of the antisqueezing in the $S_3$ direction. By summing over all the values of $S$ we obtain the total $Q (\mathbf{n})$ function, which is a probability distribution over the Poincar\'e unit sphere and is properly normalized. 

The Radon reconstruction requires a large set of measured data to get
a reasonably accurate representation of the state. There are two main
reasons for this: integrals are approximated by finite sums and the
kernel (\ref{kersing}) is singular, so some \emph{ad hoc} filtering of
the raw data is needed. Acquiring such large data sets may be unwise,
for they demand long measurement times. Ensuring the proper stability
of the setup is thus essential and might be difficult depending on the
quantum state being measured. This limitation may be circumvented by
adopting a statistically motivated method, such as maximum
likelihood estimation~\cite{lnp:2004uq}.  

For a broad class of states, the registered tomograms have Gaussian statistics,
which seems to call for a Gaussian maximum likelihood
reconstruction~\cite{Rehacek:2009aa}. The Gaussianity is used as prior
information about the signal, which helps to drastically reduce the
number of free parameters, as experimentally verified in
Ref.~\cite{Muller:2012ys}.

\section{Polarization measures}
\label{sec:desiderata}

In classical optics, the Stokes parameters and the classical degree of polarization, as defined in Eq.~(\ref{eq:defPcl}), are sufficient for the characterization of most beams, as they are Gaussian states of light. Such a description in terms of first-order moments of Stokes variables can be naively extended to the quantum domain through 
\begin{equation}
  \mathbb{P}_{s} = \frac{\lvert \langle \hat{\mathbf{S}} \rangle \rvert}
  {\langle\hat{S}_0 \rangle} = 
  \frac{\sqrt{\langle\hat{S}_1 \rangle^{2}+ \langle\hat{S}_2\rangle^{2}+
      \langle\hat{S}_3 \rangle^{2}}}{\langle\hat{S}_0\rangle} \, .
  \label{eq:Ps}
\end{equation}
We refer to this definition as the semiclassical degree of polarization. It has the advantage that it can be measured with a traditional polarimeter, such as those described in Sect.~\ref{subsec:poltrans}. Another commonly used method is the scrambling method, in which a polarization scrambler is placed in front of a polarizer and a detector. The difference between the maximum and minimum power levels $P_{\mathrm{max}} - P_{\mathrm{min}}$ is precisely the total polarized portion.  The unpolarized portion, on the other hand, is unaffected by the scrambler, except for a global reduction by a factor $1/2$. Because the contribution of the polarized portion is zero at $P_{\mathrm{min}}$, then $P_{\mathrm{min}} = P_{\mathrm{unpol}}/2$ and thus
\begin{equation}
\mathbb{P}_{s} = \frac{P_{\mathrm{max}} - P_{\mathrm{min}}}
{P_{\mathrm{max}} + P_{\mathrm{min}}} \, .
\end{equation} 
Therefore, $ \mathbb{P}_{s}$ can be determined by simply measuring the maximum and minimum power levels at the detector.

However, we might expect $ \mathbb{P}_{s}$ to be incomplete. This is confirmed when examining the two extreme situations $ \mathbb{P}_{s} = 1$ and $\mathbb{P}_{s} = 0$, representing fully polarized and fully unpolarized light, respectively. For example, all SU(2) coherent states have $ \mathbb{P}_{s} = 1$ and this holds true for any superposition of them in different polarization sectors, such as, e.g., the Glauber coherent states $|\alpha_{+}, \alpha_{-} \rangle$, as well as convex combinations thereof \cite{Goldberg:2017}.  But this means that $ \mathbb{P}_{s} = 1$ for states arbitrarily close to  the two-mode vacuum state $|0_{+},0_{-}\rangle$, which is a strange result. 

On the other hand, there are states with $ \mathbb{P}_{s} = 0$ that can hardly be regarded as unpolarized. This gives rise to the phenomenon of hidden polarization.  As heralded before, one example of hidden polarization is the state $|1_{H}, 1_{V} \rangle $, which is $|1,0\rangle$ in the angular momentum basis. A rotation by 45$^{\circ}$ degrees around its axis of propagation transforms this state into $[(\sqrt{2}+i)|1,-1\rangle+(\sqrt{2}-i)|1,1\rangle]/\sqrt{6}$, which is orthogonal to $|1,0\rangle$, and then perfectly distinguishable from the unrotated state. However, according to the semiclassical degree of polarization, this state is unpolarized, which implies an invariance when undergoing rotations. This is due to the fact that the rotated state cannot be distinguished from the unrotated state by any linear combination of the Stokes operators, as this requires higher-order field correlation measurements. For this reason, perhaps it would be better to say that such states have higher-order polarization.

\subsection{\emph{Desiderata} for polarization measures}

From the previous discussion one can conclude that an appropriate
measure of polarization assigning a number $\mathbb{P} (\hat{\varrho})$ to the density operator $\hat{\varrho}$ must satisfy some requirements that capture the properties of the Stokes operators and the polarization transformations.  Before proceeding any further, we stress that a physically meaningful reference for any degree of polarization is provided by unpolarized light. Indeed, fully unpolarized light states can be suitably defined as the states invariant under any polarization transformation. This demands that the whole probability distribution be SU(2) invariant~\cite{Prakash:1971fr,Agarwal:1971zr}; that is,
\begin{equation}
  \label{eq:2}
  [\hat{\varrho}_{\mathrm{unpol}}, \hat{\mathbf{S}} ] = 0 \, ,
\end{equation}
wherefrom it follows that in every Fock layer the state is maximally mixed~\cite{Soderholm:2001ay} 
\begin{equation}
  \label{eq:unp}
  \hat{\varrho}_{\mathrm{unpol}}^{(S)} =\frac{1}{2S+1} \,
  \hat{\openone}_{2S+1} \, .
\end{equation}
We thus require~\cite{Bjork:2010rt}

\begin{enumerate}[label={C\arabic*.--}]
\item 
$\mathbb{P}(\hat{\varrho})=0$ iff $\hat{\varrho}$ is
  unpolarized.

\item
  $\mathbb{P}(\hat{\varrho})= \mathbb{P}(\hat{U} \, 
  \hat{\varrho} \, \hat{U}^\dagger)$ for any unitary polarization
  transformation $U$.

\item $\mathbb{P}(\hat{\varrho})$ should not depend on the coherences
  between different Fock layers.
\end{enumerate}

Condition C1 rules out various possibilities such as the semiclassical degree of polarization, for this degree considers as unpolarized states that do have higher-order polarization correlations, including states that are fully polarized in different directions in each Fock layer~\cite{Goldberg:2017}. It similarly rules out the proposed second-order measure~\cite{AlodjantsArakelian1999} 
\begin{equation}
  \mathbb{P}_{s,2} = 
  \frac{\lvert  \langle \hat{\mathbf{S}} \rangle \rvert}
  {\sqrt{\langle\hat{\mathbf{S}}^2 \rangle}} = 
  \sqrt{1 - \frac{\var{\hat{\mathbf{S}}}}{\langle\hat{\mathbf{S}}^2 \rangle}} \, .
\end{equation}
This conditions also precludes the definition of a degree of polarization solely in terms of the purity of a state $\mathcal{P}=\Tr (\hat{\varrho}^2)$, as unpolarized quantum states span the whole purity scale ---there are examples of unpolarized states among the pure states (including but not limited to the two-mode vacuum), and there are also unpolarized states that are partially or maximally mixed.

Requirement C2 is a statement of SU(2) invariance, which is a desirable characteristic of any \emph{bona fide} measure. The second-order measure does not fulfill C2 either; it can be made SU(2)-invariant by~\cite{Klimov:2010uq}
\begin{equation}
  \mathbb{P}_{s,2}^{\prime}  = \sqrt{1- 3 \inf_{\mathbf{n}}
  \frac{\var{\hat{\mathbf{S}}\cdot\mathbf{n}}}{\langle\hat{\mathbf{S}}^2 \rangle}} \, ,
\end{equation}
but this still does not capture higher-order polarization correlations.

The C3 requirement is also suitable, since polarization transformations do not produce coherences between Fock layers; the Stokes operators are photon-number preserving, so any measurement of a linear combination of these operators will be independent of any coherences between Fock layers. We can rephrase C3 in a more quantitative way. To this end, let us introduce the quantum channel
\begin{equation}
  \mathcal{R} [ \hat{\varrho} ]  = \sum_{S=0}^\infty
  \hat{\openone}_S \, \hat{\varrho} \, \hat{\openone}_S =
  \hat{\varrho}_{\mathrm{pol}} \, ,
  \label{eq:chan}
\end{equation}
which can be viewed as a randomization of the phases between superpositions between states in different Fock layers.  The states $\hat{\varrho}$ and $\hat{\varrho}_{\mathrm{pol}}$ cannot be distinguished from each other in polarization measurements, as discussed in Sec.~\ref{sec:polsec}.  So, we can reformulate C3 in the equivalent form

\medskip

\indent C3$^{\prime}.$--
$\mathbb{P}(\hat{\varrho}) = \mathbb{P}(\mathcal{R} [\hat{\varrho}]) =
\mathbb{P} ( \hat{\varrho}_{\mathrm{pol}})$.

\medskip
A conventional condition is also that $0\leq \mathbb{P}(\hat{\varrho})\leq 1$.  However, some candidate polarization measures, such as the entropy, are only positive definite ($0 \leq \mathcal{S} < \infty$). In these cases, the ordering of the states is usually more important than the numerical value of the measure. For this reason, a remedy is the normalization $\mathbb{P}=\mathcal{S}/(1+\mathcal{S})$, which is a rescaling that keeps the ordering of the states intact.

Apart for the basic conditions C1--C3, the measure should be operational, easily measurable, and easy to compute. These conditions are, however, difficult to meet at the same time. 

In this section, we explore several proposals for a quantum degree of polarization. It is important to stress, though, that these measures induce different orderings between the states, as they stem from different concepts.

\subsection{Distance-based measures}
\label{sec:distances}

Quantum polarization can be quantified in terms of distance measures. The main idea is to define a degree of polarization as the shortest distance between a state and the set $\mathcal{U}$ of unpolarized states, as given in  Eq.~(\ref{eq:unp}). Other notions such as nonclassicality~\cite{Hillery:1987aa,Dodonov:2000aa,Marian:2002aa}, entanglement~\cite{Vedral:1997aa}, localization~\cite{Maassen:1988cr,Anderson:1993nx,Gnutzmann:2001aa}, and quantum information~\cite{Schumacher:1995aa,Brukner:1999aa,Schack:1999aa,Childs:2000aa, Gilchrist:2005aa} have been systematically formulated in terms of distances to a given set of states. In a way, the distance determines the distinguishability of a state with respect to that set.

Therefore, it seems sensible to quantify the degree of polarization by
\begin{equation}
  \label{DoP}
  \mathbb{P} (\hat{\varrho}) \propto
  \inf_{\hat{\sigma} \in \mathcal{U}}
  D(\hat{\varrho} \mid \hat{\sigma} ) \, ,
\end{equation}
where $D(\hat{\varrho} \mid \hat{\sigma} )$ is any measure of distance (not necessarily a metric) between the density matrices $\hat{\varrho}$ and $\hat{\sigma}$ and $\mathcal{U}$ is the set of unpolarized states, such that $\mathbb{P} (\hat{\varrho})$ satisfies the requirements C1--C3.

There are numerous nontrivial choices for $D(\hat{\varrho} \mid \hat{\sigma})$ (by nontrivial we mean that the choice is not a simple scale transformation of any other distance measure). None of them could be said to be more important than any other \textit{a priori}; the significance of each candidate must be evaluated based on its physical implications in the particular context.  For the case of polarization several distances may be considered~\cite{Klimov:2005aa,Sanchez-Soto:2006}, such as the Hilbert-Schmidt, trace, Bures, and Chernoff~\cite{Ghiu:2010aa} distances:
\begin{equation}
\begin{array}{lll}
  & \displaystyle 
  \mathbb{P}_{\mathrm{HS}}(\hat{\varrho}) =  
  \inf_{\sigma\in \mathcal{U}} \Tr[ (\hat{\varrho}-\hat{\sigma})^2]\, ,
  \qquad \qquad 
  & \displaystyle 
  \mathbb{P}_{\mathrm{T}}(\hat{\varrho})  =  
  \inf_{\sigma\in \mathcal{U}} \Tr[ \lvert \hat{\varrho}-\hat{\sigma} \rvert] \, , \\
 &  &  \\
  & \displaystyle
   \mathbb{P}_{\mathrm{B}}(\hat{\varrho} ) =  1-\sup_{\sigma \in \mathcal{U}}\sqrt{F(\hat{\varrho} \mid \hat{\sigma})}\, ,
  \qquad \qquad
  & \displaystyle
  \mathbb{P}_{\mathrm{C}}(\hat{\varrho})  = 
  1-\sup_{\sigma \in \mathcal{U}}\left[\inf_{t\in[0,1]}\Tr(\hat{\varrho}^t\hat{\sigma}^{1-t})\right] \, ,
\end{array}
\end{equation}
where the infimum in the Chernoff distance is taken over a function that
is continuous with respect to $t$, and the fidelity $F$ in the Bures distance
is~\cite{Uhlmann:1976aa,Jozsa:1994aa,Alberti:2003aa}
\begin{equation}
  F(\hat{\varrho} \mid \hat{\sigma})= 
  \{ \Tr[(\hat{\sigma}^{1/2}\hat{\varrho}\hat{\sigma}^{1/2})^{1/2}]\}^2 \, .
  \label{eq:fidel}
\end{equation}
As they stand, these degrees do not satisfy the requirement C3; i.e., these measures are sensitive to coherences between different Fock layers. One can bypass this drawback by simply defining the  distance not to the state, but to its block diagonal form or polarization sector; that is, $\mathbb{P}_{X}(\hat{\varrho}) = \mathbb{P}_X ( \mathcal{R} [\hat{\varrho}])$ for $ X \in \{\mathrm{HS,T,B,C}\}$.

Since $\mathcal{R} [\hat{\varrho}]$ and $\hat{\sigma}$ commute, we find the
following general expressions
\begin{equation}
\begin{array}{lll}
  & \displaystyle 
  \mathbb{P}_{\mathrm{HS}}(\hat{\varrho}) = 
    \sum_{S=0}^\infty w_{S}^{2}  
    \left(\xi_S^{(2)}-\frac{1}{2S+1}\right) \, ,
    \qquad
   & \displaystyle
   \mathbb{P}_{\mathrm{T}}(\hat{\varrho}) = 
   \sum_{S=0}^\infty w_{S} \left(\sum_{n=0}^{M_{S}}\lambda_{S,n}-
   \frac{M_{S}+1}{2S+1}\right)
     \, ,  \\
  & & \\
  & \displaystyle
    \mathbb{P}_{\mathrm{B}}(\hat{\varrho}) = 
    1-\left[\sum_{S=0}^\infty   \frac{w_S}{2S+1} 
    \left(\xi_S^{(1/2)}\right)^2\right]^{1/2}
    \, ,
    \qquad 
    & \displaystyle
    \mathbb{P}_{\mathrm{C}} (\hat{\varrho}) = 1 
    -\inf_{t\in [0,1]}\left[\sum_{S=0}^\infty w_{S} \;
    (2S+1)^{1-1/t} \left (\xi_S^{(t)} \right )^{1/t}\right]^{t} \, ,
     \label{eq:Pdismeas}
\end{array}
\end{equation}
where $w_{S}$ encompasses the photon statistics, $\lambda_{S,n}$ are the eigenvalues of $\hat{\varrho}^{(S)}$ taken in decreasing order, $\xi_S^{(t)}=\sum_{n=0}^{2S} \lambda_{S,n}^{t}$, and $M_{S}$ is the largest integer satisfying $\lambda_{S,M_{S}}\geq 1/(2S+1)$. As we can see, all of them require the knowledge of the complete polarization sector.

For states living in the Fock layer with spin $S$, the maximum polarization is reached for pure states $| \Psi^{(S)}\rangle$, with values
\begin{equation}
  \mathbb{P}_{\mathrm{HS}}( |\Psi^{(S)}\rangle ) =
  \mathbb{P}_{\mathrm{T}}( |\Psi^{(S)}\rangle ) =
  \mathbb{P}_{\mathrm{C}}( |\Psi^{(S)}\rangle )  =
  \frac{S}{S+1/2} \, , 
  \qquad 
  \mathbb{P}_{\mathrm{B}}( |\Psi^{(S)}\rangle )  =
  1- \frac{1}{\sqrt{2 S+1}}  \, ,
\end{equation}
so all of them tend to unity when $S$ is sufficiently large.  We note, in passing, that $\mathbb{P}_{\mathrm{B}}(\hat{\varrho}) \leq \mathbb{P}_{\mathrm{C}}(\hat{\varrho})$. 

Likewise, for quadrature coherent states $| \alpha_{+} , \alpha_{-}\rangle$, with average number of photons $\bar{N}$, we have
\begin{equation}
  \mathbb{P}_{\mathrm{HS}} = 1 -
  \frac{I_{1} (2 \bar{N})}{\bar{N}} 
  e^{-2 \bar{N}} \simeq 
  1- \frac{1}{2\sqrt{\pi } \bar{N}^{3/2}}  \, ,
\end{equation}
where $I_{n} (x)$ is the modified Bessel function~\cite{NIST:DLMF} and the last expression holds when $\bar{N} \gg 1$.  This again tends to unity, but with a scaling of the form $\bar{N}^{-3/2}$. The same scaling can be found for the Bures and Chernoff degrees of polarization. 

It is possible to find the states that, for a given average number of photons, reach the maximal degree of polarization. Using a numerical optimization procedure, one discovers that such an optimal value  is~\cite{Sanchez-Soto:2007aa}
\begin{equation}
  \label{pq2}
  \mathbb{P}_{\mathrm{HS}}^{\mathrm{opt}} = 1 -
  \frac{3}{(2 \bar{N} +1 ) ( 2 \bar{N} + 3)}
  \sim 1 - \frac{3}{4 \bar{N}^2} \, ,
\end{equation}
with similar scalings for the other degrees. To an excellent approximation, a highly squeezed vacuum can be taken as maximally polarized.

\subsection{Phase-space measures}

The ideas discussed in the previous subsection can be straightforwardly translated into the phase-space picture.  The degree of polarization of a state can be defined as the distance between its Husimi $Q$-function and the corresponding one for unpolarized light.  In this context, unpolarized light is defined  by a uniform distribution~\cite{Luis:2007aa}
\begin{equation}
  Q_{\mathrm{unpol}} (\mathbf{n} ) = 1 \, ,
\end{equation}
which agrees with (\ref{eq:unp}).  One can then define the distance~\cite{Luis:2002ul,Luis:2007kx} 
\begin{equation}
  D_{Q}(\hat{\varrho}) = \frac{1}{4\pi} 
  \int_{\mathcal{S}_{2}} d\mathbf{n} \; [ Q(\mathbf{n} ) - 1 ]^2  = 
  \frac{1}{4\pi} \left [  \int_{\mathcal{S}_{2}} d\mathbf{n} \;  
 Q^2(\mathbf{n}) \right ]  - 1 \, .
  \label{eq:DQ}
\end{equation}
The relevant term on the right-hand side of (\ref{eq:DQ}) can be
expressed in terms of the multipoles as
\begin{equation}
       \int_{\mathcal{S}_{2}} d\mathbf{n} \;  
 Q^2(\mathbf{n}) = 4\pi \sum_{S=0}^\infty  
 (2S+1)^2 \sum_{K=0}^{2S} \lvert C_{SS,K0}^{SS} \rvert^2 
 \sum_{q=-K}^K \lvert \varrho_{Kq}^{(S)} \rvert^2 \, ,
    \label{eq:QsquaredMultipoles}
\end{equation} 
and regarded as a particular instance of a general class of measures of
localization~\cite{Maassen:1988cr,Anderson:1993nx,Gnutzmann:2001aa} 
\begin{equation}
M_{r} = \left [ 
 \int_{\mathcal{S}_{2}} d\mathbf{n} \; Q^{1+r} ( \mathbf{n} ) 
\right ]^{1/r}\, ,
\end{equation}
whose mathematical properties have been studied in great detail~\cite{Hardy:1952aa}.  Since 
\begin{equation} 
\lim_{r \rightarrow 0} M_r = 
 \int_{\mathcal{S}_{2}} d\mathbf{n} \; Q ( \mathbf{n} ) \, 
\ln Q  (  \mathbf{n} ) \, ,
\end{equation}  
they include the Wehrl entropy~\cite{Wehrl:1978aa,Wehrl:1991aa} as a
limiting case.   

In physical terms, the spread of the $Q$-function gives an indication of the polarization properties of the state. For states that are highly spread over the unit sphere the degree of polarization is small, as in the case of unpolarized states. For states whose $Q$-functions are highly peaked around some point in the sphere, the degree of polarization is expected to be high. Note that, as per Eq.~(\ref{eq:QsquaredMultipoles}), $D_{Q} (\hat{\varrho})$ involves \emph{all} the multipoles and its experimental determination thus requires a full tomography.

We briefly mention that the Wigner function has been used as a measure of the area occupied by a quantum state in the case of continuous variables~\cite{Heller:1987dq}.  However, in the case of the sphere, the integral of $W^{2} (\mathbf{n})$ simply gives the purity~\cite{Varilly:1989ud}. This is one of
the compelling reasons to use the Husimi $Q$-function instead of the Wigner function throughout this review when analyzing polarization. 

Since $D_{Q}$ ranges from 0 to $\infty$, the associated degree of polarization, according to our previous discussion, is
\begin{equation}
  \mathbb{P}_{Q} (\hat{\varrho}) =\frac{D_Q  (\hat{\varrho})}
  {D_Q(\hat{\varrho})+1}\, .
  \label{eq:DQnor}
\end{equation}
The only states with $\mathbb{P}_{Q} = 0$ are the fully unpolarized states. In
contrast to the classical definition, there are  states with $\langle \hat{\mathbf{S}} \rangle = 0$ and $\mathbb{P}_Q\neq 0$. This occurs because $\mathbb{P}_Q$ is a function of all the moments of the Stokes operators and not only of the first ones. In addition, the definition of $\mathbb{P}_Q$ is invariant under SU(2) transformations applied to the field state.  This means that the
degree of polarization depends on the form of the $Q$-function but not on its orientation on the Poincar\'e sphere.

For the SU(2) coherent state $|S, \mathbf{n} \rangle$ the  $Q$-function is given in  Eq.~(\ref{eq:QCS}). If, for definiteness, we take $\mathbf{n}$  to point to the north pole, the associated degree is
\begin{equation}
  \mathbb{P}_{Q} (|S, \mathbf{n} \rangle) = \left  ( 
    \frac{2S}{2S+1} \right )^2  {\xrightarrow[S \gg 1]{}} 1 \, .
\end{equation}
In fact, the maximally polarized states according to this degree of polarization are the SU(2) coherent states. This is because these states have their $Q$-functions highly peaked at some point of the sphere, making them minimum-uncertainty polarization states and states that are maximally different from those with a uniformly-distributed $Q$-function.

\subsection{Operational approaches}

It is clear that a state that is not invariant under all possible linear
polarization transformations has a finite degree of quantum polarization.  Therefore, we can use the distinguishability of a state under all possible polarization transformations as a degree of polarization.  If we take $\Tr(\hat{\varrho}_1 \hat{\varrho}_2)$ as the distinguishability for mixed states, one could define~\cite{Bjork:2000aa,Bjork:2002aa}
\begin{equation}
  \mathbb{P}_\mathrm{d} (\hat{\varrho})= 
  \left [1- \inf_{\hat{U} \in \mathrm{SU(2)}} \;\;
    \sum_{S=0}^\infty w_{S} \; \Tr \left (\hat{\varrho}^{(S)} \;
    \hat{U}\, \hat{\varrho}^{(S)} \,  \hat{U}^\dagger \right )
  \right]^{1/2} \,. 
  \label{eq:dis}
\end{equation}
This is the minimal averaged overlap between a state and all of its SU(2)
transformed partners. Hence, it gives the maximum visibility one can achieve by using a polarization interferometer~\cite{Bjork:2014aa}. The drawback of this definition is that it is not trivial to determine the degree of polarization for a given state, since there is, in general, no obvious way to find the $\hat{U}$ maximizing the polarimetric visibility. This measure of polarization is different from the previous ones because it assigns degree of polarization unity to states with a finite number of excitations. It has been proven~\cite{Sehat:2005wd} that all pure states with an odd number of photons reach degree of polarization unity under this definition. It is also true that single photon states, the definition implies that any pure state is fully polarized, since a transformation $\hat{U}$ can transform the state to an orthogonal state, located diametrically opposite on the sphere. One can conjecture that pure states with an even number of excitations, except the two-mode vacuum, are also fully polarized according to this measure of polarization, but there is still no proof for this. 

The last speculation makes it tempting to define a degree of polarization such that the pure states with a given number of excitations are maximally polarized; this is a degree of polarization in terms of purities in every excitation manifold~\cite{Bjork:2010rt}
\begin{equation}
  \mathbb{P}_\mathrm{p} ( \hat{\varrho})=
  \sum_{S= \frac{1}{2}}^\infty   \frac{1}{2S}
  [(2S+1) \, \Tr( \hat{\varrho}^{(S)\; 2}) -1 ] \, ,
  \end{equation}
with the additional definition $\mathbb{P}_\mathrm{p}\left(|0,0\rangle\right)=0$. Again, the maximally polarized states are the pure states in every excitation manifold.  The only property of the Stokes operators used in this measure is its definition as a direct sum over the excitation manifolds with
distinct number of photons. As a consequence, this measure quantifies a distinguishability under general energy-preserving unitary transformations rather that the distinguishability under the most concrete case of polarization transformations. This measure also critically relies on the conjecture that the pure states in every excitation manifold are fully polarized. Like many of the other polarization degrees, an experimental determination of $\mathbb{P}_\mathrm{p}$ is difficult, as purities require polarization tomography to be assessed.

\subsection{Higher-order degrees of polarization}

At this stage, it should be clear that many of the difficulties in defining a proper degree of polarization can be traced to classical polarization being built on first-order moments of the Stokes variables; whereas, higher-order moments can play a major role for quantum fields. A full understanding of the subtle polarization effects arising in the quantum realm requires a characterization of higher-order polarization fluctuations, as is done in coherence theory, where one needs, in general, a hierarchy of correlation functions~\cite{Mandel:1995qy}.

As we have seen, multipoles contain the higher-order fluctuations information, sorted in the proper way. This suggests looking at the cumulative distributions~\cite{Hoz:2013om}
\begin{equation}
  \label{eq:cum}
  \mathcal{A}^{(S)}_{M} = \sum_{K = 1}^{M} \sum_{q=-K}^{K}
  \lvert \varrho_{Kq}^{(S)} \rvert^{2} \, ,
\end{equation}
which convey all of the information up to order $M$. Its experimental determination requires then reconstructing the corresponding multipoles, which can be accomplished using the methods discussed in Sec.~\ref{sec:tomo}.

\begin{figure}
  \centering{\includegraphics[width=0.85\columnwidth]{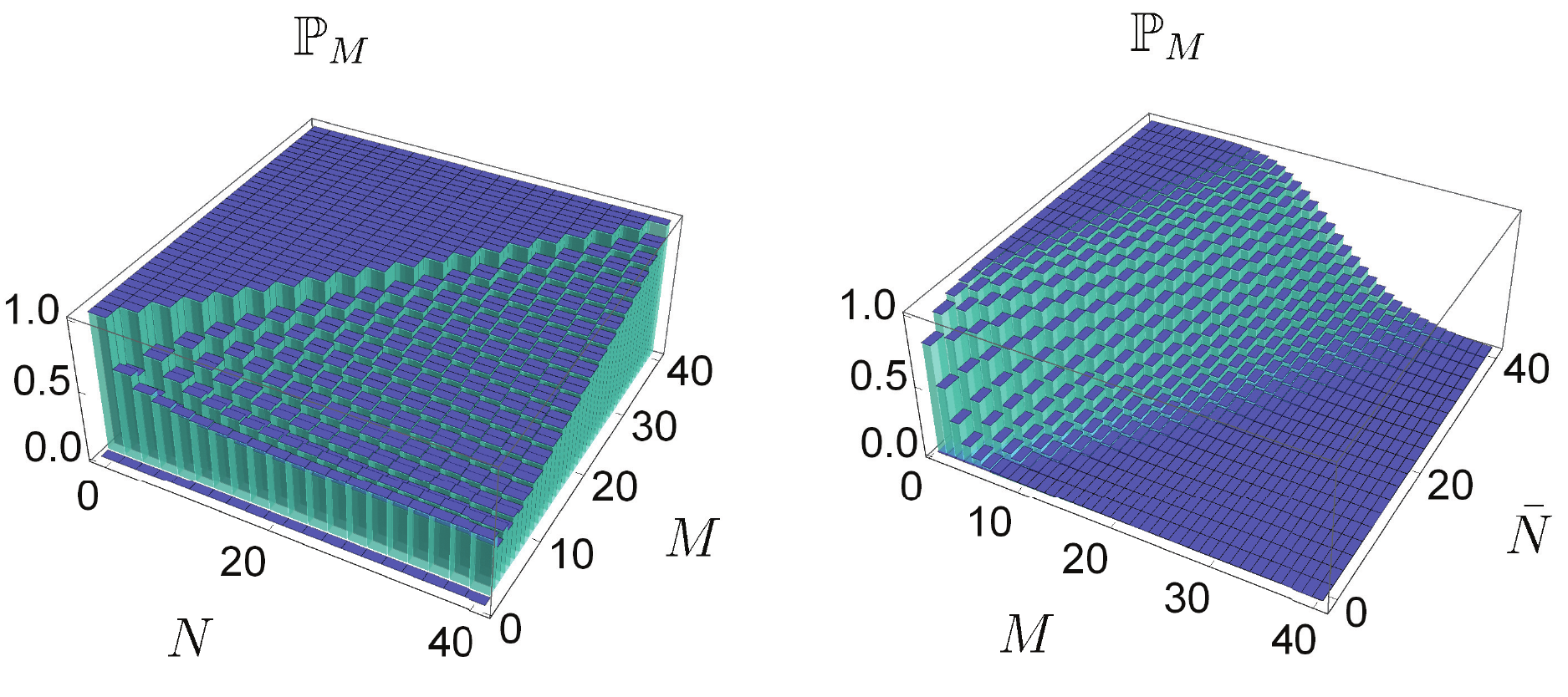}}
  \caption{(Color online)  Degree of polarization $\mathbb{P}_{M}$ as a
    function of the multipole order $M$ for the state $|S, 0 \rangle$
    (left) and a quadrature coherent state $| \alpha_{+}, \alpha_{-}
    \rangle$ with  average number of photons $\bar{N} = \lvert \alpha_{+} \rvert^{2} + \lvert \alpha_{-} \rvert^{2}$ (right).  
    \label{fig:PK}}
\end{figure}

Note that the monopole $\varrho_{00}^{(S)}$, being a trivial constant, is not included in the sum. We know from probability theory that cumulative distributions have remarkable properties~\cite{Jaynes:2003uq}. Moreover, our previous discussion in Section~\ref{sec:tomo} provides clear evidence that to obtain the $K$th multipole, one needs to determine all of the previous moments.  As with any cumulative distribution, $ \mathcal{A}^{(S)}_{M} $ is a monotonic, nondecreasing function of the multipole order, with $\mathcal{A}^{(S)}_{2S}$ being the state purity
minus the monopolar contribution.

For SU(2) coherent states, we get
\begin{equation}
  \label{eq:Aksu2}
  \mathcal{A}^{(S)}_{M,\mathrm{SU(2)}} = \frac{2S}{2S +1} -
  \frac{[\Gamma (2S + 1)]^{2}}{\Gamma (2S-M) \Gamma (2S + M +2)} \, ,
\end{equation}
and it has been proven~\cite{Bjork:2015aa} that this value is indeed maximal for every $M$ for all states in the subspace $\mathcal{H}_{S}$. This strongly suggests the definition of a hierarchy of degrees of polarization that  quantify the polarization information when measuring the Stokes operators up to order $M$:
\begin{equation}
  \label{eq:PK}
  \mathbb{P}_{M} =
  \sum_{S} w_{S} \; \left (
    \frac{\mathcal{A}^{(S)}_{M}}{\mathcal{A}^{(S)}_{M,\mathrm{SU(2)}}}
\right )^{1/2}   \, .
\end{equation}
According to (\ref{eq:PK}), $\mathbb{P}_{M}= 1$, regardless of $M$, for SU(2) coherent states, which is compatible with the idea that they are the most localized states over the sphere.   

For the significant case of the first-order degree ($M=1$), Eq.~(\ref{eq:PK}) gives
\begin{equation}
  \mathbb{P}_{1}= \sum_{S}^\infty w_{S}  
  \frac{ \sqrt{\langle \hat{S}_{1} \rangle^2+ \langle \hat{S}_{2} \rangle^2 +
      \langle \hat{S}_{3} \rangle^2}}{\langle \hat{S}_{0} \rangle} \, ,
\end{equation}
and the average values are calculated in every Fock layer $S$. Interestingly, this definition has recently been proposed as a way to circumvent the shortcomings of the standard degree of polarization~\cite{Kothe:2013fk}; in our approach, it emerges quite naturally.

For quadrature coherent states the result reads
\begin{equation}
  \mathbb{P}_{M} (|\alpha_{+}, \alpha_{-} \rangle ) = \sum _{S=M/2}^{\infty } 
  \frac{e^{-\bar{N}}\bar{N}^{2S}}{(2S)!} \simeq 
  \frac{1}{2} \text{erfc} 
  \left(\frac{M-\bar{N}}{\sqrt{2 \bar{N}}} \right ) \, ,
\end{equation}
where $\bar{N}$ is the average number of photons and the approximation in terms of the complementary error function is valid for $\bar{N} \gg 1$. From the properties of this function, we can estimate that the multipoles that contribute effectively are, roughly speaking, from 1 to $\bar{N}$. In Fig.~\ref{fig:PK} we plot $\mathbb{P}_{M}$ for the states $|S, 0 \rangle$ and $| \alpha_{+}, \alpha_{-} \rangle$.

For $M=2$, this gives 
\begin{equation}
  \mathbb{P}_{2} (|\alpha_{+}, \alpha_{-} \rangle ) =
  1 -  (1+ \bar{N}) \exp (- \bar{N}) \, ,
\end{equation}
which tends to unity when the average number of photons $\bar{N}$ becomes large enough, in agreement with previous second-order approaches~\cite{Klimov:2010uq,Singh:2013aa}. 

In principle, one would naively expect that all the approaches discussed in this Section  should be compatible with classical polarization, much in the spirit of the correspondence principle. Nevertheless, potential applications of quantum polarization seem to involve states without classical analog. In addition, inconsistencies and difficulties can be expected when characterizing something as complicated as quantum polarization by a single number. From this viewpoint, we believe that the hierarchy (\ref{eq:PK}) is the most proper and sensible way to deal with the problem.

\section{Polarized light and quantum complementarity}
\label{sec:complementarity}

\subsection{SU(2) complementarity}

In the classical domain, the relative phase between the modes determines the shape of the polarization ellipse. It is thus natural to try a description of quantum polarization in terms of such a variable. A glance at the Stokes operators suggests that this information is encoded in $\hat{S}_{\pm}$, so we can perform a polar decomposition~\cite{Luis:1993aa,Sanchez-Soto:1994aa}
\begin{equation}
\hat{S}_{-} = \hat{E} \, \sqrt{\hat{S}_{+} \hat{S}_{-}} \, ,
\end{equation}
where $\hat{S}_{+} \hat{S}_{-}$ is a Hermitian positive operator and the unitary operator $\hat{E}$ represents the exponential of the relative phase.

One can verify that $[\hat{E}, \hat{N}] = 0$, so we have $\hat{E} = \sum_{S} \hat{E}^{(S)}$ and we can work out the solution in $\mathcal{H}_{S}$. In each of these Fock layers, there are $2S+1$  orthonormal states $| \delta_{r}^{(S)} \rangle$ ($r = -S, \ldots, S $), given by
\begin{equation}
\label{eq:comES}
| \delta_{r}^{(S)} \rangle = \frac{1}{\sqrt{2S+1}} \sum_{m=-S}^{S}
e^{- i m \delta_{r}^{(S)}} |S, m \rangle \, , 
\qquad \qquad
\delta_{r}^{(S)} = \frac{2\pi}{2S+1} r \, ,
\end{equation}
such that 
\begin{equation}
\hat{E}^{(S)} | \delta_{r}^{(S)} \rangle =  e^{ i \delta_{r}^{(S)}}
| \delta_{r}^{(S)} \rangle \, ,
\end{equation}
so this is a good quantum description of the relative phase. 

From Eq.~\eqref{eq:comES} we see that the eigenstates of $\hat{E}$ are related to the angular momentum basis  $|S,m \rangle$ by a finite Fourier transform~\cite{Levy:1976aa,Guise:2012aa}. This means that the relative phase $\hat{E}$ and the number difference $\hat{S}_{3}$ are complementary observables and, in consequence, the corresponding eigenvectors $| \delta_{r}^{(S)} \rangle $ and $|S,m \rangle $ are mutually unbiased~\cite{Wootters:1989aa,Durt:2010aa}. This purely quantum effect has been experimentally tested~\cite{Tsegaye:2000aa,Usachev:2001aa}, confirming in a crystal-clear manner the differences between classical and quantum polarization properties.

On the other hand, since this complementary pair $\hat{E}$-$\hat{S}_{3}$ has discrete spectra, one could imagine representing polarization as a discrete grid~\cite{Vourdas:2004aa,Vourdas:2007aa,Bjork:2008aa}. This approach fulfills many of the desirable properties and the resulting discrete analogs of the $Q$-function can be experimentally determined by so-called weak measurements~\cite{Salvail:2013aa}.

\subsection{Degree of polarization and complementarity}

The superposition principle for quantum states is the very bedrock of quantum theory. Indeed, Young’s double slit experiment was central to the discussions that laid the foundations~\cite{Feynman:2006aa} of this discipline. It incorporates in a natural way Bohr’s principle of complementarity~\cite{Bohr:1928aa}, which can be concisely stated by saying that a quantum system possesses real but mutually exclusive properties. 

The most popular formulation of this principle is probably the time-honored wave–particle duality, which restricts the coexistence of wave and particle qualities of quantum objects~\cite{Wootters:1979aa,Scully:1991aa,Zeilinger:1999aa}, in the sense that interferometric \emph{welcher Weg} information is complementary to the visibility of intensity fringes~\cite{Jaeger:1993aa,Jaeger:1995aa, Englert:1996aa}. This was later completed~\cite{Jakob:2010aa} by showing that concurrence $\mathcal{C}$~\cite{Wootters:1998aa}, a well-known measure of entanglement, naturally emerges as a quantity complementary to single-partite properties in a bipartite qubit system. The single-partite properties consist of two local but mutually exclusive realities given by visibility the $\mathcal{V}$ and predictability $\mathcal{P}$, forming what is commonly termed as wave–particle duality. Thus, the complementarity relation for bipartite qubits naturally contains three mutually exclusive quantities, namely
\begin{equation}
\mathcal{C}^{2} + \mathcal{V}^{2} + \mathcal{P}^{2} = 1 \, .
\end{equation}
In the same vein, Mandel examined the question of whether the degree of coherence can be understood in terms of quantum complementarity and he established a quantitative relationship between the degree of coherence and the \emph{welcher Weg} information~\cite{Mandel:1991aa}; a convincing experimental demonstration was also presented~\cite{Zou:1991aa}. 

Since the classical explanations of partial coherence and partial polarization are very similar in structure~\cite{Wolf:2007aa}, one might ask whether partial polarization can also be connected to quantum complementarity. This question was addressed recently~\cite{Lahiri:2011aa,Zela:2014aa,Eberly:2017aa,Lahiri:2017aa,Qian:2018aa,Kanseri:2019aa,Sanchez:2019aa,Qian:2020aa}, where much of this work has focused on single photons. This is somewhat special: for single photons, the semiclassical degree of polarization is equivalent to the purity of the state. Alternatively, the density matrix of a single-photon state contains only the first-order moment (i.e., the dipole), so, from the perspective of polarization, represents a classical state.  However,  the results have been appropriately extended to arbitrary quantum states~\cite{Norrman:2020aa}; viz.,
\begin{equation}
\mathbb{D}^{2} + \mathbb{V}^{2} = \mathbb{P}^{2} \, ,
\end{equation}
where the intensity distinguishability $\mathbb{D}$ and the Stokes visibility $\mathbb{V}$ are given by
\begin{equation}
\mathbb{D} = \frac{\lvert \langle \hat{N}_{+} \rangle - \langle \hat{N}_{-} \rangle \rvert}{\langle \hat{N}_{+} \rangle + \langle \hat{N}_{-} \rangle} \,, 
\qquad \qquad
\mathbb{V} = \frac{2 \sqrt{\langle \hat{N}_{+} \rangle \langle \hat{N}_{-} \rangle}}{\langle \hat{N}_{+} \rangle + \langle \hat{N}_{-} \rangle} \lvert g_{\pm} \rvert \, ,
\end{equation}
and the associated degree of polarization here turns out to be
\begin{equation}
\mathbb{P} = \left [ 1 - \frac{4 \langle \hat{N}_{+} \rangle \langle \hat{N}_{-} \rangle  (1 - \mid g_{\pm} \mid^{2} )}{( \langle \hat{N}_{+} \rangle + \langle \hat{N}_{-} \rangle)^{2}} \right ]^{1/2} \, .
\end{equation}
In these expressions $g_{\pm}$ is a mode correlation coefficient defined as
\begin{equation}
g_{\pm} = \frac{\langle \hat{a}_{+}^{\dagger} \hat{a}_{-} \rangle}{\sqrt{\langle \hat{N}_{+} \rangle \langle \hat{N}_{-} \rangle}} \, .
\end{equation}
This shows that all perfectly polarized quantum states~\cite{Goldberg:2017}; i.e., states with $\mathbb{P}= 1$, obey strong complementarity. If the intensity distinguishability (or path predictability for a single photon) is zero, we obtain $\mathbb{V} = \mathbb{P}$; that is, the Stokes visibility is exactly determined by the degree of polarization. This result unifies the interpretations for the degree of polarization established in classical two-way interferometry, where the degree of polarization has been connected to intensity visibility  and to polarization modulation~\cite{Leppanen:2014aa}.  Therefore, in addition to its role as a complementarity measure, the degree of polarization can be viewed as an intrinsic quantity that characterizes the ability of light to exhibit intensity and polarization-state variation.

\section{Unpolarized light}
\label{sec:unpol}

\subsection{Higher-order unpolarized states}

The constitution of unpolarized light has been investigated since the dawn of modern optics.  Indeed, Verdet~\cite{Verdet:1869ve} already offered a lucid characterization of what was known as \emph{natural} light by using the projections of the intensity onto the axes of a rotated Cartesian coordinate system. Unpolarized states are those that remain invariant under any rotation of that coordinate system and under any phase shift between its rectangular components.

In classical optics, the field components of unpolarized light are modeled by zero-mean, uncorrelated, stationary Gaussian random processes~\cite{Kampen:2007qr}. The previous invariance conditions thus determine the entire probabilistic structure of the projected intensities~\cite{Barakat:1989fp}. In contrast, the standard theory is limited to first-order moments, presenting unpolarized light as having a zero-mean Stokes vector, which in geometrical terms means that the Stokes vector reduces to the origin of the Poincar\'e sphere.

At the quantum level, the invariance requirement fixes once and for all the structure of the density matrix, so it specifies the probability distribution for and, as a result, all the moments of the Stokes variables, as we have already discussed. However, one could think of extending this notion: when all of the multipoles up to a given order (say $M$) vanish, the state lacks of polarization information up to that order and hence will be called \emph{$M$th-order unpolarized}. The classical picture matches the first-order theory, corresponding to a certain set of quantum states that are only invariant to first-order \cite{Goldberg:2017}; whereas, the quantum condition implies that all the multipoles are identically null.

In consequence, using the previous notion of higher-order polarization degrees, a state is $M$th-order unpolarized when $\mathbb{P}_{M}^{(S)}=0$, which obviously implies $\mathcal{A}_M^{(S)}=0$; i.e., all the multipoles up to order $M$ vanish.  We will denote these states as $\hat{\varrho}_{\mathrm{unpol},M}^{(S)}$. In more physical terms, the condition of $M$th-order unpolarization amounts to imposing that the moments $\langle \hat{\mathbf{S}}_{\mathbf{n}}^{\ell} \rangle$ be independent of the direction $\mathbf{n}$ for $\ell = 1, \ldots, M$ (i.e., that they be isotropic). Therefore, all of the moments up to order $M$ do not show any directional structure, while higher-order moments do. 
Because of this, $M$th-order unpolarized states do carry polarization information when one inspects higher-order moments. As we have said, this is referred to as {hidden} polarization, according to the terminology coined by Klyshko~~\cite{Klyshko:1992wd,Klyshko:1997yq,Klyshko:1998wd}, although we will say that such states display higher-order polarization~\cite{Gupta:2011lq}.

 We illustrate this point with a few examples, starting with single-photon states ($S=1/2$).  The multipole expansion of an arbitrary single-photon state reads
\begin{equation}
  \hat{\varrho}^{(1/2)} = \varrho_{00}^{(1/2)} \, \hat{T}_{00}^{(1/2)}   + 
  \sum_{q}  \varrho_{1q}^{(1/2)} \, \hat{T}_{1q}^{(1/2)} \, .
\end{equation} 
Since the state only has dipolar components, the quantum and classical descriptions coincide and these states can only be first-order unpolarized.  Positivity constrains the possible values of the dipole to the range $ 0 \leq \mathcal{A}_{1}^{(1/2)} \leq 1/2$.  The condition $\mathcal{A}_{1}^{(1/2)} =0$ immediately fixes the unpolarized states; viz.,
\begin{equation}
  \varrho_{\mathrm{unpol},1}^{(1/2)} =  \tfrac{1}{2} \; \openone_{2} \, .
\end{equation}
We stress, though, that like all quantum objects, these states can only be considered as elements of an ensemble~\cite{Peres:2002oz}. 

\begin{figure}
  \centering
  \includegraphics[height=5.5cm]{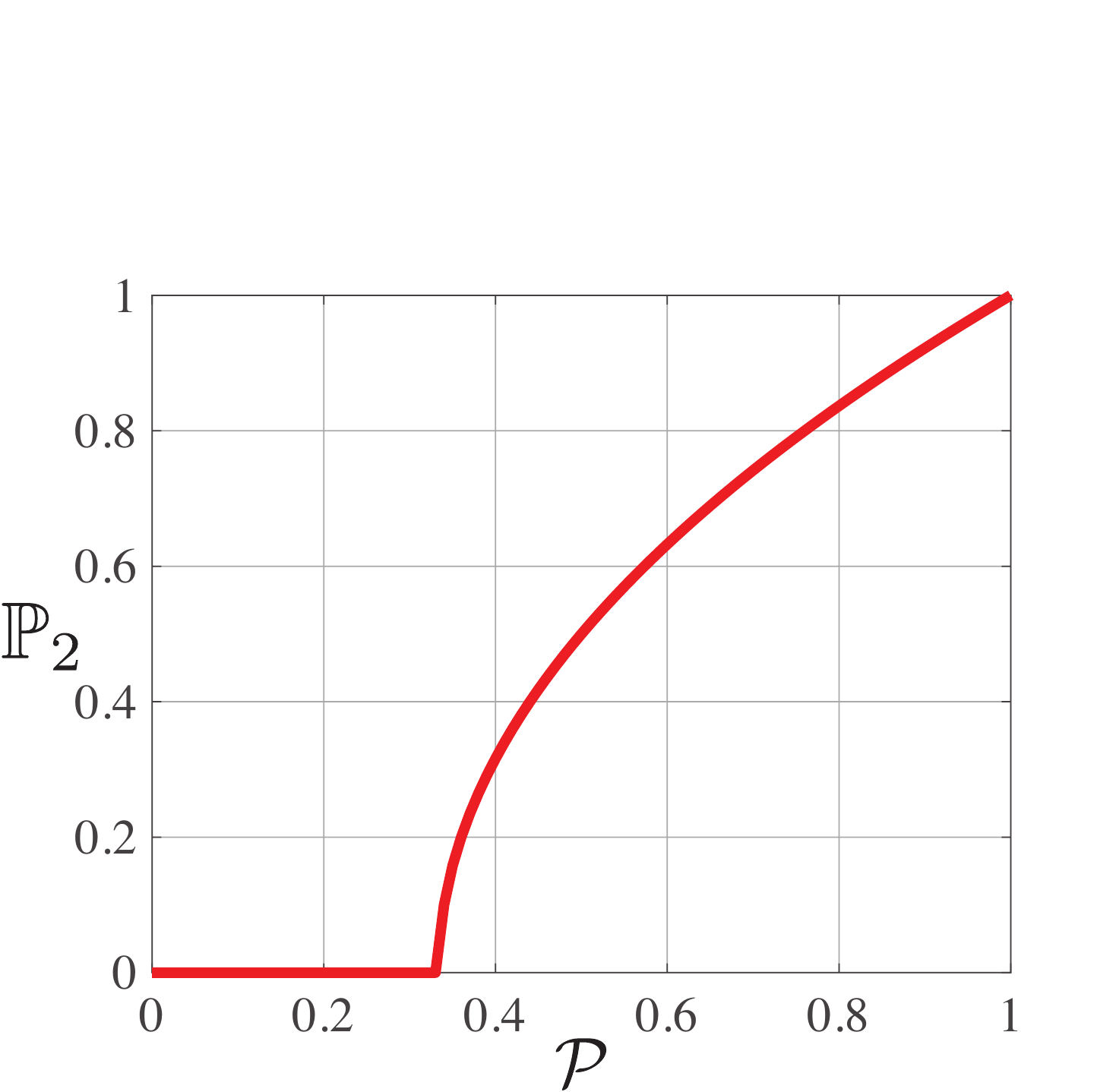}
  \caption{Second-order degree of polarization as a function of
    purity, for the first-order unpolarized states \eqref{eq:diag2}.}
  \label{fig:Pvspur}
\end{figure}

For two-photon states, there are first-order (classical) and second-order (quantum) unpolarized states. The general condition for first-order unpolarization is that the dipolar term vanishes; i.e., 
\begin{equation}
  \hat{\varrho}_{\mathrm{unpol},1}^{(1) } = 
  \varrho_{00}^{(1)} \, \hat{T}_{00}^{(1)} 
  +\sum_{q} \varrho_{2 q}^{(1)} \, \hat{T}_{2 q}^{(1)} \, ,
  \label{eq:un2ph}
\end{equation}
with the extra constraint of positivity.  Let us assume that the density matrix can be diagonalized via an SU(2) transformation (which is true for a broad class of axially symmetric states~\cite{Blum:1981rb}), so that it can be expressed as 
\begin{equation}
  \hat{\varrho}_{\mathrm{diag}}^{(1)}  = \mathrm{diag}(\lambda_{1}, \lambda_{2}, \lambda_{3})  =   \frac{1}{\sqrt{3}}  \hat{T}_{00}^{(1)} +
  \frac{\lambda_{1}-\lambda_{3}}{\sqrt{2}} \;  \hat{T}_{10}^{(1)} + 
  \frac{1-3\lambda_{2}}{\sqrt{6}} \; \hat{T}_{20}^{(1)} \, ,
  \label{eq:diag}
\end{equation}
where $\mathrm{diag} (d_{1}, \ldots, d_{n})$ represents a diagonal matrix  whose diagonal elements are those in the argument.  The state is first-order unpolarized when $\lambda_{1}
=\lambda_{3}$. Since $\Tr (\hat{\varrho}_{d}) = 1$, we can write
\begin{equation}
  \hat{\varrho}^{(1)}_{\mathrm{unpol},1} = 
  \mathrm{diag} ( \lambda, 1-2\lambda, \lambda )
  \label{eq:diag2}
\end{equation}
and positivity enforces $ 0 \leq \lambda \leq 1/2$; i.e., $0 \leq \mathcal{A}_2^{(1)} \leq 2/3$. Both the purity $\mathcal{P} = \Tr [ (\hat{\varrho}_{\mathrm{unpol},1}^{(1)})^{2} ]$ and the second-order degree $ \mathbb{P}_{2}$ depend on $\lambda$
\begin{equation}
  \mathcal{P}^{(1)} = 6 \lambda^{2} - 4 \lambda + 1 \, , 
  \qquad \qquad \qquad 
  \mathbb{P}_{2}^{(1)} = \sqrt{(3 \lambda -1)^2} \, ,
\end{equation}
while $\mathbb{P}_{1}^{(1)} =0$, as anticipated. This can be concisely recast
as 
\begin{equation} 
  \mathbb{P}_{2}^{(1)} = \sqrt{\tfrac{1}{2} [ 3 \mathcal{P}^{(1)} -1 ]} \, ,
\end{equation}
which is plotted in Fig.~\ref{fig:Pvspur}.  The maximum degree $\mathbb{P}_{2}^{(1)}$ is
attained for the pure states 
\begin{equation}
  \label{eq:pSon}
  |\Psi_{\mathrm{unpol,1}}^{(1)} \rangle =  \frac{1}{\sqrt{2}} 
  \sin \beta    \left ( e^{i\alpha} |1, 1 \rangle - e^{-i\alpha}
  |1,-1\rangle \right ) + \cos \beta  |1,0\rangle \, ,
\end{equation}
which are transformed versions of the state $|1,0 \rangle$ under SU(2)
transformations.  These states have served as the conduit to experimentally verify the existence of hidden polarization~\cite{Usachev:2001ve,Sehat:2005wd}; for a more detailed analysis of these states, the reader is referred to Ref.~\cite{Hoz:2014kq}.

\subsection{Majorana representation}
\label{sec:Majorana}

The SU(2) coherent states correspond as nearly as possible to a classical spin vector pointing in a particular direction. It is irresistible to ask which
pure states are, in a sense, \emph{the opposite} of SU(2) coherent states and therefore the most quantum ones.  This idea has been recently pursued~\cite{Zimba:2006fk,Crann:2010qd,Bannai:2011pi} as those states that \emph{point nowhere}; i.e., the average Stokes vector vanishes and the fluctuations up to order $M$ are isotropic: they have been dubbed \emph{anticoherent states}.

To investigate this point it is advantageous to use the Majorana stellar representation~\cite{Majorana:1932ul}, which allows us to uniquely depict a spin state state living in $\mathcal{H}_{S}$ by $2S$ points on the unit sphere~\cite{Bengtsson:2006aa}. Several decades after its conception, this representation has recently attracted a great deal of attention in several fields~\cite{Hannay:1998aa,Hannay:1998ab,Ribeiro:2007aa,Makela:2010aa,Lamacraft:2010aa,Bruno:2012aa,Lian:2012aa,Devi:2012aa,Cui:2013aa,Yang:2015aa,Liu:2016aa,Chryssomalakos:2018aa,Goldberg:2018aa,Chabaud:2020aa}.

The Majorana representation is a direct generalization of the Bloch sphere from spin-$1/2$ particles to spin-$S$ particles, making use of the coherent-state representation of the wave function $\Psi (\mathbf{n} ) = \langle \mathbf{n} | \Psi \rangle$ (to lighten notation, we will skip the label $S$, henceforth restricting to a fixed $S$). Once we insert the identity in terms of the states $|S,m\rangle $ and take into account the overlap $\langle \mathbf{n} | S, m\rangle $, the wavefunction can be written as 
\begin{equation}
  \Psi( \mathbf{n} )  =  \frac{1}{(1 + \lvert \zeta \rvert^{2})^{S}} 
  \sum_{m=-S}^S \sqrt{\frac{(2S)!}{(S-m)!(S+m)!}} \Psi_m \, \zeta^{S+m}\, , 
  \label{eq:Majpol}
\end{equation}
where $\zeta = \tan (\theta/2) e^{i \phi}$ is the stereographic projection mapping a point on the unit sphere with angle $(\theta, \phi)$ to the point $\zeta \in \mathbb{C}$.  Apart from an unessential positive factor, this wave function is a polynomial of order $2S$; thus, $| \Psi \rangle$ is determined by
the set $\{ \zeta_{i} \}$ of the $2S$ complex zeros of 
$\Psi ( \mathbf{n} )$~\cite{Bengtsson:2006aa}
\begin{equation}
\Psi ( \mathbf{n} ) = \frac{Z^{\ast}_{2S}}{(1 + | \zeta |^{2})^{S}} 
(\zeta - \zeta_{1}) \ldots (\zeta - \zeta_{2S})\, ,
\end{equation} 
with the resulting
set suitably completed by points at infinity if the degree of $\Psi( \mathbf{n} )$ is less than $2S$. Here $Z^{\ast}_{2S}$ is the final component of the vector of coefficients of $| \Psi \rangle $ in the angular momentum basis. The corresponding configuration of points on the unit sphere is called the Majorana constellation associated with the state $|\Psi\rangle$.  

There is a complementary way to look at the Majorana polynomial that helps us gain further insights.  Any pure state $|\Psi \rangle \in \mathcal{H}_{S}$ can be factorized in terms of the bosonic operators $\hat{a}_{\pm}$ as 
\begin{equation}
  \label{Eq: Majorana}
  | \Psi \rangle = \frac{1}{\sqrt{\cal{N}}} \prod_{m=1}^{2S}
  [    \cos ( {\theta_m}/{2} ) \hat{a}_+^\dagger +
    e^{i \phi_m}  \sin ( {\theta_m}/{2}  ) \hat{a}_-^\dagger ]
  |0_{+}, 0_{- } \rangle \, ,
\end{equation}
where $\cal{N}$ is a normalization factor, $|0_{+}, 0_{- } \rangle$ is
the two-mode vacuum and the angles $\theta_m$ and $\phi_m$ satisfy the natural constraints $0 \leq \theta_m \leq \pi$ and $0 \leq \phi_m < 2 \pi$. Thus, each factor in \eqref{Eq: Majorana} can be visualized as a point on the unit sphere. Since the operators $\hat{a}^{\dagger}_{+}$ and $\hat{a}^{\dagger}_{-}$ create an excitation in right- and left-hand circularly polarized modes, respectively, each of the factors in (\ref{Eq: Majorana}) can also be naively thought of as creating an \emph{excitation component} of a polarization state corresponding to its position on the sphere~\cite{Bengtsson:2006aa}.

An SU(2) rotation simply corresponds to a rigid rotation of the Majorana constellation and, consequently, it does not affect the degree of polarization: states with the same constellation, irrespective of their relative orientations, have the same polarization invariants. 

For SU(2) coherent states, the Majorana constellation collapses to a
single point on the unit sphere. Intuitively, one would guess that states
with the most isotropic polarization moments would have the most symmetric constellations
that are possible. We further develop this idea in the subsequent section.

\subsection{Extremal polarizaton states}

To formalize the idea of the most quantum states we will use the cumulative distribution $\mathcal{A}_{M}$. The idea is to identify which states minimize $\mathcal{A}^{(S)}_{M}$ for each order $M$. We shall be considering only pure states, which we expand as $|\Psi \rangle =\sum_{m=-S}^{S} \Psi_{m} \, |S,m\rangle$. We then have
that 
\begin{equation}
  \label{eq:AMS}
  \mathcal{A}_{M}  =\sum_{K=1}^{M} \sum_{q=-K}^{K}
  \frac{2K+1}{2S+1}
  \left |  \sum_{m,m^{\prime }=-S}^{S} C_{Sm,Kq}^{Sm^{\prime} }
    \Psi_{m^{\prime}} \Psi_{m}^{\ast } \right |^{2} \, .
\end{equation}
As we have discussed, SU(2) coherent states maximize $\mathcal{A}_{M}$ for all orders $M$.

\begin{figure*}
   \centerline{\includegraphics[width=0.90\columnwidth]{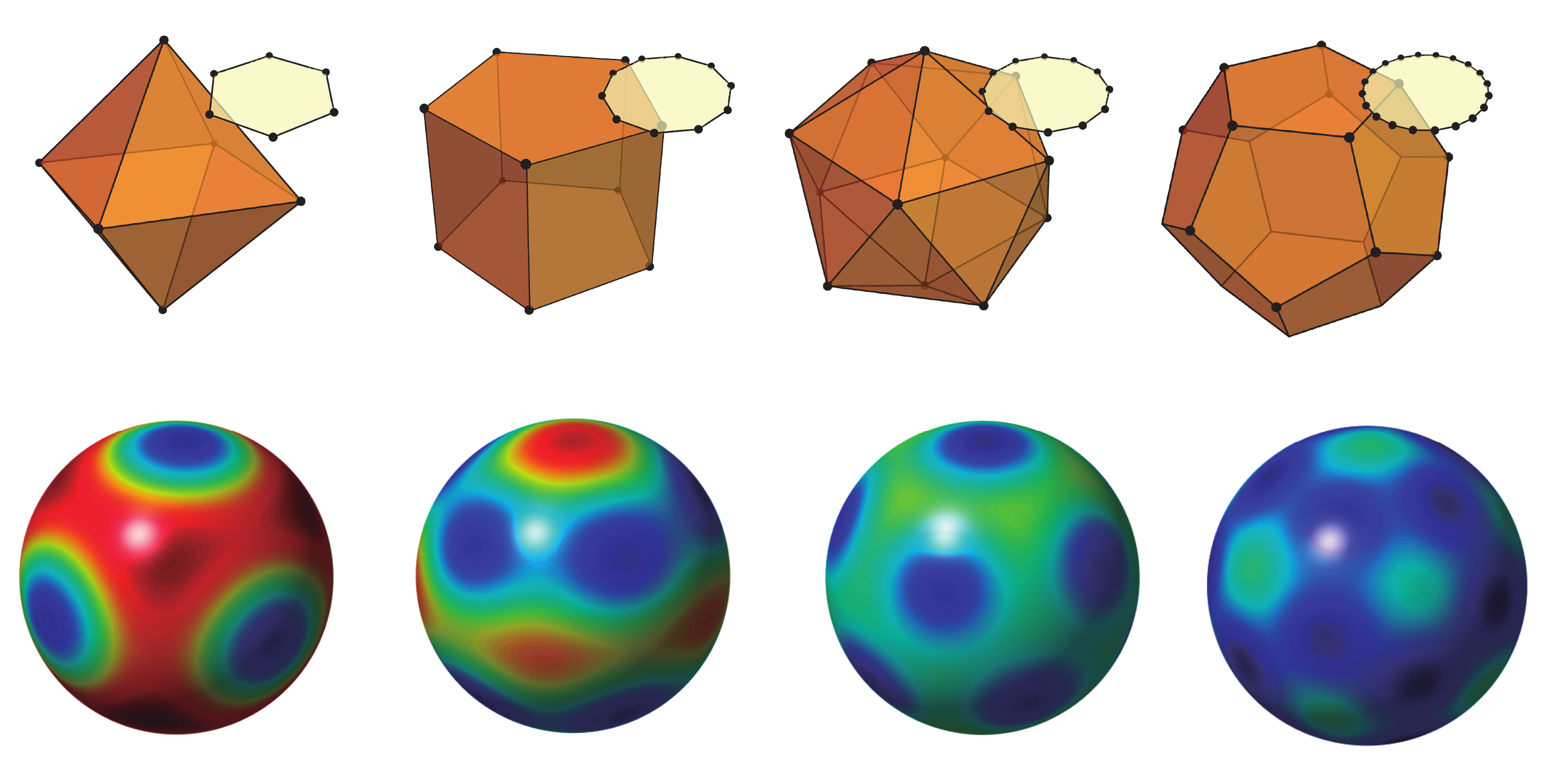}}
  \caption{Density plots of the SU(2) $Q$-functions for
    the optimal states in Table I for the cases
    $S = 5/2, 3, 7/2, 9/2, 5$, and $7$ (from left to right, blue
    indicates the zero values and red maximal ones). On top, we sketch
    the Majorana constellation for each of them.}
  \label{fig:Qfunc}
\end{figure*}

The strategy to minimize $\mathcal{A}_{M}$ is simple to state: starting from a set of unknown normalized state amplitudes in Eq.~(\ref{eq:AMS}), which we write as $\Psi_{m} = a_{m}+i b_{m}$ ($a_{m}, b_{m} \in \mathbb{R}$), we try to obtain $\mathcal{A}_{M} =0$ for the highest possible $M$.  This yields a system of polynomial equations of degree two for $a_{m}$ and $b_{m}$, which we solve using Gr{\"o}bner bases~\cite{Adams:1994ru} implemented in the computer algebra system {\sc magma}~\cite{Bosma:1997xp}. Since the orientation of the constellation is irrelevant, one can reduce the number of variables by fixing one of the points to be at, say, the north pole and another to lie in the $S_2$-$S_3$ plane. In this way, we get exact algebraic expressions and we can detect when there is no feasible solution. Table~\ref{table1} lists the resulting states (which, in some cases, are not unique) for selected values of $S$.  We also indicate the solutions' associated Majorana constellations. For completeness, in Fig.~\ref{fig:Qfunc} we plot the constellations as well as the $Q$-functions for some of these states.  A complete list can be found in \cite{Markus:2015pr} and a more detailed discussion can be found in \cite{Bjork:2015ab}. The resulting states systematize the notion of anticoherent states (of note is the fierce objection of Roy Glauber, the father of coherent states, to such a term) they have been called the Kings of Quantumness to avoid a disputable denomination.

\begin{table}[t]
  \caption{States for which $\mathcal{A}_{M}$ vanishes for the
    indicated values of $S$. In the second column, we indicate the
    order $M$, which we conjecture is the highest possible. We give 
    the nonzero state components $\Psi_{m}$ ($m=-S, \ldots, S$) and 
    describe the corresponding Majorana  constellation.}
  \label{table1}
  \bigskip

  \centering
  \begin{tabular}{ccll}
    \hline
    $S$ & $M$ &  State & Constellation \\
    \hline
    1 & 1 & $\Psi_{0} = 1$ & radial line \\
    $\frac{3}{2}$ & 1 & 
    $\Psi_{\scriptsize{\pm \frac{3}{2}}} = 
    \scriptsize{\frac{1}{\sqrt{2}}}$ &      
    equatorial triangle  \\
    2 & 2 & $\Psi_{-1} = \scriptsize{\frac{2}{\sqrt{3}}} , 
    \quad  \Psi_{2} = \scriptsize{\sqrt{\frac{1}{3}}} $ &      
    tetrahedron  \\
    $\frac{5}{2}$ & 1 & $\Psi_{\scriptsize{\pm\frac{3}{2}}} = 
    \scriptsize{\frac{1}{\sqrt{2}}}$ &  
    equatorial triangle  + poles \\
    3 & 3 & $\Psi_{\pm 2} = \scriptsize{\frac{1}{\sqrt{2}}}$  & 
    octahedron  \\
    $\frac{7}{2}$ & 2 & $ \Psi_{- \scriptsize{\frac{5}{2}}}=
    \Psi_{\scriptsize{\frac{1}{2}}} =   
    \scriptsize{\sqrt{\frac{7}{18}}},
    \quad  \Psi_{ \scriptsize{\frac{7}{2}}} = 
    \scriptsize{\sqrt{\frac{2}{9}}} $ &  two triangles + pole  \\
    4 & 3 & $\Psi_{\pm 4} = \scriptsize{\sqrt{\frac{5}{24}}}, \quad
    \Psi_{0} = \scriptsize{\sqrt{\frac{7}{12}}} $ & cube   \\
    $\frac{9}{2} $ & 2 & $\Psi_{\pm  \scriptsize{\frac{9}{2}}} =  
    \scriptsize{\frac{1}{\sqrt{6}}} ,  \quad 
     \Psi_{\pm  \scriptsize{\frac{3}{2}}} 
     = \scriptsize{\frac{1}{\sqrt{3}}} $    & three triangles  \\
    5 & 3 & $\Psi_{\pm 5}=  \scriptsize{\frac{1}{\sqrt{5}}} ,
            \quad  \Psi_{0} =  \scriptsize{\frac{3}{\sqrt{5}}}$ & 
      pentagonal prism  \\
    $\frac{11}{2} $ & 3 & 
    $\Psi_{\pm \scriptsize{\frac{11}{2}}} =  
    \scriptsize{\frac{\sqrt{17}}{12}}  ,
     \quad  \Psi_{\pm \scriptsize{\frac{5}{2}}} =
    i \scriptsize{\frac{\sqrt{55}}{12}}$  & 
    pentagon + two triangles \\ 
    6 & 5 &$\Psi_{\pm 5}= \pm \scriptsize{\frac{\sqrt{7}}{5}}, 
    \quad \Psi_{0} =-  \scriptsize{\frac{\sqrt{11}}{5}} $ & 
   icosahedron   \\
   7 & 4 & $\Psi_{\pm 6}=   \scriptsize{\sqrt{\frac{854}{3645}}},  \quad
   \Psi_{\pm 3} =   \scriptsize{\sqrt{\frac{637}{13420}} + i 
  \sqrt{\frac{512603}{9783180}}} $ & 
  three squares + poles \\
        &  &  $\Psi_{0} = \scriptsize{\sqrt{\frac{12561757}{163053000}}} -
   \scriptsize{i \sqrt{\frac{512603}{2013000}}}$ &  \\
    10 & 5 & $\Psi_{\pm 10}= \scriptsize{\sqrt{\frac{187}{1875}}} , \quad 
  \Psi_{\pm 5}=  \pm \scriptsize{\sqrt{\frac{209}{625}}} ,  \quad 
  \Psi_{0} =\scriptsize{\sqrt{\frac{247}{1875}}}$ & 
  deformed dodecahedron 
  \end{tabular}
\end{table}

From a physical perspective one would expect these constellations to have their points distributed as symmetrically as possible on the unit sphere.   For some values of $S$, such as 4, 6, 8, 12 and 20, one can guess a maximally unpolarized constellation, in each case corresponding to the vertices of a Platonic solid. For other numbers it is not easy to guess an optimal, exact constellation, but solving the system of polynomial equations, as described before, yields exact algebraic expressions for the coefficients $\Psi_{m}$, from which one can easily compute the points of the Majorana constellation with arbitrary numerical precision.

The problem of distributing $N$ points on a sphere in the \emph{most symmetric} fashion has a long history and has many different solutions depending on the cost function one tries to optimize~\cite{Saff:1997aa,Brauchart:2015aa}. Here, we shall only discuss a few of the formulations: spherical $t$-designs~\cite{Delsarte:1977dn}, the Thomson problem~\cite{Thomson:1904qp,Ashby:1986bk,   Edmundson:1992uf,Melnyk:1977gm} and the Queens of quantumness~\cite{Giraud:2010db}. 

Spherical $t$-designs are configurations of $N$ points on a sphere such that the average value of any polynomial of degree at most $t$ is the same when evaluated over the $N$ points as over the entire sphere.  Thus, the $N$ points can be seen to give a representative average value of all polynomials up to degree $t$. It has been conjectured that a state is unpolarized to order $t$ if and only if its Majorana constellation is a spherical $t$-design~\cite{Crann:2010qd}. However, while the statement is true for some $t$-designs, such as those represented by the Platonic solids, the general conjecture has been disproven~\cite{Bannai:2011pi}.

It is clear that there should be some connection between the number of points $N$ and the maximal degree $t$ for which  a spherical $t$-design exists. The configurations that maximize $t$ for a given $N$ are called optimal designs, and we use $t$ to denote the degree of an optimal $N$-point design in the following. No analytical relationship is known between $N$ and $t$: it is known that for a $t$-design, the number of points $N$ is at least proportional to $t^2$; whereas, for some orders $t$ the only known constructions have $N$ scaling proportionally with $t^3$.  As a function of $N$, the order $t$ is non-monotonic. The results for  $1 \leq N \leq 100$ are summarized in~\cite{Hardin:1996bv}.

From the numerical data thus far one can conclude that the maximum value of $M$and $t$ coincide.  We therefore conjecture that if an optimal spherical design of order $t$ exists for some $N$, then one can find an $M$th-order unpolarized $N$-photon state with $M=t$. 

The next thing one can note is that an optimal $t$-design does not necessarily give a $t$th-order unpolarized state. Quite often the configurations are similar, e.g. regular polygons with their surface normals along the polar axis, but displaced from each other along the axis by certain distances. However, these distances often need to be fine-tuned for an optimal $t$-design to become a maximally-unpolarized state.  The Platonic solids are exceptions to this observation. That  the optimal configurations for $t$-designs and maximally unpolarized states do not always coincide underscores the mystery that the optimal $t$ and maximal $M$ always seem to be equal for any $N$.

The Thomson problem consists of arranging $N$ identical point charges
on the surface of a sphere so that the electrostatic potential energy
of the entire configuration is minimized.  The problem can be generalized to potential energies that are proportional to $r^{-d}$, where $r$ is the Euclidean distance between the charges. The choice $d=1$ is the Thomson problem, corresponding to the usual Coulomb potential; the case $d \rightarrow \infty$ is called Tammes problem~\cite{Tammes:1930rc}.  For small $S$, up to 3, the Thomson  configurations are identical to the  optimal spherical $t$-design and to the maximally unpolarized states. For larger $S$, they differ in general and the degree of unpolarization of the Thomson states is lower than the maximum. Different from the two previous cases, the solution to the Thomson problem appears to be unique for every $S$~\cite{Erber:1991lq}.

The Queens of quantumness are the states that maximize the Hilbert-Schmidt distance to the closest point of the convex hull of the mixed SU(2) coherent states~\cite{Giraud:2010db}. This convex hull defines the subspace of classical states. Therefore, the states maximizing the distance to the nearest point on this hull can be thought of as having maximally quantum characteristics. In Ref.~\cite{Giraud:2010db} it is claimed that the Queens can be seen as the least classical (or most quantum) of all states given this metric. Although we have used another figure of merit, our approach and that in \cite{Giraud:2010db} share the view that the states \emph{most different} from SU(2) coherent states are the most quantum.

When we interpret our subspace $\mathcal{H}_{S}$ as the symmetric subspace of a system of $S$ qubits, the Kings seem to also be closely linked to other intriguing problems, such as maximally entangled symmetric states~\cite{Aulbach:2010jw,Giraud:2015oj} and $k$-maximally mixed states~\cite{Arnaud:2013hm,Goyeneche:2014so}.

\subsection{Metrological applications}

The main goal of quantum metrology is to measure a physical magnitude
with surprising precision by exploiting quantum resources~\cite{Giovannetti:2011aa}. In particular, tailoring polarization states to better detect SU(2) rotations is quite a relevant problem with direct applications to
magnetometry~\cite{Wasilewski:2010aa,Sewell:2012aa,Muessel:2014aa}, polarimetry~\cite{Meyer:2001aa,DAmbrosio:2013aa}, and metrology in general~\cite{Rozema:2014aa}.

The salient feature of maximally unpolarized states is their ability to detect small, but arbitrary, SU(2) transformations with optimal resolution. This was already anticipated in~\cite{Kolenderski:2008mo}, where the authors specifically found that, for photon numbers 4, 6, 8, 12 and 20, the states corresponding to regular polyhedron Majorana constellations best signaled misalignments between two Cartesian reference frames.  To understand this, it is instructive to look at related states, namely the NOON states. Such NOON states are known to have the highest sensitivity to small rotations about the $\hat{S}_3$-axis among all states with a fixed excitation $S$ ~\cite{Bollinger:1996am}. This can be easily understood by looking at their Majorana constellation, which is apparent from Fig.~\ref{fig:cohNoon} and consists of $2S$ equidistantly placed points around the equator. A rotation around the $\hat{S}_3$-axis is described by the unitary operator $\hat{U}(\phi)=\exp(- i \phi \hat{S}_3)$; therefore, for $\phi=\pi/(2 S)$ the states $|\mathrm{NOON}\rangle$ and $\hat{U}(\phi )|\mathrm{NOON}\rangle$ are orthogonal, while for $\vartheta= q\pi/S$ they are parallel, where $q$ is an integer. Thus, it should not come as a surprise that NOON states are optimal for detecting small rotations around the $\hat{S}_3$-axis, in the interval $0 \leq \vartheta \leq \pi/(2 S)$. If the rotation axis lies in the equatorial plane, then a rotation of $\pi$ is needed to obtain a parallel state, irrespective of $S$. When $S$ is a half integer this only happens when the axis intersects one of the Majorana points, and when $S$ is an integer this happens if the axis either intersects either a point or is the intersector between two points. Thus, the rotation resolution is
highly directional for a NOON state.

This is precisely the advantage of maximally unpolarized states: having a high degree of spherical symmetry, they resolve rotations around any axis approximately equally well. It may not be obvious from their appearance that they have high sensitivity to small rotations about an arbitrary axis. To substantiate this claim, recall that the action $\tau$ needed to make a state $|\Psi\rangle$ evolve so that $|\langle\Psi| \exp(i \hat{A} \tau) |\Psi\rangle|^2 = 1-\epsilon$, where $\epsilon$ is a small, positive, real number, and $\hat{A}$ is Hermitian, is inversely proportional to the variance $\var{\hat{A}}$~\cite{Mandelshtam:1945aa}. The relation connecting the evolution speed $d \epsilon/ d \tau$ and the variance is sometimes called the quantum speed limit~\cite{Taddei:2013aa,Campo:2013aa}. A NOON state in the $\hat{S}_3$ basis has maximal variance $\langle \var{\hat{S}_3} \rangle=S^2$ for a fixed $S$ and is thus the state with maximal sensitivity to a rotation around the $\hat{S}_3$ axis. However, the variances $\var{\hat{S}_1}$ and $\var{\hat{S}_2}$ are only $S/2$ and thus the state is rather insensitive to rotations around those axes (or around any rotation axis in the $\hat{S}_1$-$\hat{S}_2$ plane). In contrast, all the Kings of Quantumness have isotropic variances equal to $S(S+1)/3$; namely, close to the maximum possible variance.  The quantum speed limit theorem thus asserts that, having a large and isotropic variance of the Stokes operator, these states are rather sensitive to rotations around any axis $\hat{S}_{{\bf n}}$~\cite{Chryssomalakos:2017aa,Martin:2020aa}.

Another way of explaIning the sensitivity to a rotation around an arbitrary axis is to observe that, since these states have maximal spherical symmetry, they become parallel, or almost parallel, for relatively small rotations around several axes. To quantify this statement one could use the Fisher information and the Cram\'{e}r-Rao bound to assess the uncertainty in estimating the rotation direction and the rotation angle~\cite{Taddei:2013aa,Campo:2013aa}.

\section{Conclusions}
\label{sec:conclusions}

Quantum polarization provides an excellent landscape to flaunt nonclassical  features. This is because the polarization degree of freedom is easily accessible both theoretically and experimentally. Moreover, we have shown that this variable finds parallels with many different topics of quantum physics.

As polarization is the manifestation of photon spin, it in turn contributes, together with the orbital angular momentum (OAM), to the total angular momentum of light. It should come as no surprise that the methods developed for polarization characterization lend themselves, virtually without modification, to the analysis of OAM. In fact, since it is at present experimentally easier to generate specific OAM states than the corresponding polarization states (which are, in general, highly entangled), some of the theory developed for polarization has actually been tested with OAM as the model system. 

Light has been an excellent laboratory for research in quantum theory. We hope that the results summarized in this contribution show how this is particularly true  for light polarization. 

\noindent \textbf{Acknowledgments.} Over the years, the ideas in this paper have been further developed and expanded with questions, suggestions, criticism, and advice from many colleagues. Particular thanks for help in various ways goes to G. S. Agarwal, F. Bouchard, R. W. Boyd, A. S. Chirkin, J. H. Eberly, A. T. Friberg, R. J. Glauber, D. F. V. James, V. P. Karassiov, Y. H. Kim, E. Karimi, N. Korolkova, Ch. Marquardt, C. M\"{u}ller, A. Normann, \L. Rudnicki, Ch. Silberhorn, W. P. Schleich, and J. S\"{o}derholm. 

\noindent \textbf{Funding.} 
Ministerio de Ciencia e Innovaci\'on (PGC2018-099183-B-I00);
Consejo Nacional de Ciencia y Tecnologia de Mexico (254127);
Fundacja na rzecz Nauki Polskiej (2018/MAB/5);
Canada National Sciences and Engineering Research Council;
Walter C. Sumner Foundation; 
Lachlan Gilchrist Fellowship Fund; 
Michael Smith Foreign Study Supplement; 
Mitacs Globalink;
Ministry of Education and Science of the Russian Federation (Mega Grant 14.W03.31.0032).

\noindent\textbf{Disclosures.} 
The authors declare no conflicts of interest.
\appendix

\section{Coherent states}
\label{sec:cs}

The coherent-state approach is not just a convenient mathematical tool, but it also helps one to understand how physical properties of a system are reflected by the geometrical structure of the related phase space. The reader interested can check more details in the pertinent monographs~\cite{Gazeau:2009aa,Perelomov:1986ly}

Let $G$ be a Lie group (connected and simply connected, with finite dimension $n$), which is the dynamical group of a given quantum system. Let $T$ be a unitary irreducible representation (irrep) of $G$ acting on the Hilbert space $\mathcal{H}$: 
\begin{equation}
  |\Psi_g\rangle= \hat{T} (g)| \Psi_0\rangle \, , \qquad \qquad 
  g \in G\, ,
  \label{eq:defCS}
\end{equation}
where $|\Psi_0\rangle$ is a fixed vector in $\mathcal{H}$ of the representation $\hat{T}(g)$.  The subgroup $H \in G$ such that
\begin{equation}
  \hat{T} (h)| \Psi_0\rangle = e^{i \alpha (h)} | \Psi_{0} \rangle \, , 
  \qquad \qquad \forall h \in H\, ,
  \label{eq:defIso}
\end{equation}
is called the isotropy group (or little group) of the state $| \Psi_{0} \rangle$. Since the states of a quantum system are defined up to multiplication by a global phase, any element of the orbit $ | \Psi_{g} \rangle$ of the state $| \Psi_{0} \rangle$ can be put in correspondence with an element of the coset space $X = G/H$.  We  thus define the set of coherent states for the group $G$ with respect to the fiducial vector $| \Psi_{0} \rangle $ by 
\begin{equation}
| \Omega \rangle = \hat{T} (\Omega) | \Psi_{0} \rangle \, ,
\end{equation}
where $\Omega \in X$; that is,  $g=\Omega h$.

An essential property is that the coherent states form an (overcomplete) basis in the state space of the system
\begin{equation}
  \int_X d\mu (\Omega) \, 
  |\Omega \rangle \langle\Omega| = \hat{\openone}
  \label{eq:resolId} \, ,
\end{equation}
where $d\mu(\Omega)$ is the invariant integration measure on $X$. 

In principle, coherent states can be generated from any state $| \Psi_{0} \rangle$. However,  the states $| \Psi_{0} \rangle$  having the largest isotropy groups generate coherent states closest to the classical states in the sense that they minimize the uncertainty relations. 

An important class of coherent states corresponds to the coset spaces
$X$ that are homogeneous K\"{a}hler manifolds. Then, a natural symplectic structure can be introduced on $X$, so that it can be considered the phase space of a classical dynamical system~\cite{Klimov:2009aa}.  

Let us first consider the canonical example of the Heisenberg-Weyl group $W_{1}$, which is the dynamical symmetry group for a mode of the quantized radiation field. The associated Lie algebra $\mathfrak{w}_{1}$ is defined by the canonical commutation relations 
\begin{equation} 
[\hat{a}, \hat{a}^\dagger]=\hat{\openone} \, .
  \label{eq:HW1}
\end{equation}
A general element of this algebra has the form $t \hat{\openone} + i  (\alpha^{\ast} \hat{a} - \alpha \hat{a}^\dagger)$, where $t \in \mathbb{R}$  and $\alpha \in \mathbb{C}$. Therefore, the elements of the group are obtained  via the exponential map  
\begin{equation}
  \hat{T}(t,\alpha)= e^{it} \hat{D}(\alpha) \, , 
  \qquad \qquad 
  \hat{D}(\alpha)=\exp (\alpha \hat{a}^\dagger-\alpha^*\hat{a} )
  \label{eq:disp} \, .
\end{equation}
An element  of this group is specified by $g= (t, \alpha)$ and $\alpha\in \mathbb{C}$ and the operators $\hat{T}(g)= \hat{T}(t,\alpha)$ are an irrep of the Hilbert space of Fock states $\{ | n\rangle \}$. The phase space is the complex plane $\mathbb{C} = W_1/U(1)$, and the (quadrature) coherent states are:
\begin{equation}
  |\alpha\rangle= \hat{D} (\alpha) | 0 \rangle,  
  \label{eq:CScan}
\end{equation} 
with the vacuum $|0 \rangle$ being the fiducial state.  

The expansion in the Fock basis can be easily obtained by disentangling the displacement operator $\hat{D} (\alpha)$; the result is  
\begin{equation}
  |\alpha\rangle= e^{-|\alpha|^2/2 } 
  \sum_{n=0}^{\infty} \frac{\alpha^n}{\sqrt{n!}}|n\rangle
  \label{eq:CSStat} \, ,
\end{equation} 
which exhibits the well-known Poissonian statistics of a radiation field with average value $\bar{N} = \lvert \alpha \rvert^{2}$. 

If we consider the transformation of the operators of the algebra $\mathfrak{w}_1$ by the elements of the group $W_1$, it is not difficult to show that:
\begin{equation}
  \label{a1}
   \hat{D}(\alpha) \, \hat{a} \, \hat{D}^\dagger(\alpha) = 
   \hat{a} - \alpha \; \hat{\openone} \, ,
\end{equation}
and by applying \eqref{a1} to the coherent states $|\alpha\rangle$ we see 
\begin{equation}
  \hat{a} | \alpha \rangle= \alpha | \alpha\rangle
  \label{eq:nose} ,
\end{equation}
which is another equivalent definition of the coherent states. 

The resolution of unity in terms of $|\alpha \rangle $ is 
\begin{equation}
  \int_{\mathbb{C}}  d\mu (\alpha ) \; 
  |\alpha\rangle\langle\alpha| =\hat{\openone}, 
  \qquad \qquad 
  d\mu(\alpha)=\frac{1}{\pi}\, d\alpha \, ,
  \label{eq:eu} 
\end{equation}
and, as a consequence, we can expand an arbitrary state using \eqref{eq:eu}:
\begin{equation}
  | \Psi \rangle= \int_{\mathbb{C}} d\mu (\alpha) \; 
  \Psi (\alpha^{\ast}) | \alpha\rangle, 
  \qquad \qquad 
  \Psi (\alpha^{\ast}) = \langle\alpha | \Psi \rangle \, .
  \label{eq:GSexp} 
\end{equation}
The function $\Psi (\alpha^{\ast})$ is the coherent-state wave function of the state $|\Psi\rangle$.

The coherent states saturate the Heisenberg uncertainty relation
\begin{equation}
  \var_{\alpha}{\hat{x}} \; \var_{\alpha}{\hat{p}} \geq 1/2
  \label{eq:MUS}
\end{equation}
where the canonical position-momentum operators are implicitly defined by $\hat{a} = (\hat{x} + i \hat{p})/\sqrt{2}$. For this reason, coherent states are often considered the \emph{most classical} states of the harmonic oscillator.  

Next, we consider the SU(2) symmetry, whose Lie algebra $\mathfrak{su}(2)$ is spanned by the operators $\{\hat{S}_1,\hat{S}_2,\hat{S}_3\}$, with commutation relations
\begin{equation} 
[\hat{S}_{k}, \hat{S}_{\ell} ]=i \epsilon_{klm} \hat{S}_{m}  \, .
  \label{eq:SU2CR}
\end{equation}
The irreps are labeled by the index $S$ $(S=0,\tfrac{1}{2},1,\dots)$ and the  carrier space $\mathcal{H}_S$ is spanned by the standard  basis $\{ |S,m\rangle \mid m=S, \dots,-S) \}$.  The highest weight state is $|S,S\rangle$ and is annihilated by the ladder operator $\hat{S}_{+}$. The isotropy subgroup for this state consists of all the elements of the form $\exp( i \psi \hat{S}_{3})$, so it is isomorphic to U(1). The coset space is then SU(2)/U(1), which is the unit sphere $\mathcal{S}_{2}$, and it is exactly the same as the classical phase space, the natural arena to describe the dynamics.

Since the irreps can be written in terms of the Euler angles as 
\begin{equation}
\hat{T}^{(S)} (\phi, \theta, \psi) = 
\exp(-i\phi \hat{S}_3)\exp(-i\theta \hat{S}_2)\exp(-i\psi \hat{S}_3) \,, 
\end{equation} 
it is clear that the elements of SU(2)/U(1) can be represented as in \eqref{eq:dissph}; i.e.,
\begin{equation}
\hat{D}(\mathbf{n}) =  
e^{- i \phi \hat{S}_{3}} e^{- i \theta \hat{S}_{2}}
= \exp \left [ \tfrac{1}{2} \theta (\hat{S}_{+} e^{-i \phi} - 
\hat{S}_{-} e^{i \phi} )\right] \, ,
\end{equation}
so they depend on the spherical angles $(\theta, \phi)$ determining the unit vector $\mathbf{n}$.  The system of coherent states is thus
\begin{equation}
  |S,\mathbf{n} \rangle= \hat{D} (\mathbf{n}) |S,S \rangle \, .
  \label{eq:CSSU2} 
\end{equation}

One can write the operator $\hat{D}(\mathbf{n} )$ in a disentangled form to obtain  another useful parametrization of the SU(2) coherent states:
\begin{equation}
  |S, \zeta \rangle = (1 + | \zeta|^{2})^{-S} 
  \exp( \zeta\hat{S}_{+} ) |S, -S \rangle \, ,
\end{equation}
with $\zeta =  \tan (\theta/2) e^{i \phi}$ being the stereographic projection from the north pole of the sphere $\mathcal{S}^2$ onto the complex plane $\mathbb{C}$, tangent to the south pole.

Expanding the exponential, we obtain the  expression for the SU(2)
coherent states in terms of the angular momentum basis $|S,m\rangle$
\begin{equation}
  |S, \mathbf{n} \rangle = \sum_{m=-S}^S  \binom{2S}{S+m}^{1/2} 
  [ \sin ( \theta / 2 )  ]^{S-m} \, 
  [ \cos ( \theta/2)]^{S+m} \, \exp [ - i  ( S+m ) \phi] |S,m\rangle \, . 
\end{equation}
SU(2) coherent states are not orthogonal; their overlap is 
\begin{equation}
\lvert \langle \mathbf{n}_1 | \mathbf{n}_2 \rangle \rvert^{2} =
\left [ 
    \tfrac{1}{2} (1 + \mathbf{n}_{1} \cdot \mathbf{n}_{2} ) 
  \right ]^{2S} \, ,
\end{equation}
which has been used in \eqref{eq:QCS}. 

The resolution of the identity now reads
\begin{equation}
  \int_{\mathcal{S}_{2}} d\mu ( \mathbf{n}) \, 
 |S,  \mathbf{n} \rangle \langle S, \mathbf{n} | = 
 \hat{\openone}, 
 \qquad \qquad d\mu (\mathbf{n} ) = \frac{2S+1}{4\pi} 
 \sin\theta d\theta d\phi \, ,
  \label{eq:SU2resun} 
\end{equation}
which allows for an expansion 
\begin{equation}
  |\Psi\rangle = \int d\mu(\mathbf{n}) \; \Psi(\mathbf{n} ) |S, \mathbf{n} \rangle \, ,
  \label{eq:SU2CSExp} 
\end{equation}
where $\Psi (\mathbf{n} )$ is the coherent-state wave function. 

Finally, as discussed in Section~\ref{sec:uncrel}, the SU(2) coherent  states minimize the Heisenberg uncertainty relation.

\section{Quasiprobability distributions on the sphere}
\label{app:phsp}

Presenting quantum mechanics as a statistical theory on a classical phase space has attracted a great deal of attention since the very early days of this discipline. 

The main ingredient for any successful phase-space method is a \emph{bona fide} mapping that relates operators with functions defined on a smooth manifold $\mathcal{M}$, the phase space of the system, endowed with a very precise mathematical structure~\cite{Kirillov:2004aa}. This mapping, first suggested by Weyl~\cite{Weyl:1928aa}, is not unique. In fact, a whole family of $s$-parametrized functions can be assigned to each operator and the choice of a particular element of the family depends on  its convenience for each problem. In particular, the time-honored quasiprobability distributions are the functions connected with the density operator. The most common choices of $s$ are $+1$, 0, and $-1$, which correspond to the $P$ (Glauber-Sudarshan)~\cite{Glauber:1963aa,Sudarshan:1963aa}, $W$~(Wigner)~\cite{Wigner:1932uq}, and $Q$~(Husimi)~\cite{Husimi:1940aa} functions, respectively.

For the relevant case of the SU(2) dynamical symmetry, the $s$-parametrized Weyl-Stratanovich map
\begin{equation}
  \hat{A}  \mapsto  W_{A}^{(s)} (\mathbf{n} ) =
  \Tr [  \hat{A} \, \hat{w}^{(s)}(\mathbf{n} ) ]  ,
  \label{map}
\end{equation}
puts in one-to-one correspondence each operator $\hat{A}$ invariantly acting on $\mathcal{H}_{S}$ with a function on the sphere $\mathcal{S}_{2}$. The corresponding kernels  $\hat{w}^{(s)}$ are defined as~\cite{Berezin:1975mw,Agarwal:1981bd,Varilly:1989ud,Klimov:2002cr}   
\begin{equation}
  \hat{w}^{(s)}(\mathbf{n} )= \sqrt{\frac{4\pi}{2S+1}}
  \sum_{K=0}^{2S} \sum_{q=-K}^{K}
  ( C_{SS,K0}^{SS} )^{-s} \; Y_{Kq}^{\ast}( \mathbf{n} ) \,
  \hat{T}_{Kq}^{S}\,,
  \label{ker}
\end{equation}
where $Y_{Kq}(\mathbf{n})$ are the spherical harmonics, $C_{S_{1}m_{1},S_{2}m_{2}}^{Sm}$ the Clebsch-Gordan coefficients and $\hat{T}_{Kq}^{S}$ the irreducible tensor operators
\begin{equation}
  \hat{T}_{Kq}^{S} = \sqrt{\frac{2K+1}{2S+1}}
  \sum_{m,m^{\prime}=-S}^{S} C_{Sm,Kq}^{Sm^{\prime}} \,
  |S,m^{\prime}\rangle \langle S,m| \, .
\end{equation}
As expected, the kernels are properly normalized 
\begin{equation}
  \Tr [ \hat{w}^{(s)} ( \mathbf{n} ) ] = 1 \, ,
  \qquad
  \frac{2S+1} {4\pi}\int_{\mathcal{S}_{2}}
  d\mathbf{n}\; \hat{w}^{(s)} (\mathbf{n})=\openone \,,
\end{equation}
with $d\mathbf{n} =\sin \theta\ d\theta\ d\phi $ the invariant measure on the sphere.

Consequently, the symbol of $\hat{A}$ can be concisely  expressed as
\begin{equation}
  W_{A}^{(s)} ( \mathbf{n} ) = \sqrt{\frac{4\pi}{2S+1}}
  \sum_{K=0}^{2S} \sum_{q=-K}^{K}  ( C_{SS,K0}^{SS} )^{-s}
  A_{Kq} \; Y_{LM}^{\ast} (\mathbf{n} ) \, ,
\label{eq:symb}
\end{equation}
where $A_{Kq}=\Tr ( \hat{A} \hat{T}_{Kq}^{S \dagger} )$. 

The traditional SU(2) quasiprobability distributions are simply the $s$-symbols of the density operator $\hat{\varrho}$. The value $s=0$ corresponds to the standard Wigner function, while $s=\pm 1$ leads to the $P$ and $Q$-functions respectively, defined as dual coefficients in the basis of spin coherent states $|S, \mathbf{n} \rangle$~\cite{Perelomov:1986ly}, according to
\begin{equation}
  Q(\mathbf{n} ) = \langle \mathbf{n} |\hat{\varrho}|\mathbf{n} \rangle \, ,
  \qquad \qquad
    \hat{\varrho} =\frac{2S+1}{4\pi}\int_{\mathcal{S}_{2}}d\mathbf{n} \;
  P(\mathbf{n}) \; |\mathbf{n}\rangle \langle \mathbf{n} | \, .
\end{equation}
The symbols $W_{A}^{(s)} (\mathbf{n} )$ are covariant under SU(2) transformations and provide the overlap relation
\begin{equation}
  \Tr (\hat{\varrho} \hat{A}) = \frac{2S+1}{4\pi}
  \int_{\mathcal{S}_{2}} d\mathbf{n} \, W_{\varrho}^{(s)}(\mathbf{n} )
  \,W_{A}^{(-s)}(\mathbf{n}) \, .
\end{equation}



\end{document}